\documentclass[12pt]{article}
\usepackage{amsmath,amssymb,epsfig}
\makeatletter \@addtoreset{equation}{section} \makeatother
\renewcommand{\theequation}{\thesection.\arabic{equation}}
\newcommand{\ba}{\begin{array}}
\newcommand{\ea}{\end{array}}
\newcommand{\beq}{\begin{equation}}
\newcommand{\eeq}{\end{equation}}
\newcommand{\bea}{\begin{eqnarray}}
\newcommand{\eea}{\end{eqnarray}}

\def\bce{\begin{center}}
\def\ece{\end{center}}

\def\nonu{\nonumber}

\def\be{\beta}

\def\eps6{{\displaystyle \mathop{\epsilon}^{6}}{}}

\def\nab6{{\displaystyle \mathop{\nabla}^{6}}{}}

\def\0{{\sst{(0)}}}
\def\1{{\sst{(1)}}}
\def\2{{\sst{(2)}}}
\def\3{{\sst{(3)}}}
\def\4{{\sst{(4)}}}
\def\5{{\sst{(5)}}}
\def\6{{\sst{(6)}}}
\def\7{{\sst{(7)}}}
\def\8{{\sst{(8)}}}

\def\ba{\begin{array}}
\def\ea{\end{array}}
\def\beq{\begin{equation}}
\def\eeq{\end{equation}}
\def\be{\begin{equation}}
\def\ee{\end{equation}}

\def\eps{\epsilon}

\def\ba{\begin{array}}
\def\ea{\end{array}}
\def\beq{\begin{equation}}
\def\eeq{\end{equation}}
\def\be{\begin{equation}}
\def\ee{\end{equation}}

\def\eps{\epsilon}

\newcommand{\bean}{\begin{eqnarray*}}
\newcommand{\eean}{\end{eqnarray*}}

\begin{document}
\thispagestyle{empty} \addtocounter{page}{-1}
\begin{flushright}
{\tt arXiv:yymm.nnnn}\\
\end{flushright}

\vspace*{1.3cm}

\centerline{ \Large \bf Meta-Stable Brane Configurations of   }
\vspace{.3cm} 
\centerline{ \Large \bf   Multiple Product
  Gauge Groups with Orientifold 6 Plane} 
\vspace*{1.5cm}
\centerline{{\bf Changhyun Ahn} 
} 
\vspace*{1.0cm} 
\centerline{\it 
Department of Physics, Kyungpook National University, Taegu
702-701, Korea} 
\vspace*{0.8cm} 
\centerline{\tt ahn@knu.ac.kr} 
\vskip2cm

\centerline{\bf Abstract}
\vspace*{0.5cm}

Starting from an ${\cal N}=1$ supersymmetric electric gauge theory 
with the gauge group $SU(N_c) \times SU(N_c') \times SU(N_c'')$ 
with fundamentals for each 
gauge group, the bifundamentals, a symmetric flavor and a conjugate
symmetric flavor for $SU(N_c)$, we apply Seiberg dual 
to each gauge group, obtain the ${\cal N}=1$
supersymmetric 
dual magnetic gauge theories with dual matters including the
gauge singlets, and 
describe the intersecting brane configurations of
type 
IIA string theory corresponding to the meta-stable nonsupersymmetric 
vacua of this gauge theory.
We also discuss the case where a symmetric flavor is replaced by an
antisymmetric flavor.
Next we apply to 
the case for ${\cal N}=1$ supersymmetric electric gauge theory 
with the gauge group $SO(N_c) \times SU(N_c') \times SU(N_c'')$ 
with flavors for each 
gauge group and the bifundamentals. 
Finally, we describe the case where the orientifold 
6-plane charge is reversed.

\baselineskip=18pt
\newpage
\renewcommand{\theequation}
{\arabic{section}\mbox{.}\arabic{equation}}

\section{Introduction}

The nonsupersymmetric meta-stable vacua exist in ${\cal N}=1$ super QCD
with massive fundamental quarks where 
the classical flat directions can be 
lifted by quantum corrections which generate positive mass terms for the 
pseudomoduli \cite{ISS}.   
See also the review paper \cite{IS} for 
the recent developments of dynamical supersymmetry breaking.
Turning on the quark masses in the 
electric theory corresponds to deform the magnetic superpotential by adding 
a linear term of a singlet field in the dual magnetic theory. 
The misalignment of D4-branes(01236) 
connecting NS5'-brane(012389) in the type IIA string theory, corresponding to
 magnetic theory, can be 
analyzed as the nontrivial F-term conditions
providing nonzero vacuum expectation values of
dual quarks and some of D4-branes connecting NS5'-brane can move 
in other two directions freely. 
See
the review paper \cite{GK98} for the gauge theory and the brane dynamics.

When an orientifold 6-plane(0123789) is added into the standard brane
configuration for a single unitary gauge group, consisting of 
a NS5-brane(012345), a NS5'-brane, D4-branes and D6-branes(0123789),   
there exist two possible brane configurations. Either NS-brane 
overlaps with an orientifold 6-plane at $x^6=0$ or there is no overlap between
NS-brane and an orientifold 6-plane at $x^6=0$.
The former has three NS-branes and the gauge theory is 
described by \cite{ILS,LL,LLL,Ahn07} or 
\cite{LLL1,BHKL,EGKT,Ahn07-1}
depending on whether the NS5-brane overlaps with O6-plane or the
NS5'-brane overlaps with O6-plane. The latter has two NS-branes and
the gauge theory and corresponding brane configuration are studied by \cite{CSST}. 

As an orientifold 6-plane is added into the standard brane
configuration for the product of two unitary gauge groups
\cite{ILS,BH,BIWW}, 
consisting of 
three NS-branes, D4-branes and D6-branes,   
there exist also two possible brane configurations.
Either five NS-branes where an O6-plane has common value at $x^6=0$
with NS-brane or four NS-branes where at $x^6=0$ there is no NS-brane.
The former brane configuration 
is described by \cite{Ahn07-4} while the latter brane configuration is
described by \cite{LO,Ahn07-3}.

When we add an orientifold 6-plane to the standard brane
configuration for the product of three unitary gauge groups \cite{BH,Ahn07-8}, 
consisting of 
four NS-branes, D4-branes and D6-branes,   
there exist also two possible brane configurations.
Either seven NS-branes where an O6-plane has common value at $x^6=0$
with NS-brane or six NS-branes where at $x^6=0$ there is no NS-brane.
We'll study these brane configurations in details in the context of
nonsupersymmetric meta-stable brane configurations.

In this paper, we continue to study for the meta-stable brane
configurations of type IIA string theory 
in the context of three(and multiple) product gauge groups in the
presence of an orientifold 6-plane.  
When the total number of NS-branes is seven, the gauge group is
described by the product of three unitary gauge groups while 
for the case where the total number of NS-branes is six, 
one of the gauge groups contains 
orthogonal or symplectic gauge group as well as two unitary gauge
groups, depending whether O6-plane charge
is positive or negative.
In particular, the Seiberg dual for the
middle gauge group has a new feature that
there exist two possible
magnetic brane configurations as in \cite{Ahn07-8}. 
Each of them has the different 
magnetic superpotential and contains the different matter contents. 
One can generalize the meta-stable brane configurations 
to the ones corresponding to a multiple product of gauge groups.
Then one takes the Seiberg dual for the first gauge group factor, the
last gauge group factor or for any gauge group factor except the
first and last gauge group factors. One can write 
down the magnetic superpotentials in terms of the interactions
between the gauge singlets and dual matters. 

In section 2,
we describe the type IIA brane configuration corresponding
to the electric theory based on the ${\cal N}=1$ $SU(N_c) \times
SU(N_c') \times SU(N_c'')$ 
gauge theory 
with fundamentals, bifundamentals, a symmetric flavor and a conjugate
symmetric flavor for $SU(N_c)$, 
and deform this theory by adding the mass term
for the quarks for each gauge group. 
Then we construct the Seiberg dual magnetic theories for each gauge
group with corresponding dual
matters as well as additional gauge singlets. 
After that, the
nonsupersymmetric brane configurations are  found by recombination and
splitting for the flavor D4-branes.
One generalizes to the meta-stable brane configurations corresponding
to a multiple product of gauge groups and describe them
very briefly.

In section 3, 
we consider the type IIA brane configuration corresponding
to the electric theory based on the ${\cal N}=1$ $SU(N_c) \times
SU(N_c') \times SU(N_c'')$ 
gauge theory 
with fundamentals, bifundamentals, eight-fundamentals, 
an antisymmetric flavor and a conjugate
symmetric flavor for $SU(N_c)$, 
and deform this theory by adding the mass terms
for the quarks for each gauge group. 
We present the Seiberg dual magnetic theories for each gauge
group with corresponding dual
matters as well as additional gauge singlets. 
Finally, the
nonsupersymmetric brane configurations are  found.
One also generalizes to the meta-stable brane configurations corresponding
to a multiple product of gauge groups.

In section 4,
we study the type IIA brane configuration corresponding
to the electric theory based on the ${\cal N}=1$ $SO(N_c) \times
SU(N_c') \times SU(N_c'')$ 
gauge theory 
with fundamentals, vectors, and  bifundamentals, 
and deform this theory by adding the mass term
for the quarks for each gauge group.
Explicitly we construct the Seiberg dual magnetic theories for each gauge
group factor with corresponding dual
matters as well as extra gauge singlets and the
nonsupersymmetric brane configurations are  found.
The generalization to a multiple product of gauge groups is also described. 

In section 5,
we explain the type IIA brane configuration corresponding
to the electric theory based on the ${\cal N}=1$ $Sp(N_c) \times
SU(N_c') \times SU(N_c'')$ 
gauge theory 
with fundamentals, and  bifundamentals, 
and deform this theory by adding the mass term
for the quarks for each gauge group.
After that we describe the Seiberg dual magnetic theories for each gauge
group factor with corresponding dual
matters as well as extra gauge singlets. 
Finally, the
nonsupersymmetric brane configurations are  found from the magnetic
brane configurations.
The generalization to a multiple product of gauge groups is discussed. 

Finally, in section 6, 
we summarize what we have found in this paper and 
make some comments for the future direction.

\section{Meta-stable brane configurations
of multiple product gauge theories}


\subsection{Electric theory}

Let us describe the gauge theory with triple product gauge groups 
$SU(N_c) \times SU(N_c') \times SU(N_c'')$ where the symmetric and a
conjugate symmetric tensors are present in addition to the
fundamentals and bifundamentals.
The matter contents 
are characterized by 

$\bullet$
$N_f$-chiral multiplets $Q$ are  in the
representation $({\bf N_c, 1, 1
})$, and 
$N_f$-chiral multiplets $\widetilde{Q}$ are in  
the representation $({\bf \overline{N_c}, 1, 1})$,
under the gauge group

$\bullet$
$N_f'$-chiral multiplets $Q'$ are  in the
representation $({\bf 1, N_c', 1})$, and 
$N_f'$-chiral multiplets $\widetilde{Q}'$ are in  
the representation $({\bf 1,\overline{N_c'}, 1})$,
under the gauge group

$\bullet$
$N_f''$-chiral multiplets $Q''$ are  in the
representation $({\bf 1, 1, N_c''
})$, and 
$N_f''$-chiral multiplets $\widetilde{Q}''$ are in  
the representation $({\bf 1, 1, \overline{N_c''}})$,
under the gauge group

$\bullet$
The flavor-singlet field $F$ is 
in the bifundamental representation $({\bf N_c, \overline{N_c'}, 1 })$, 
and its conjugate field $\widetilde{F}$
 is 
in the bifundamental representation $({\bf \overline{N_c}, N_c', 1})$, 
under the gauge group

$\bullet$
The flavor-singlet field $G$ is 
in the bifundamental representation $({\bf 1, N_c', \overline{N_c''} })$, 
and its conjugate field $\widetilde{G}$
 is 
in the bifundamental representation $({\bf 1, \overline{N_c'}, N_c''})$, 
under the gauge group

$\bullet$ The flavor-singlet field $S$, which is 
in a symmetric tensor representation under the $SU(N_c)$, is in the
representation $({\bf \frac{1}{2} N_c(N_c+1), 1, 1})$, and
its conjugate field $\widetilde{S}$ is in the 
representation $({\bf \overline{\frac{1}{2} N_c(N_c+1)}, 1, 1})$, under the
gauge group

If we ignore the symmetric and conjugate symmetric tensors $S$ and
$\widetilde{S}$, 
this
theory was studied in \cite{BH,Ahn07-8}. If we put to $Q'',
\widetilde{Q}'', G, \widetilde{G}, S$ and $\widetilde{S}$ zero, then 
this becomes the product gauge group theory with fundamentals and
bifundamentals
\cite{ILS,BIWW,BH,AT97}. Morover, if we 
 put to $Q', \widetilde{Q}', Q'',
\widetilde{Q}'', F, \widetilde{F}, G$, and $ \widetilde{G}$ zero,
then this theory is described by a single gauge group with
fundamentals, a symmetric tensor, and a conjugate symmetric tensor 
\cite{ILS,LL,LLL,Ahn07}.


Now it is easy to check from the matter contents above that 
the coefficient of the beta function of the first gauge group 
is given by
$
b_{SU(N_c)}=3N_c-N_f-N_c'-(N_c+2)
$
where the index of the symmetric representation of 
$SU(N_c)$ corresponding to $S$ and $\widetilde{S}$ is equal
to $\frac{1}{2}(N_c+2)$.
On the other hand,
the coefficient of the beta function of the second gauge group  
is given by
$
b_{SU(N_c')}=3N_c'-N_f'-N_c-N_c''$.
Moreover, the coefficient of the beta function of the third gauge group  
is given by
$
b_{SU(N_c'')}=3N_c''-N_f''-N_c'$.
This theory is asymptotically free when the condition $b_{SU(N_c)} >
0$ is satisfied for the
$SU(N_c)$
gauge group, when the condition $b_{SU(N_c')} > 0$ is satisfied 
for the $SU(N_c')$ gauge group, and 
when the condition $b_{SU(N_c'')} > 0$ is satisfied 
for the $SU(N_c'')$ gauge group.
We'll see how these coefficients change in the magnetic theory.
We denote the
strong coupling scales for $SU(N_c)$ as $\Lambda_1$, for $SU(N_c')$
as $\Lambda_2$ and for $SU(N_c'')$
as $\Lambda_3$ respectively.  

The electric superpotential by adding the mass terms for $Q, Q'$ and
$Q''$ 
is given by
\bea
W_{elec} & = & 
\left( \mu A^2 +  Q A \widetilde{Q} + S A \widetilde{S} + 
\widetilde{F} A F +
\mu' A'^2 +  Q' A' \widetilde{Q}' + \widetilde{F} A' F +
\widetilde{G} A' G \right. \nonu \\
&+& \left.
 \mu'' A''^2 +  Q'' A'' \widetilde{Q}'' + \widetilde{G} A'' G \right)
+ m Q \widetilde{Q} + m' Q' \widetilde{Q}' + m'' Q'' \widetilde{Q}''.
\nonu
\eea
After integrating the adjoint fields 
$A$ for $SU(N_c)$, $A'$ for $SU(N_c')$ and $A''$ for $SU(N_c'')$ 
and taking $\mu, \mu'$ and $\mu''$ to infinity limit which is
equivalent to take any two NS-branes be perpendicular to each other(in
other words, the nearest NS-branes for given $NS5$-brane are
NS5'-branes while those for NS5'-brane are NS5-branes),
the mass-deformed electric superpotential becomes 
$
W_{elec}  = 
m Q \widetilde{Q} + m' Q' \widetilde{Q}' + m'' Q'' \widetilde{Q}''$.

The type IIA brane configuration for this mass-deformed theory 
can be described by as follows. 
We introduce the two complex coordinates 
\bea
v \equiv x^4 + i x^5  \qquad \mbox{and} \qquad 
w \equiv x^8 + i x^9
\nonu
\eea
for convenience.
The $N_c$-color 
D4-branes (01236) are suspended between the $NS5_M$-brane (012345)
located at $x^6=0$ 
and the $NS5_L'$-brane (012389) along positive $x^6$
direction,
together with $N_f$ D6-branes (0123789) 
which are parallel to $NS5_L'$-brane and have nonzero $v$ direction.
The NS5-brane 
is located at the right hand side of
the $NS5_L'$-brane along the positive $x^6$ direction and 
there exist $N_c'$-color D4-branes
suspended 
between them, with  $N_f'$ D6-branes which have nonzero $v$ direction. 
Moreover, 
the $NS5_R'$-brane 
is located at the right hand side of
the NS5-brane along the positive $x^6$ direction and there 
exist $N_c''$-color D4-branes
suspended 
between them, with  $N_f''$ D6-branes which have nonzero $v$ direction.
There exists an orientifold 6-plane (0123789) at the origin $x^6=0$
and it acts as $(x^4, x^5, x^6) \rightarrow (-x^4, -x^5, -x^6)$. 
Then the mirrors of above branes appear in 
the negative $x^6$ region and are denoted by bar on the corresponding branes.
From the left to the right, there are $\overline{NS5_R'}$-,
$\overline{NS5}$-, 
$\overline{NS5_L'}$-, $NS5_M$-,
$NS5_L'$-, $NS5$-, and $NS5_R'$-branes.

We summarize the ${\cal N}=1$ supersymmetric electric brane
configuration in type IIA string theory as follows:

$\bullet$ Three
NS5-branes in (012345) directions 

$\bullet$ Four
NS5'-branes in (012389) directions

$\bullet$ Two sets of
$N_c(N_c')[N_c'']$-color D4-branes in (01236) directions 
  
$\bullet$ Two sets of
$N_f(N_f')[N_f'']$ D6-branes in (0123789) directions

$\bullet$
O6-plane in (0123789) directions with $x^6=0$

Now we draw this electric brane configuration in Figure 1 and we put
the coincident $N_f(N_f')[N_f'']$ D6-branes with positive $x^6$ in 
the nonzero $v$ direction in general. 
This brane configuration can be obtained from the brane configuration
of \cite{Ahn07-4} by adding the two outer NS5'-branes(i.e., 
$\overline{NS5_R'}$-brane and $NS5_R'$-brane), two sets of $N_c''$ D4-branes
and two sets of $N_f''$ D6-branes or from the one of \cite{BH}
with the gauge theory of triple product gauge groups
by adding O6-plane and the extra NS-branes, D4-branes and D6-branes.
Then the mirrors with 
negative $x^6$ can be constructed by using the action of O6-plane and
are located at the positions by changing (456) directions for original
branes with minus signs.
The quarks $Q(Q')[Q'']$ and
$\widetilde{Q}(\widetilde{Q}')[\widetilde{Q}'']$ 
correspond to strings
stretching between the
$N_c(N_c')[N_c'']$-color D4-branes with $N_f(N_f')[N_f'']$ D6-branes.
The bifundamentals $F(G)$ and $\widetilde{F}(\widetilde{G})$ 
correspond to   strings 
stretching between the
$N_c(N_c')$-color D4-branes with $N_c'(N_c'')$-color D4-branes. 
The symmetric and a conjugate symmetric tensors $S$ and
$\widetilde{S}$ correspond to strings
stretching between $N_c$ D4-branes with positive $x^6$ and its mirror
$N_c$ D4-branes with negative $x^6$.

\begin{figure}[ht]
   \epsfxsize=3.0in 
\centerline{\epsffile{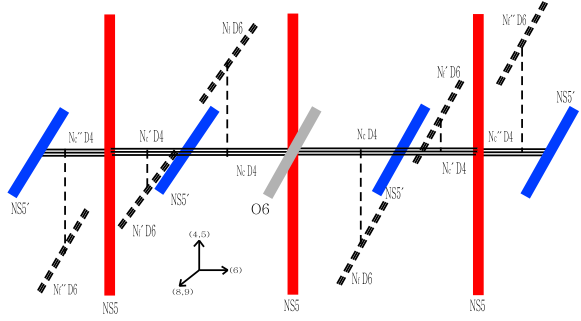}}
   \caption[FIG. \arabic{figure}.]{ 
The ${\cal N}=1$ supersymmetric electric brane configuration with
$SU(N_c) \times SU(N_c') \times SU(N_c'')$ gauge group with
fundamentals $Q(Q')[Q'']$ and
$\widetilde{Q}(\widetilde{Q}')[\widetilde{Q}'']$ 
for each gauge group and bifundamentals $F(G)$,  
$\widetilde{F}(\widetilde{G})$, a symmetric tensor $S$, and a conjugate
symmetric tensor $\widetilde{S}$. The O6-plane is located at the
origin $x^6=0$.
The  two NS5'-branes with positive $x^6$ coordinates
are denoted by $NS5_{L,R}'$-branes. The mirror
branes located at the negative region of $x^6$ 
are preserved under the O6-plane action. The mass terms for the quarks
are realized geometrically 
by the displacement of D6-branes to the $v$ direction.}
\end{figure}

\subsection{Magnetic theory with dual for first gauge group}

By brane motion \cite{GK98}, one gets the Seiberg dual theory
for the gauge group $SU(N_c)$ and dualized gauge group's dynamical
scale is far above that of the gauge groups $SU(N_c')$ and $SU(N_c'')$.
From the magnetic brane configuration which is shown in Figure 2A,  
the linking number \cite{HW} of $NS5_L'$-brane(which is
the mirror of $NS5_L'$-brane in an electric theory) can
be computed and is
$L_5=\frac{N_f}{2}-\widetilde{N}_c+N_f+N_c'$
as in
the situation of \cite{Ahn07}.
On the other hand, 
the linking number of $\overline{NS5_L'}$-brane from the
electric brane configuration in Figure 1 can be read off and is 
given by $L_5=-\frac{N_f}{2}+N_c-N_c'$. 
Then the number of dual color $\widetilde{N}_c$, by linking number
conservation, 
is given by
\bea
\widetilde{N}_c = 2(N_f + N_c')-N_c.
\nonu
\eea

Let us draw this magnetic brane configuration in Figure 2A and recall
that we put
the coincident $N_f$ D6-branes in the nonzero $v$-direction in the
electric theory and consider massless flavors for $Q'$ and $Q''$ by
putting $N_f'$ and $N_f''$ D6-branes at $v=0$.
The $N_f$ created D4-branes connecting between
D6-branes and $NS5_L'$-brane can move freely in the $w$-direction.
Moreover, since $N_c'$ or $N_c''$ D4-branes are suspending between two 
$NS5'_{L,R}$-branes located at different $x^6$ coordinate, these D4-branes
can slide along the $w$-direction also.
If we ignore the NS5-brane, $N_c'$ D4-branes, $N_f'$ 
D6-branes, the $NS5_R'$-brane, $N_c''$ D4-branes and $N_f''$ 
D6-branes(detaching these
branes from Figure 2A), 
then this brane configuration 
leads to the ${\cal N}=1$ magnetic theory with gauge group 
$SU(\widetilde{N}_c=2N_f-N_c)$ with
$N_f$ massive fundamental 
flavors plus symmetric, conjugate symmetric flavors 
and gauge singlets \cite{FGU,Ahn07}.
On the other hand, when 
we ignore the $NS5_R'$-brane, $N_c''$ D4-branes and $N_f''$ 
D6-branes(detaching these
branes from Figure 2A), 
then this brane configuration 
leads to the ${\cal N}=1$ magnetic theory with gauge group 
$SU(\widetilde{N}_c=2N_f+2N_c'-N_c) \times SU(N_c')$ with
fundamental 
flavors, bifundamentals, symmetric, conjugate 
symmetric flavors and gauge singlets in Figure 4 of \cite{Ahn07-4}.

Now let us recombine $\widetilde{N}_c$ flavor D4-branes among $N_f$
flavor 
D4-branes(connecting between D6-branes and $NS5_L'$-brane) 
with the same number of 
color D4-branes(connecting between $NS5_M$-brane and $NS5_L'$-brane) and push
them in $+v$ direction from Figure 2A. We assume that $N_c \geq N_f+2N_c'$. 
After this procedure, there are no color D4-branes between 
$NS5_M$-brane and $NS5_L'$-brane.
For the flavor D4-branes, we are left with only 
$(N_f-\widetilde{N}_c)=N_c-N_f-2N_c'$ flavor D4-branes
connecting between D6-branes and $NS5_L'$-brane in Figure 2B.  

\begin{figure}[ht]
   \epsfxsize=4.0in 
\centerline{\epsffile{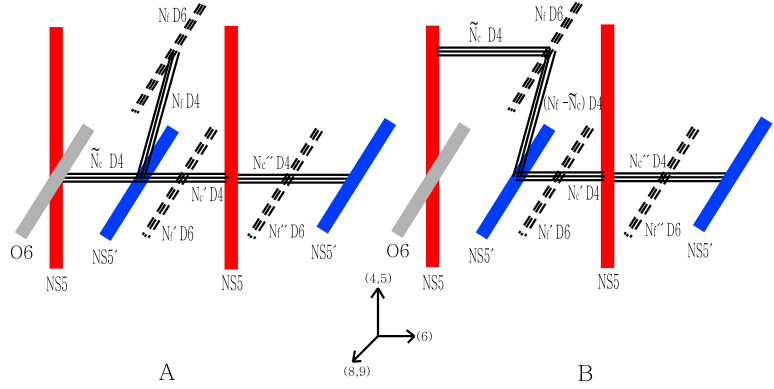}}
   \caption[FIG. \arabic{figure}.]{ 
The ${\cal N}=1$ supersymmetric magnetic brane configuration with
$SU(\widetilde{N}_c = 2N_f + 2N_c'-N_c) \times SU(N_c') 
\times SU(N_c'')$ gauge group
with fundamentals $q(Q')[Q'']$ and
$\widetilde{q}(\widetilde{Q}')[\widetilde{Q}'']$ 
for each gauge group and bifundamentals $f(G)$ and
$\widetilde{f}(\widetilde{G})$, $s$ and $\widetilde{s}$, and 
gauge singlets in Figure 2A. In
Figure 2B, the nonsupersymmetric minimal energy brane configuration
with the same gauge group and matter contents above 
for massless  $Q'(Q'')$ and
$\widetilde{Q}'(\widetilde{Q}'')$ is given. Compared with the 
Figure 2A, there
exists a misalignment of the flavor D4-branes. Although there exist
the mirrors in the negative $x^6$ region in Figures 2A and 2B(and also  
in the magnetic brane
configurations we'll present), we do not draw the mirrors 
for simplicity. 
}
\end{figure}

Then the dual magnetic gauge group is  
$SU(\widetilde{N}_c) \times SU(N_c') \times SU(N_c'')$ 
and the matter contents 
are as follows:

$\bullet$ 
$N_f$-chiral multiplets $q$ are  in the
representation $({\bf \widetilde{N}_c, 1, 1})$,
$N_f$-chiral multiplets $\widetilde{q}$ are in the representation 
$({\bf \overline{\widetilde{N}_c}, 1, 1})$,
under the gauge group

$\bullet$
$N_f'$-chiral multiplets $Q'$ are  in the
representation $({\bf 1, N_c', 1})$, and 
$N_f'$-chiral multiplets $\widetilde{Q}'$ are in  
the representation $({\bf 1,\overline{N_c'}, 1})$,
under the gauge group

$\bullet$
$N_f''$-chiral multiplets $Q''$ are  in the
representation $({\bf 1, 1, N_c''
})$, and 
$N_f''$-chiral multiplets $\widetilde{Q}''$ are in  
the representation $({\bf 1, 1, \overline{N_c''}})$,
under the gauge group

$\bullet$
The flavor-singlet field $f$ is 
in the bifundamental representation $({\bf \widetilde{N}_c, 
\overline{N_c'}, 1 })$, 
and its complex conjugate field $\widetilde{f}$
 is 
in the bifundamental representation $({\bf \overline{\widetilde{N}_c}, 
N_c', 1})$, 
under the gauge group

$\bullet$
The flavor-singlet field $G$ is 
in the bifundamental representation $({\bf 1, N_c', 
\overline{N_c''} })$, 
and its complex conjugate field $\widetilde{G}$
 is 
in the bifundamental representation $({\bf 1, \overline{N_c'}, 
N_c''})$, 
under the gauge group

$\bullet$ The flavor-singlet field $s$, which is 
in a symmetric tensor representation under the $SU(\widetilde{N}_c)$, 
is in the
representation $({\bf \frac{1}{2} \widetilde{N}_c(\widetilde{N}_c+1),
  1, 1})$, and
its conjugate field $\widetilde{s}$ is in the 
representation $({\bf \overline{\frac{1}{2} 
\widetilde{N}_c(\widetilde{N}_c+1)}, 1, 1})$, under the
gauge group

There are also $(N_f+N_c')^2$ gauge-singlets in the first dual gauge group
factor
as follows:

$\bullet$
$N_f$-fields $X'$ are  in the representation $({\bf 1, N_c', 1 })$, 
and its complex conjugate
$N_f$-fields $\widetilde{X}'$ are in the representation 
$({\bf 1, \overline{N_c'}, 1 })$, 
under the gauge group

$\bullet$
$N_f^{2}$-fields $M$ are in the representation $({\bf 1, 1, 1})$ under the
gauge group

$\bullet$
The $N_c^{'2}$-fields 
$\Phi'$ is in the representation $({\bf 1, N_c^{'2}-1, 1}) \oplus ({\bf
  1, 1, 1
})$ 
under the gauge group  

Moreover, there are additional $N_f(2N_f+1)$ gauge-singlets

$\bullet$
$N_f^{2}$-fields $N$ are in the representation $({\bf 1, 1, 1})$ under the
gauge group

$\bullet$ $\frac{1}{2} N_f(N_f+1)$-fields $P$ are 
 in the representation $({\bf 1, 1, 1})$, and its 
conjugate $\frac{1}{2} N_f(N_f+1)$-fields 
 $\widetilde{P}$  are 
 in the representation $({\bf 1, 1, 1})$,
under the
gauge group

The additional $N_f$-$SU(N_c')$ fundamentals $X'$ and $N_f$-$SU(N_c')$
antifundamentals $\widetilde{X}'$ 
are originating from 
the $SU(N_c)$ chiral mesons $\widetilde{F} Q$ and $F
\widetilde{Q}$ 
respectively. Therefore, 
there are free indices for a single color and a single flavor. 
Then the strings stretching between the $N_f$ $D6$-branes and $N_c'$
D4-branes will give rise to these additional $N_f$-$SU(N_c')$
fundamentals and  $N_f$-$SU(N_c')$
antifundamentals.
The gauge singlet $M$ corresponds to the $SU(N_c)$ 
chiral meson $Q \widetilde{Q}$
where the color indices are contracted.
Because the
$N_f$ D6-branes are parallel to the $NS5'_L$-brane from Figure 2A, 
the newly created $N_f$-flavor D4-branes can slide along the plane
consisting of these $N_f$ $D6$-branes and  $NS5'_L$-brane
freely.
The fluctuations of the gauge-singlet $M$ correspond to the motion of $N_f$
flavor D4-branes along (789) directions in Figure 2A.
As we will see later, for the nonsupersymmetric brane configuration,  
a misalignment for the $N_f$-flavor D4-branes arises and some of the
vacuum expectation value of $M$ is fixed and the remaining
components are arbitrary.
The $\Phi'$ corresponds to the $SU(N_c)$ chiral meson $F \widetilde{F}$
where the color indices for the first gauge group 
are contracted each other.
The fluctuations of the singlet $\Phi'$ correspond to the motion of
$N_c'$ D4-branes suspended two $NS5_{L,R}'$-branes along the (789)
directions in Figure 2A.  
Although the gauge singlets 
$N, P$ and $\widetilde{P}$
appear in the dual magnetic
superpotential for the general rotation angles of NS-branes and D6-branes,
the case we are considering 
does not contain these gauge singlets, as observed in \cite{Ahn07}. 

The coefficient of the beta function of the first dual gauge group factor, 
as done in electric theory, is given by
$
b_{SU(\widetilde{N}_c)}^{mag}=3\widetilde{N}_c-N_f-N_c'-(\widetilde{N}_c+2)
$
and 
the coefficient of the beta function of the second gauge group factor 
is given by
$
b_{SU(N_c')}^{mag}=3N_c'-N_f'-\widetilde{N}_c-N_c''-N_f-N_c'$.
Finally, the coefficient of the beta function of the third dual gauge group
factor
is
$
b_{SU(N_c'')}^{mag}=3N_c''-N_f''-N_c'=b_{SU(N_c'')}$.
Then both $SU(\widetilde{N}_c)$,  
$SU(N_c')$, and $SU(N_c'')$ gauge couplings are IR free
by requiring the negativeness of the coefficients of beta function.
One relies on the perturbative calculations at low energy 
for this magnetic IR free region with $b_{SU(\widetilde{N}_c)}^{mag} < 0$,  
$b_{SU(N_c')}^{mag} < 0$, and $b_{SU(N_c'')}^{mag} < 0$.

The dual magnetic superpotential coming from \cite{Ahn07}  
is given by
\bea
W_{dual}= \left(M q \widetilde{s} s \widetilde{q} +  m M \right)+ 
\widetilde{f} X' q +
f \widetilde{q} \widetilde{X}' + \Phi' f \widetilde{f}. 
\nonu
\eea
Then, $q \widetilde{s} s \widetilde{q}$ has rank 
$\widetilde{N}_c$ while $m$ has a
rank $N_f$.  Therefore, the derivative of the 
superpotential $W_{dual}$ with respect to $M$, cannot be satisfied 
if the rank $N_f$ exceeds $\widetilde{N}_c$ and the supersymmetry is broken.    
The classical moduli space of vacua can be obtained from F-term
equations and one gets \cite{Ahn07}
\bea
 q \widetilde{s} s \widetilde{q} +  m & = & 0, \qquad
\widetilde{s} s \widetilde{q} M + \widetilde{f} X'   =  0, \nonu \\
s \widetilde{q} M q & = & 0, \qquad
\widetilde{q} M q \widetilde{s}  =  0, \nonu \\
  M q \widetilde{s} s + \widetilde{X}' f & = & 0, \qquad
X' q + \Phi' f  =  0,
\nonu \\
q \widetilde{f} & = & 0, \qquad
\widetilde{q} \widetilde{X}' + \widetilde{f} \Phi'  =  0, \nonu \\
f \widetilde{q} & = & 0, \qquad
f \widetilde{f}  =  0. 
\nonu
\eea
Some of F-term equations are satisfied if one takes the zero vacuum
expectation values for the fields $f, \widetilde{f}, X'$ and 
$\widetilde{X}'$. 

Then the gauge group and matter contents we consider 
are summarized as follows:
\bea
 & \mbox{gauge group}:& \;\;\;\;\;   SU(\widetilde{N}_c) \times SU(N_c') \times
 SU(N_c'')  \nonu
\\
\mbox{matter}: 
 &q_f \oplus \widetilde{q}_{\widetilde{f}}& \;\;\;\;\;\;\;\;\; 
\;\;\; {(\bf \Box, 1, 1) \oplus (\overline{\Box}, 1, 1)}
\;\;\;\;\; (f, \widetilde{f}=1,  \cdots, N_f) 
\nonu \\
 &Q'_{f'} \oplus \widetilde{Q}'_{\widetilde{f}'}& \;\;\;\;\;\;\;\;\;\;
\;\; {(\bf 1, \Box, 1) \oplus ( 1, \overline{\Box}, 1)}
\;\;\;\;\; (f', \widetilde{f}' =1,  \cdots, N_f') 
\nonu \\
 &Q''_{f''} \oplus \widetilde{Q}''_{\widetilde{f}''}& \;\;\;\;\;\;\;\;\;\; 
\;\; {(\bf 1, 1, \Box) \oplus ( 1, 1, \overline{\Box})} 
\;\;\;\;\; (f'', \widetilde{f}'' =1,  \cdots, N_f'')
\nonu \\
 &f \oplus \widetilde{f}& \;\;\;\;\;\;\;\;\; 
\;\; {(\bf \Box, \overline{\Box},1) \oplus (\overline{\Box}, \Box, 1)} 
\nonu \\
 &G \oplus \widetilde{G}& \;\;\;\;\;\;\;\;\; 
\;\; {(\bf 1, \Box, \overline{\Box}) \oplus ( 1, \overline{\Box}, \Box)} 
\nonu \\
 &s \oplus \widetilde{s}& \;\;\;\;\;\;\;\;\; 
\;\; {(\bf symm, 1, 1) \oplus (\overline{symm}, 1, 1)} 
\nonu \\ 
& (X_{n}' \equiv) \widetilde{F} Q \oplus F \widetilde{Q} (\equiv 
\widetilde{X}_{\widetilde{n}}') & 
 \;\;\;\;\;\;\;\;\;\;
\;\; {(\bf 1, \Box, 1) \oplus ( 1, \overline{\Box}, 1)}
\;\; (n, \widetilde{n} =1,  \cdots, N_f) 
\nonu \\
&  (M_{f,\widetilde{g}} \equiv) Q \widetilde{Q} & 
 \;\;\;\;\;\;\;\;\;\;\;\;\;\;\;\;\;\;\;
\;\; {(\bf 1, 1, 1)} \;\;\;\;\;\;\;\;\;
\;\; (f, \widetilde{g} =1,  \cdots, N_f) 
\nonu \\
& (\Phi' \equiv) F \widetilde{F} & 
 \;\;\;\;\;\;\;\;\;\;
\;\; {(\bf 1, adj, 1) \oplus ( 1, 1, 1)}
\nonu \\
& (P_{f, g} \equiv)  Q \widetilde{S} Q \oplus \widetilde{Q} S 
\widetilde{Q} (\equiv 
\widetilde{P}_{\widetilde{f}, \widetilde{g}}) & 
 \;\;\;\;\;\;\;\;\;\;
\;\; {(\bf 1, 1, 1) \oplus ( 1, 1, 1)}
\;\; (f, \widetilde{f}, g, \widetilde{g} =1,  \cdots, N_f) 
\nonu \\
& (N_{f, \widetilde{g}} \equiv) Q \widetilde{S} S \widetilde{Q} & 
 \;\;\;\;\;\;\;\;\;\;
\;\; \;\;\;\;\;\; \;\;\; {(\bf 1, 1, 1)}  \;\;
(f, \widetilde{g} =1,  \cdots, N_f) 
\nonu
\eea

Then, it is easy to see that 
$
s \widetilde{q} M =0= M q \widetilde{s}, 
 q \widetilde{s} s \widetilde{q} +  m  =  0$.
Then the solutions can be written as
\bea
<q \widetilde{s}>  & = &  \left(
\begin{array}{c}
\sqrt{m} e^{\phi} {\bf 1}_{\widetilde{N}_c}  \\
0
\end{array}
\right),  
<s \widetilde{q}> =
 \left(
\begin{array}{cc}
\sqrt{m} e^{-\phi}  {\bf 1}_{\widetilde{N}_c}   &
0
\end{array}
\right), 
<M>  =
 \left(
\begin{array}{cc}
0  & 0 
 \\
0 & M_0  {\bf 1}_{N_f-\widetilde{N}_c} 
\end{array}
\right),
\nonu \\
<f> & = & <\widetilde{f}> = <X'> = <\widetilde{X}'>= 0.
\nonu
\eea
Let us expand around on a point on the vacua, as done in
\cite{ISS}. 
Then the remaining relevant terms of superpotential are given by
$
W_{dual}^{rel}  =   M_0 \left( \delta \hat{\varphi}  
\; \delta \hat{\widetilde{\varphi}} + m \right) +
  \delta Z \; \delta \hat{\varphi} \; s_0 \; \widetilde{q}_0 
+ \delta \widetilde{Z} \; q_0 \; \widetilde{s}_0 \;
\delta \hat{\widetilde{\varphi}}
$
by following the fluctuations for the various fields in \cite{Ahn07}.
Note that there exist three kinds of terms, 
the vacuum  $<q>$ multiplied by 
$\delta \widetilde{f} \delta X'$,  
the vacuum  $<\widetilde{q}>$ multiplied by $\delta \widetilde{X}' 
\delta f$, and 
the vacuum  $<\Phi'>$ multiplied by $\delta f 
\delta \widetilde{f}$.
By redefining these,  
they do not enter the 
contributions for the one loop result, up to quadratic order. 
As done in \cite{Ahn07}, the defining function ${\cal F}(v^2)$ \cite{Shih}
can be
computed
and $m_{M_0}^2$ will contain $(\log 4 -1) > 0$ implying that these
are stable.

The minimal energy supersymmetry breaking brane configuration is
shown in Figure 2B.
If we ignore the NS5-brane, $N_c'$ D4-branes,
$N_f'$ D6-branes, $NS5_R'$-brane, $N_c''$ D4-branes and $N_f''$ 
D6-branes(detaching these from Figure 2B), 
as observed already, 
then this brane configuration 
corresponds to  the minimal energy supersymmetry breaking brane
configuration
for the ${\cal N}=1$ SQCD with the magnetic gauge group 
$SU(\widetilde{N}_c=2N_f-N_c)$ with
$N_f$ massive flavors \cite{FGU,Ahn07}.

The nonsupersymmetric minimal energy brane configuration Figure 2B
with a replacement $N_f$ D6-branes by 
the NS5'-brane(neglecting  the
$NS5_R'$-brane, $N_f''$ D6-branes and $N_c''$ D4-branes 
and $N_f'$ D6-branes)
leads to 
the Figure 5B of \cite{Ahn07-7} with a rotation of NS5'-brane by 
$\frac{\pi}{2}$ angle.

In \cite{Witten,LL}, the Riemann surface 
describing a set of NS-branes 
with D4-branes suspended between them and 
in a background space of $x t = (-1)^{N_f+N_f'+N_f''} v^{2N_f'+2N_f''+4}
(v^2 -m^2)^{N_f}$
was found.
Since we are dealing with seven NS-branes, the magnetic M5-brane 
configuration in Figure 2 with equal mass for $q$ and massless 
for $Q'(Q'')$ 
can be characterized by the following seventh order equation for $t$ 
as follows:
\bea
& & t^7 + \left[v^{N_c''} \right] t^6 + \left[ v^{N_c'+N_f''} \right] t^5
 + \left[ v^{\widetilde{N}_c+2N_f''+N_f'} (v -m)^{N_f}\right]  t^4 
\nonu \\
&& + \left[ (-1)^{\widetilde{N}_c}
v^{\widetilde{N}_c +3N_f''+2N_f'+ 2} 
(v -m)^{2N_f} \right] t^3  + 
\left[ (-1)^{N_c'} v^{N_c' +4N_f''+3N_f'+6}  
(v -m)^{3N_f} \right] t^2 \nonu \\
&& + \left[(-1)^{N_f+N_f'+N_c''}
v^{10+5N_f'+5N_f''+N_c''}  (v -m)^{4N_f} 
 (v +m)^{N_f}\right] t  \nonu \\
&& + \left[(-1)^{N_f''} v^{14+7N_f'+7N_f''} (v-m)^{5N_f} 
(v+m)^{2N_f}\right] = 0.
\nonu
\eea

At nonzero string coupling constant, 
the NS5-branes bend due to their interactions with the D4-branes and
D6-branes.
Now the asymptotic regions of various NS-branes 
can be determined by reading off the first two terms of the seventh order
curve above giving the $\overline{NS5_R'}$-brane
asymptotic region, next two terms giving 
the $\overline{NS5}$-brane asymptotic region, next two terms
giving the $\overline{NS5_L'}$-brane asymptotic region, next two terms 
giving $NS5_M$-brane asymptotic region, 
 next two terms 
giving $NS5_L'$-brane asymptotic region, 
 next two terms 
giving NS5-brane asymptotic region, 
and 
final two terms giving $NS5_R'$-brane asymptotic region.
Then the behavior of the supersymmetric M5-brane curves can be
summarized 
as follows:

1. $v \rightarrow \infty$ limit implies
\bea
w & \rightarrow & 0, \quad y \sim    v^{N_c'+N_f''-N_c''} + \cdots \quad
\mbox{$\overline{NS5}$ 
asymptotic region}, \nonu \\
w & \rightarrow  & 0, \quad y \sim    
v^{N_f''+N_f+N_f'+2} + \cdots \quad
\mbox{$NS5_{M}$ asymptotic region}, 
\nonu \\
w & \rightarrow  & 0, \quad y \sim    
v^{N_f''+2N_f+2N_f'-N_c'+N_c''+4} + \cdots \quad
\mbox{NS5 asymptotic region}.   
\nonu
\eea

2.  $w \rightarrow \infty$ limit implies
\bea
v & \rightarrow &   -m, \quad 
y \sim  w^{\widetilde{N}_c-N_c'+N_f''+N_f+N_f'}
 +\cdots
\quad \mbox{$\overline{NS5_{L}'}$ asymptotic region}, 
\nonu
\\
v & \rightarrow &  -m, \quad  
y \sim w^{N_c''}
+\cdots
\quad \mbox{$\overline{NS5_{R}'}$ asymptotic region}, 
\nonu \\
v & \rightarrow &   +m, \quad 
y \sim  w^{-\widetilde{N}_c+N_c'+N_f''+N_f+N_f'+4}
 +\cdots
\quad \mbox{$NS5_{L}'$ asymptotic region}, 
\nonu
\\
v & \rightarrow &  +m, \quad  
y \sim w^{-N_c''+2N_f''+2N_f+2N_f'+4}
+\cdots
\quad \mbox{$NS5_{R}'$ asymptotic region} 
\nonu
\eea
where we denote the mirror branes by writing the bar on the
corresponding NS-brane.
The two $NS5_{L,R}'$-branes 
are moving in the $ +v$ direction respectively holding everything 
else fixed instead of moving D6-branes in the $+v$ direction 
\cite{BGHSS,Ahn06-1}.
The corresponding mirrors of D4-branes are moved appropriately.
The harmonic function in the Tau-NUT space, sourced by $2N_f$
D6-branes, O6-plane, $2N_f'$ D6-branes and $2N_f''$ D6-branes, 
can be determined once we
fix the $x^6$ position for these branes. Then the first order
differential equation for the $g(s)$ where the absolute value of
$g(s)$ is equal to the absolute value of $w$ can be solved exactly with the
appropriate boundary conditions on $NS5_L'$ or $NS5_R'$ asymptotic
region from above. Since the extra terms in the harmonic
function contribute to the $g(s)$ as a multiplication factor,   
the contradiction with the correct statement that 
$y$ should vanish only if $v=0$
implies that there exists the instability from a new M5-brane mode at
some point from the transition of SQCD-like theory description 
to M-theory description.

\subsection{Magnetic theory with dual for second gauge group}

By moving the NS5-brane in Figure 1
to the left all the way past the  
$NS5'_L$-brane, one arrives at the Figure 3A.
The linking number of NS5-brane from Figure 3A
is 
$
L_5 = -\frac{N_f'}{2} +\widetilde{N}_c'-N_c
$ 
while
the linking number of NS5-brane from Figure 1
is
$
L_5 = \frac{N_f'}{2} + N_c'' -N_c'$. 
From these two relations, one obtains
the number of colors of dual magnetic theory
\bea
\widetilde{N}_c' = N_f' + N_c''+ N_c-N_c'.
\nonu
\eea

Let us draw this magnetic brane configuration in Figure 3A and recall
that we put
the coincident $N_f'$ D6-branes in the nonzero $v$-direction in the
electric theory and consider massless flavors for $Q$ and $Q''$ by
putting $N_f$ and $N_f''$ D6-branes at $v=0$.
The $N_f'$ created D4-branes connecting between
D6-branes and $NS5_L'$-brane can move freely in the $w$-direction.
Moreover, 
since $N_c$ D4-branes are suspending between the
$NS5_M$-brane located at the origin $x^6=0$ 
and NS5-brane located at different $x^6$ coordinate, 
these D4-branes
can slide along the $v$-direction.
Then this brane configuration 
leads to the standard ${\cal N}=1$ magnetic gauge theory 
$SU(N_c) \times SU(\widetilde{N}_c'=N_f'+N_c+N_c''-N_c') \times
SU(N_c'')$ 
with fundamentals and
bifundamentals, and singlets in Figure 3 of \cite{Ahn07-8}.

Now let us recombine $\widetilde{N}_c'$ flavor D4-branes among $N_f'$
flavor 
D4-branes(connecting between D6-branes and $NS5_L'$-brane) with 
the same number of 
color D4-branes(connecting between NS5-brane and $NS5_L'$-brane) and push
them in $+v$ direction from Figure 3A. 
For the flavor D4-branes, we are left with only 
$(N_f'-\widetilde{N}_c')=N_c'-N_c''-N_c$ flavor D4-branes
connecting between D6-branes and $NS5_L'$-brane in Figure 3B.  

\begin{figure}[ht]
   \epsfxsize=4.0in 
\centerline{\epsffile{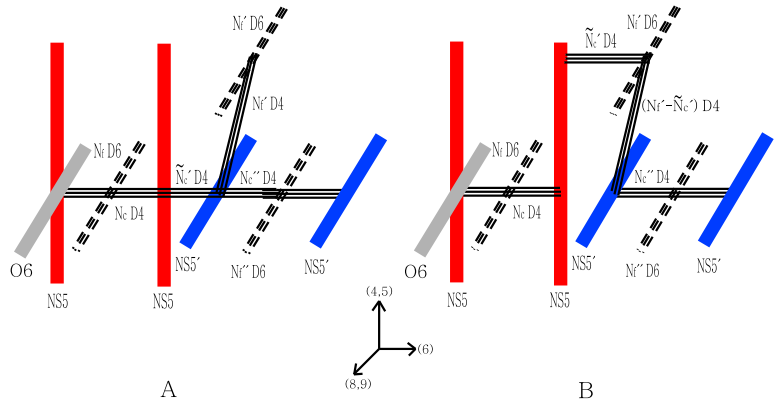}}
   \caption[FIG. \arabic{figure}.]{ 
The ${\cal N}=1$ supersymmetric magnetic brane configuration with
$SU(N_c) \times SU(\widetilde{N}_c'=N_f' + N_c''+ N_c-N_c') 
\times SU(N_c'')$ gauge group
with fundamentals $Q(q')[Q'']$ and
$\widetilde{Q}(\widetilde{q}')[\widetilde{Q}'']$ 
for each gauge group and bifundamentals $F(g)$ and
$\widetilde{F}(\widetilde{g})$, $S$ and $\widetilde{S}$,  
and gauge singlets in Figure 3A. In
Figure 3B, the nonsupersymmetric minimal energy brane configuration
with the same gauge group and matter contents above 
for massless  $Q(Q'')$ and
$\widetilde{Q}(\widetilde{Q}'')$ is given. 
}
\end{figure}

The additional $N_f'$-$SU(N_c'')$ fundamentals $X''$ and $N_f'$-$SU(N_c'')$
antifundamentals $\widetilde{X}''$ 
are originating from 
the $SU(N_c')$ chiral mesons $\widetilde{G} Q'$ and $G
\widetilde{Q}'$ 
respectively. 
Then the strings stretching between the $N_f'$ $D6$-branes and $N_c''$
D4-branes will give rise to these additional $N_f'$-$SU(N_c'')$
fundamentals and  $N_f'$-$SU(N_c'')$
antifundamentals.
The gauge singlet $M'$ corresponds to the $SU(N_c')$ 
chiral meson $Q' \widetilde{Q}'$
where the color indices are contracted.
The $\Phi''$ corresponds to the $SU(N_c')$ chiral meson $G \widetilde{G}$
where the color indices for the second gauge group 
are contracted each other.

The coefficient of the beta function of the first gauge group factor 
is given by
$
b_{SU(N_c)}^{mag}=3N_c-N_f-\widetilde{N}_c' -
(N_c+2)
$
and 
the coefficient of the beta function of the second gauge group factor 
is given by
$
b_{SU(\widetilde{N}_c')}^{mag}=3\widetilde{N}_c'-N_f'-N_c-N_c''$.
The coefficient of the beta function of the third gauge group factor is 
$
b_{SU(N_c'')}^{mag}=3N_c''-N_f''-\widetilde{N}_c'-N_f'-N_c''$.

Then the gauge group and matter contents we consider 
are summarized as follows:
\bea
 & \mbox{gauge group}:& \;\;\;\;\;   SU(N_c) \times SU(\widetilde{N}_c') \times
 SU(N_c'')  \nonu
\\
\mbox{matter}: 
 &Q_f \oplus \widetilde{Q}_{\widetilde{f}}& \;\;\;\;\;\;\;\;\; 
\;\;\; {(\bf \Box, 1, 1) \oplus (\overline{\Box}, 1, 1)}
\;\;\;\;\; (f, \widetilde{f}=1,  \cdots, N_f) 
\nonu \\
 &q'_{f'} \oplus \widetilde{q}'_{\widetilde{f}'}& \;\;\;\;\;\;\;\;\;\;
\;\; {(\bf 1, \Box, 1) \oplus ( 1, \overline{\Box}, 1)}
\;\;\;\;\; (f', \widetilde{f}' =1,  \cdots, N_f') 
\nonu \\
 &Q''_{f''} \oplus \widetilde{Q}''_{\widetilde{f}''}& \;\;\;\;\;\;\;\;\;\; 
\;\; {(\bf 1, 1, \Box) \oplus ( 1, 1, \overline{\Box})} 
\;\;\;\;\; (f'', \widetilde{f}'' =1,  \cdots, N_f'')
\nonu \\
 &F \oplus \widetilde{F}& \;\;\;\;\;\;\;\;\; 
\;\; {(\bf \Box, \overline{\Box},1) \oplus (\overline{\Box}, \Box, 1)} 
\nonu \\
 &g \oplus \widetilde{g}& \;\;\;\;\;\;\;\;\; 
\;\; {(\bf 1, \Box, \overline{\Box}) \oplus ( 1, \overline{\Box}, \Box)} 
\nonu \\
 &S \oplus \widetilde{S}& \;\;\;\;\;\;\;\;\; 
\;\; {(\bf symm, 1, 1) \oplus (\overline{symm}, 1, 1)} 
\nonu \\ 
& (X_{n'}'' \equiv) \widetilde{G} Q' \oplus G \widetilde{Q}' (\equiv 
\widetilde{X}_{\widetilde{n}'}'') & 
 \;\;\;\;\;\;\;\;\;\;
\;\; {(\bf 1, 1, \Box ) \oplus ( 1, 1, \overline{\Box})}
\;\; (n', \widetilde{n}' =1,  \cdots, N_f') 
\nonu \\
&  (M_{f',\widetilde{g}'}' \equiv) Q' \widetilde{Q}' & 
 \;\;\;\;\;\;\;\;\;\;\;\;\;\;\;\;\;\;\;
\;\; {(\bf 1, 1, 1)} \;\;\;\;\;\;\;\;\;
\;\; (f', \widetilde{g}' =1,  \cdots, N_f') 
\nonu \\
& (\Phi'' \equiv) G \widetilde{G} & 
 \;\;\;\;\;\;\;\;\;\;
\;\; {(\bf 1, 1, adj) \oplus ( 1, 1, 1)}
\nonu 
\eea

From the superpotential 
\bea
W_{dual}= \left( M' q' \widetilde{q}'  + m' M' \right) + 
X'' \widetilde{g} q' +
\widetilde{X}'' \widetilde{q}' g + \Phi'' g \widetilde{g} 
\nonu
\eea
one sees that $q' \widetilde{q}'$ has rank 
$\widetilde{N}_c'$ while $m'$ has a
rank $N_f'$.  If the rank $N_f'$ exceeds $\widetilde{N}_c'$, then 
the supersymmetry is broken. 
The classical moduli space of vacua can be obtained from F-term
equations.
Other F-term equations are satisfied if one takes the zero vacuum
expectation values for the fields $g, \widetilde{g}, X''$ and 
$\widetilde{X}''$.
Then, it is easy to see that 
$
\widetilde{q}' M' =0= M' q', 
 q'  \widetilde{q}' +  m'  =  0$.
Then the solutions can be written as
\bea
<q' >  & = &  \left(
\begin{array}{c}
\sqrt{m'} e^{\phi} {\bf 1}_{\widetilde{N}_c'}  \\
0
\end{array}
\right),  
< \widetilde{q}'> =
 \left(
\begin{array}{cc}
\sqrt{m'} e^{-\phi}  {\bf 1}_{\widetilde{N}_c'}   &
0
\end{array}
\right), 
<M'>  =
 \left(
\begin{array}{cc}
0  & 0 
 \\
0 & M_0'  {\bf 1}_{N_f'-\widetilde{N}_c'} 
\end{array}
\right),
\nonu \\
<g> & = & <\widetilde{g}> = <X''> = <\widetilde{X}''>= 0.
\nonu
\eea
By expanding the fields around the vacua and it 
turns out that states are stable by realizing the mass of 
$m_{M_0'}^2$ positive.

The nonsupersymmetric minimal energy brane configuration Figure 3B
(neglecting  the O6-plane and the mirrors)
leads to 
the Figure 3B of \cite{Ahn07-8} and 
the Figure 3B
with a replacement $N_f'$ D6-branes by 
the NS5'-brane(neglecting  the
$NS5_R'$-brane, $N_f''$ D6-branes and $N_c''$ D4-branes 
and $N_f$ D6-branes)
leads to 
the Figure 2B of \cite{Ahn07-7} with a rotation of NS5'-brane by 
$\frac{\pi}{2}$ angle.

The Riemann surface 
describing a set of NS5-branes 
with D4-branes suspended between them and 
in a background space of $x t = (-1)^{N_f+N_f'+N_f''} v^{2N_f+2N_f''+4}
(v^2 -m'^2)^{N_f'}$
was found.
Since we are dealing with seven NS-branes, the magnetic M5-brane 
configuration in Figure 3 with equal mass for $q'$ and massless 
for $Q(Q'')$ 
can be characterized by the following seventh order equation for $t$ 
as follows:
\bea
& & t^7 + \left[v^{N_c''} \right] t^6 + 
\left[ v^{\widetilde{N}_c'+N_f''}(v-m')^{N_f'} \right] t^5
+ \left[ v^{N_c+2N_f''} (v -m')^{2N_f'}\right]  t^4 
\nonu \\
&& + \left[ (-1)^{N_c}
v^{N_c +3N_f''+N_c+ 2} 
(v -m')^{3N_f'} \right] t^3  + 
\left[ (-1)^{\widetilde{N}_c'} v^{\widetilde{N}_c' +4N_f''+2N_c+4}  
(v -m')^{4N_f'} \right] t^2 \nonu \\
&& + \left[(-1)^{N_c''}
v^{10+5N_c+5N_f''+N_c''}  (v -m')^{5N_f'} 
 \right] t  \nonu \\
&& + \left[(-1)^{N_c+N_f'+N_f''} 
v^{14+7N_c+6N_f''} (v-m')^{6N_f'} (v+m')^{N_f'}\right] = 0.
\nonu
\eea

At nonzero string coupling constant, 
the NS-branes bend due to their interactions with the D4-branes and
D6-branes.
Now the asymptotic regions of various NS-branes 
can be determined similarly.
Then the behavior of the supersymmetric M5-brane curves can be
summarized 
as follows:

1. $v \rightarrow \infty$ limit implies
\bea
w & \rightarrow & 0, \quad y \sim    v^{N_c+N_f''-
\widetilde{N}_c'+N_f'} + \cdots \quad
\mbox{$\overline{NS5}$ 
asymptotic region}, \nonu \\
w & \rightarrow  & 0, \quad y \sim    
v^{N_c+N_f''+N_f'+2} + \cdots \quad
\mbox{$NS5_{M}$ asymptotic region}, 
\nonu \\
w & \rightarrow  & 0, \quad y \sim    
v^{N_f''+N_f'+\widetilde{N}_c'+2} + \cdots \quad
\mbox{NS5 asymptotic region}.   
\nonu
\eea

2.  $w \rightarrow \infty$ limit implies
\bea
v & \rightarrow &   -m', \quad 
y \sim  w^{\widetilde{N}_c'-N_c''+N_f''+N_f'}
 +\cdots
\quad \mbox{$\overline{NS5_{L}'}$ asymptotic region}, 
\nonu
\\
v & \rightarrow &  -m', \quad  
y \sim w^{N_c''}
+\cdots
\quad \mbox{$\overline{NS5_{R}'}$ asymptotic region}, 
\nonu \\
v & \rightarrow &   +m', \quad 
y \sim  w^{-\widetilde{N}_c'+3N_c+N_c''+N_f''+N_f'+6}
 +\cdots
\quad \mbox{$NS5_{L}'$ asymptotic region}, 
\nonu
\\
v & \rightarrow &  +m', \quad  
y \sim w^{-N_c''+2N_c+N_f''+2N_f'+4}
+\cdots
\quad \mbox{$NS5_{R}'$ asymptotic region}. 
\nonu
\eea
We denote the mirror branes by writing the bar on the
corresponding NS-brane.
The two $NS5_{L,R}'$-branes 
are moving in the $ +v$ direction respectively holding everything 
else fixed instead of moving D6-branes in the $+v$ direction.
The corresponding mirrors of D4-branes are moved appropriately.

\subsection{Magnetic theory with dual for second gauge group}

By moving the $NS5_L'$-brane in Figure 1
to the right all the way past the  
NS5-brane, one arrives at the Figure 4A.
The linking number of $NS5'_L$-brane from Figure 4A
is 
$
L_5 = \frac{N_f'}{2} -\widetilde{N}_c'+N_c''$ and
the linking number of $NS5'_L$-brane from the  
Figure 1
is
$
L_5 = -\frac{N_f'}{2} + N_c' -N_c$. 
From these two relations, one obtains
the number of colors of dual magnetic theory
\bea
\widetilde{N}_c' = N_f' + N_c''+ N_c-N_c'.
\nonu
\eea

Let us draw this magnetic brane configuration in Figure 4A and recall
that we put
the coincident $N_f'$ D6-branes in the nonzero $v$-direction in the
electric theory and consider massless flavors for $Q$ and $Q''$ by
putting $N_f$ and $N_f''$ D6-branes at $v=0$.
If we ignore the mirror branes 
and O6-plane(detaching these
branes from Figure 4A), 
then this brane configuration 
leads to the standard ${\cal N}=1$ magnetic gauge theory 
$SU(N_c) \times SU(\widetilde{N}_c'=N_f'+N_c+N_c''-N_c') 
\times SU(N_c'')$ with fundamentals and
bifundamentals in Figure 4 of \cite{Ahn07-8}.
On the other hand, when 
we ignore the $NS5_R'$-brane, $N_c''$ D4-branes and $N_f''$ 
D6-branes(detaching these
branes from Figure 4A), 
then this brane configuration 
leads to the ${\cal N}=1$ magnetic theory with gauge group 
$SU(N_c) \times SU(\widetilde{N}_c'=N_f' + N_c-N_c')$ with
fundamental 
flavors, bifundamentals, symmetric, conjugate 
symmetric flavors and gauge singlets in Figure 5 of \cite{Ahn07-4}.

Now let us recombine $\widetilde{N}_c'$ flavor D4-branes among $N_f'$
flavor 
D4-branes(connecting between D6-branes and NS5-brane) with the same number of 
color D4-branes(connecting between NS5-brane and $NS5_L'$-brane) and push
them in $+v$ direction from Figure 4A. 
For the flavor D4-branes, we are left with only 
$(N_f'-\widetilde{N}_c')=N_c'-N_c''-N_c$ flavor D4-branes
connecting between D6-branes and $NS5_L'$-brane. 

\begin{figure}[ht]
   \epsfxsize=4.0in 
\centerline{\epsffile{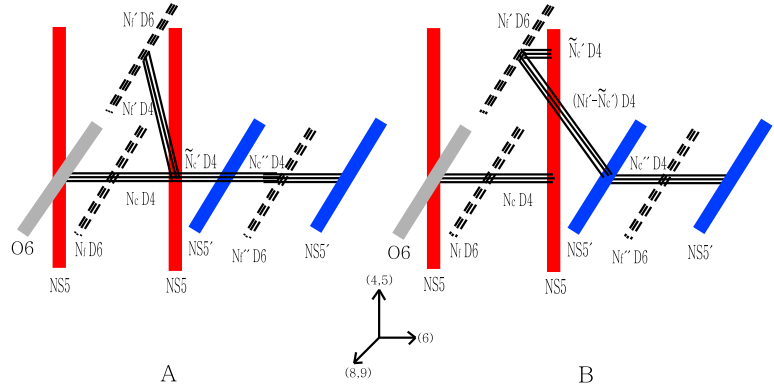}}
   \caption[FIG. \arabic{figure}.]{ 
The ${\cal N}=1$ supersymmetric magnetic brane configuration with
$SU(N_c) \times SU(\widetilde{N}_c'=N_f' + N_c''+ N_c-N_c') 
\times SU(N_c'')$ gauge group
with fundamentals $Q(q')[Q'']$ and
$\widetilde{Q}(\widetilde{q}')[\widetilde{Q}'']$ 
for each gauge group and bifundamentals $f(G)$ and
$\widetilde{f}(\widetilde{G})$, $S$ and $\widetilde{S}$, 
and gauge singlets in Figure 4A. In
Figure 4B, the nonsupersymmetric minimal energy brane configuration
with the same gauge group and matter contents above 
for massless  $Q(Q'')$ and
$\widetilde{Q}(\widetilde{Q}'')$ is given. 
}
\end{figure}

The additional $N_f'$-$SU(N_c)$ fundamentals $X$ and $N_f'$-$SU(N_c)$
antifundamentals $\widetilde{X}$ 
are originating from 
the $SU(N_c')$ chiral mesons $F Q'$ and $\widetilde{F}
\widetilde{Q}'$ 
respectively. 
The gauge singlet $M'$ corresponds to the $SU(N_c')$ 
chiral meson $Q' \widetilde{Q}'$
where the color indices are contracted.
The fluctuations of the gauge-singlet $M'$ correspond to the motion of $N_f'$
flavor D4-branes along (789) directions in Figure 4B(and their mirrors).
The $\Phi$ corresponds to the $SU(N_c')$ chiral meson $F \widetilde{F}$
where the color indices for the second gauge group 
are contracted each other.

The coefficient of the beta function of the first gauge group factor 
is given by
$
b_{SU(N_c)}^{mag}=3N_c-N_f-\widetilde{N}_c' -
(N_c+2)-N_f'-N_c$, 
the coefficient of the beta function of the second gauge group factor 
is given by
$
b_{SU(\widetilde{N}_c')}^{mag}=3\widetilde{N}_c'-N_f'-N_c-N_c''$,
and 
the coefficient of the beta function of the third gauge group factor 
is given by
$
b_{SU(N_c'')}^{mag}=3N_c''-N_f''-\widetilde{N}_c'$.

Then the gauge group and matter contents we consider 
are summarized as follows:
\bea
 & \mbox{gauge group}:& \;\;\;\;\;   SU(N_c) \times SU(\widetilde{N}_c') \times
 SU(N_c'')  \nonu
\\
\mbox{matter}: 
 &Q_f \oplus \widetilde{Q}_{\widetilde{f}}& \;\;\;\;\;\;\;\;\; 
\;\;\; {(\bf \Box, 1, 1) \oplus (\overline{\Box}, 1, 1)}
\;\;\;\;\; (f, \widetilde{f}=1,  \cdots, N_f) 
\nonu \\
 &q'_{f'} \oplus \widetilde{q}'_{\widetilde{f}'}& \;\;\;\;\;\;\;\;\;\;
\;\; {(\bf 1, \Box, 1) \oplus ( 1, \overline{\Box}, 1)}
\;\;\;\;\; (f', \widetilde{f}' =1,  \cdots, N_f') 
\nonu \\
 &Q''_{f''} \oplus \widetilde{Q}''_{\widetilde{f}''}& \;\;\;\;\;\;\;\;\;\; 
\;\; {(\bf 1, 1, \Box) \oplus ( 1, 1, \overline{\Box})} 
\;\;\;\;\; (f'', \widetilde{f}'' =1,  \cdots, N_f'')
\nonu \\
 &f \oplus \widetilde{f}& \;\;\;\;\;\;\;\;\; 
\;\; {(\bf \Box, \overline{\Box},1) \oplus (\overline{\Box}, \Box, 1)} 
\nonu \\
 &G \oplus \widetilde{G}& \;\;\;\;\;\;\;\;\; 
\;\; {(\bf 1, \Box, \overline{\Box}) \oplus ( 1, \overline{\Box}, \Box)} 
\nonu \\
 &S \oplus \widetilde{S}& \;\;\;\;\;\;\;\;\; 
\;\; {(\bf symm, 1, 1) \oplus (\overline{symm}, 1, 1)} 
\nonu \\ 
& (X_{n'} \equiv) F Q' \oplus \widetilde{F} \widetilde{Q}' (\equiv 
\widetilde{X}_{\widetilde{n}'}) & 
 \;\;\;\;\;\;\;\;\;\;
\;\; {(\bf \Box, 1, 1) \oplus ( \overline{\Box}, 1, 1)}
\;\; (n', \widetilde{n}' =1,  \cdots, N_f') 
\nonu \\
&  (M_{f',\widetilde{g}'}' \equiv) Q' \widetilde{Q}' & 
 \;\;\;\;\;\;\;\;\;\;\;\;\;\;\;\;\;\;\;
\;\; {(\bf 1, 1, 1)} \;\;\;\;\;\;\;\;\;
\;\; (f', \widetilde{g}' =1,  \cdots, N_f') 
\nonu \\
& (\Phi \equiv) F \widetilde{F} & 
 \;\;\;\;\;\;\;\;\;\;
\;\; {(\bf adj, 1, 1) \oplus ( 1, 1, 1)}
\nonu 
\eea

From the superpotential
\bea
W_{dual}= \left( M' q' \widetilde{q}'  + m' M' \right) + 
X f q' +
\widetilde{X} \widetilde{q}' \widetilde{f} + \Phi f \widetilde{f} 
\nonu
\eea
one sees that $q' \widetilde{q}'$ has rank 
$\widetilde{N}_c'$ while $m'$ has a
rank $N_f'$.  If the rank $N_f'$ exceeds $\widetilde{N}_c'$, then 
the supersymmetry is broken. 
The classical moduli space of vacua can be obtained from F-term
equations.
Then, it is easy to see that 
$
\widetilde{q}' M' =0= M' q', 
 q'  \widetilde{q}' +  m'  =  0$.
Then the solutions can be written as
\bea
<q' >  & = &  \left(
\begin{array}{c}
\sqrt{m'} e^{\phi} {\bf 1}_{\widetilde{N}_c'}  \\
0
\end{array}
\right),  
< \widetilde{q'}> =
 \left(
\begin{array}{cc}
\sqrt{m'} e^{-\phi}  {\bf 1}_{\widetilde{N}_c'}   &
0
\end{array}
\right), 
<M'>  =
 \left(
\begin{array}{cc}
0  & 0 
 \\
0 & M_0'  {\bf 1}_{N_f'-\widetilde{N}_c'} 
\end{array}
\right),
\nonu \\
<f> & = & <\widetilde{f}> = <X> = <\widetilde{X}>= 0.
\nonu
\eea
By expanding the fields around the vacua and it 
turns out that states are stable by realizing the mass of 
$m_{M_0'}^2$ positive.

The nonsupersymmetric minimal energy brane configuration Figure 4B
(neglecting  the O6-plane and the mirrors)
leads to 
the Figure 4B of \cite{Ahn07-8}.

The Riemann surface 
describing a set of NS5-branes 
with D4-branes suspended between them and 
in a background space of $x t = (-1)^{N_f+N_f'+N_f''} 
v^{2N_f+2N_f''+4} (v^2 -m'^2)^{N_f'}$
was found.
Since we are dealing with seven NS-branes, the magnetic M5-brane 
configuration in Figure 4 with equal mass for $q'$ and massless 
for $Q(Q'')$ 
can be characterized by the following seventh order equation for $t$ 
as follows:
\bea
& & t^7 + \left[v^{N_c''} \right] t^6 + 
\left[ v^{\widetilde{N}_c'+N_f''} \right] t^5
 + \left[ v^{N_c+2N_f''+2} \right]  t^4 
+ \left[ (-1)^{N_c}
v^{N_c +3N_f''+N_f+ 6} 
(v -m')^{N_f'} \right] t^3  \nonu \\
& & + 
\left[ (-1)^{\widetilde{N}_c'+N_f+N_f'} v^{\widetilde{N}_c' +4N_f''+2N_f+10}  
(v -m')^{2N_f'}(v+m')^{N_f'} \right] t^2 \nonu \\
&& + \left[(-1)^{N_c''}
v^{16+3N_f+5N_f''+N_c''}  (v -m')^{3N_f'} 
 (v +m')^{2N_f'}\right] t  \nonu \\
&& + \left[(-1)^{N_f+N_f'+N_f''} v^{22+4N_f+7N_f''} 
(v-m')^{4N_f'} (v+m')^{3N_f'}\right] = 0.
\nonu
\eea

At nonzero string coupling constant, 
the NS5-branes bend due to their interactions with the D4-branes and
D6-branes.
Then the behavior of the supersymmetric M5-brane curves can be
summarized 
as follows:

1. $v \rightarrow \infty$ limit implies
\bea
w & \rightarrow & 0, \quad y \sim    v^{N_c+N_f''+2-\widetilde{N}_c'}
+ 
\cdots \quad
\mbox{$\overline{NS5}$ 
asymptotic region}, \nonu \\
w & \rightarrow  & 0, \quad y \sim    
v^{N_f''+N_f+N_f'+4} + \cdots \quad
\mbox{$NS5_{M}$ asymptotic region}, 
\nonu \\
w & \rightarrow  & 0, \quad y \sim    
v^{N_f''+N_f+2N_f'-\widetilde{N}_c'-N_c+4} + \cdots \quad
\mbox{NS5 asymptotic region}.   
\nonu
\eea

2.  $w \rightarrow \infty$ limit implies
\bea
v & \rightarrow &   -m', \quad 
y \sim  w^{\widetilde{N}_c'-N_c''+N_f''}
 +\cdots
\quad \mbox{$\overline{NS5_{L}'}$ asymptotic region}, 
\nonu
\\
v & \rightarrow &  -m', \quad  
y \sim w^{N_c''}
+\cdots
\quad \mbox{$\overline{NS5_{R}'}$ asymptotic region}, 
\nonu \\
v & \rightarrow &   +m', \quad 
y \sim  w^{-\widetilde{N}_c'+N_c''+N_f''+N_f+2N_f'+4}
 +\cdots
\quad \mbox{$NS5_{L}'$ asymptotic region}, 
\nonu
\\
v & \rightarrow &  +m', \quad  
y \sim w^{-N_c''+2N_f''+N_f+2N_f'+6}
+\cdots
\quad \mbox{$NS5_{R}'$ asymptotic region}. 
\nonu
\eea
Here we denote the mirror branes by writing the bar on the
corresponding NS-brane.
The two $NS5_{L,R}'$-branes 
are moving in the $+v$ direction respectively holding everything 
else fixed instead of moving D6-branes in the $+v$ direction.
The corresponding mirrors of D4-branes are moved appropriately.

\subsection{Magnetic theory with dual for third gauge group}

By moving the NS5-brane with massive $N_f''$ D6-branes 
to the right all the way past the  
$NS5'_R$-brane, one arrives at the Figure 5A.
The linking number  of NS5-brane from Figure 5A
is given by 
$
L_5 = \frac{N_f''}{2} -\widetilde{N}_c''$ and
the linking number of NS5-brane from Figure 1
is
$
L_5 = -\frac{N_f''}{2} + N_c'' -N_c'$. 
From these two relations, one obtains
the number of colors of dual magnetic theory
\bea
\widetilde{N}_c'' = N_f'' + N_c'-N_c''.
\nonu
\eea

Let us draw this magnetic brane configuration in Figure 5A and recall
that we put
the coincident $N_f''$ D6-branes in the nonzero $v$-direction in the
electric theory and consider massless flavors for $Q$ and $Q'$ by
putting $N_f$ and $N_f'$ D6-branes at $v=0$.
If we ignore the mirror branes 
and O6-plane(detaching these
branes from Figure 5A), 
then this brane configuration 
leads to the standard ${\cal N}=1$ SQCD with the magnetic gauge group 
$SU(N_c) \times SU(N_c') \times 
SU(\widetilde{N}_c''=N_f''+N_c'-N_c'')$ with fundamentals,
bifundamentals, and singlets in Figure 2 of \cite{Ahn07-8}.

Now let us recombine $\widetilde{N}_c''$ flavor D4-branes among $N_f''$
flavor 
D4-branes(connecting between D6-branes and $NS5_R'$-brane) with the same number of 
color D4-branes(connecting between $NS5_R'$-brane and NS5-brane) and push
them in $+v$ direction from Figure 5A. 
For the flavor D4-branes, we are left with only 
$(N_f''-\widetilde{N}_c'')=N_c''-N_c'$ flavor D4-branes
connecting between D6-branes and $NS5_R'$-brane.  

\begin{figure}[ht]
   \epsfxsize=4.0in 
\centerline{\epsffile{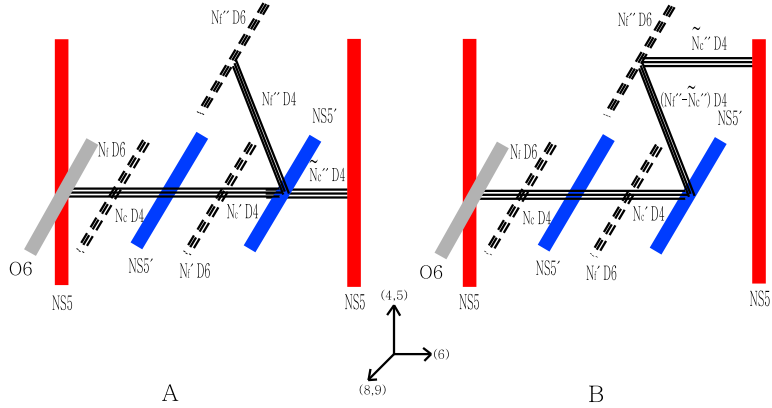}}
   \caption[FIG. \arabic{figure}.]{ 
The ${\cal N}=1$ supersymmetric magnetic brane configuration with
$SU(N_c) \times SU(N_c') \times SU(\widetilde{N}_c''=N_f'' +
N_c'-N_c'')$ 
gauge group
with fundamentals $Q(Q')[q'']$ and 
$\widetilde{Q}(\widetilde{Q}')[\widetilde{q}'']$ 
for each gauge group, bifundamentals $F(g)$ and
$\widetilde{F}(\widetilde{g})$, $S$ and $\widetilde{S}$, 
and gauge singlets in Figure 5A. In
Figure 5B, the nonsupersymmetric minimal energy brane configuration
with the same gauge group and matter contents above 
for massless  $Q(Q')$ and
$\widetilde{Q}(\widetilde{Q}')$ is given. 
}
\end{figure}

The $N_f''$-$SU(N_c')$ fundamentals $X'$ and $N_f''$-$SU(N_c')$
antifundamentals $\widetilde{X}'$ 
are originating from 
the $SU(N_c'')$ chiral mesons $G Q''$ and $\widetilde{G}
\widetilde{Q}''$ 
respectively. 
The gauge singlet $M''$ corresponds to the $SU(N_c'')$ 
chiral meson $Q'' \widetilde{Q}''$
where the color indices are contracted.
The fluctuations of the gauge-singlet $M''$ correspond to the motion of $N_f''$
flavor D4-branes along (789) directions in Figure 5A(and their mirrors).
The $\Phi'$ corresponds to the $SU(N_c'')$ chiral meson $G \widetilde{G}$
where the color indices for the third gauge group 
are contracted each other.

Then the gauge group and matter contents we consider 
are summarized as follows:
\bea
 & \mbox{gauge group}:& \;\;\;\;\;   SU(N_c) \times SU(N_c') \times
 SU(\widetilde{N}_c'')  \nonu
\\
\mbox{matter}:  
&Q_f \oplus \widetilde{Q}_{\widetilde{f}}& \;\;\;\;\;\;\;\;\; 
\;\;\; {(\bf \Box, 1, 1) \oplus (\overline{\Box}, 1, 1)}
\;\; (f, \widetilde{f}=1,  \cdots, N_f) 
\nonu \\
 &Q'_{f'} \oplus \widetilde{Q}'_{\widetilde{f}'}& \;\;\;\;\;\;\;\;\;\;
\;\; {(\bf 1, \Box, 1) \oplus ( 1, \overline{\Box}, 1)}
\;\; (f', \widetilde{f}' =1,  \cdots, N_f') 
\nonu \\
 &q''_{f''} \oplus \widetilde{q}''_{\widetilde{f}''}& \;\;\;\;\;\;\;\;\;\; 
\;\; {(\bf 1, 1, \Box) \oplus ( 1, 1, \overline{\Box})} 
\;\; (f'', \widetilde{f}'' =1,  \cdots, N_f'')
\nonu \\
&F \oplus \widetilde{F}& \;\;\;\;\;\;\;\;\; 
\;\; {(\bf \Box, \overline{\Box},1) \oplus (\overline{\Box}, \Box, 1)} 
\nonu \\
 &g \oplus \widetilde{g}& \;\;\;\;\;\;\;\;\; 
\;\; {(\bf 1, \Box, \overline{\Box}) \oplus ( 1, \overline{\Box}, \Box)} 
\nonu \\
 &S \oplus \widetilde{S}& \;\;\;\;\;\;\;\;\; 
\;\; {(\bf symm, 1, 1) \oplus (\overline{symm}, 1, 1)} 
\nonu \\ 
& (X_{n''}' \equiv) G Q'' \oplus \widetilde{G} \widetilde{Q}'' (\equiv 
\widetilde{X}_{\widetilde{n}''}') & 
 \;\;\;\;\;\;\;\;\;\;
\;\; {(\bf 1, \Box, 1) \oplus ( 1, \overline{\Box}, 1)}
\;\; (n'', \widetilde{n}'' =1,  \cdots, N_f'') 
\nonu \\
&  (M''_{f'',\widetilde{g}''} \equiv) Q'' \widetilde{Q}'' & 
 \;\;\;\;\;\;\;\;\;\;\;\;\;\;\;\;\;\;\;
\;\; {(\bf 1, 1, 1)} \;\;\;\;\;\;\;\;\;
\;\; (f'', \widetilde{g}'' =1,  \cdots, N_f'') 
\nonu \\
& (\Phi' \equiv) G \widetilde{G} & 
 \;\;\;\;\;\;\;\;\;\;
\;\; {(\bf 1, adj, 1) \oplus ( 1, 1, 1)}
\nonu
\eea

The coefficient of the beta function of the first gauge group factor 
$
b_{SU(N_c)}^{mag}=3N_c-N_f-N_c' -
(N_c+2) = b_{SU(N_c)}
$
and 
the coefficient of the beta function of the second gauge group factor 
is given by
$
b_{SU(N_c')}^{mag}=3N_c'-N_f'-N_c-\widetilde{N}_c''-N_f''-N_c'
$
and 
the coefficient of the beta function of the third gauge group factor 
is
$
b_{SU(\widetilde{N}_c'')}^{mag}=3\widetilde{N}_c''-N_f''-N_c'$.
Since $b_{SU(N_c')}-b_{SU(N_c')}^{mag} > 0$, $SU(N_c')$ is more
asymptotically free than $SU(N_c')^{mag}$.

The superpotential is given by
\bea
W_{dual}= \left( M'' q'' \widetilde{q}''  + m'' M'' \right) + 
X' g q'' +
\widetilde{X}' \widetilde{q}'' \widetilde{g} + \Phi' g \widetilde{g}. 
\nonu
\eea
Then, $q'' \widetilde{q}''$ has rank 
$\widetilde{N}_c''$ while $m''$ has a
rank $N_f''$.  The derivative of the 
superpotential $W_{dual}$ with respect to $M''$, cannot be satisfied 
if the rank $N_f''$ exceeds $\widetilde{N}_c''$ and the supersymmetry is broken. 
The classical moduli space of vacua can be obtained from F-term
equations.
Then, it is easy to see that 
$
\widetilde{q}'' M'' =0= M'' q'', 
 q''  \widetilde{q}'' +  m''  =  0$.
Then the solutions can be written as
\bea
<q'' >  & = &  \left(
\begin{array}{c}
\sqrt{m''} e^{\phi} {\bf 1}_{\widetilde{N}_c''}  \\
0
\end{array}
\right),  
< \widetilde{q}''> =
 \left(
\begin{array}{cc}
\sqrt{m''} e^{-\phi}  {\bf 1}_{\widetilde{N}_c''}   &
0
\end{array}
\right), 
<M''>  =
 \left(
\begin{array}{cc}
0  & 0 
 \\
0 & M_0''  {\bf 1}_{N_f''-\widetilde{N}_c''} 
\end{array}
\right),
\nonu \\
<g> & = & <\widetilde{g}> = <X'> = <\widetilde{X}'>= 0.
\nonu
\eea
One can analyze the one loop 
computation by expanding the fields around the vacua and it will
turn out that states are stable by realizing the mass of 
$m_{M_0''}^2$ positive.

The nonsupersymmetric minimal energy brane configuration Figure 5B
with a replacement $N_f''$ D6-branes by 
the NS5'-brane(neglecting  the
$NS5_L'$-brane, $N_f$ D6-branes and $N_c$ D4-branes 
and $N_f'$ D6-branes)
leads to 
the Figure 4B of \cite{Ahn07-7}.

The Riemann surface 
describing a set of NS5-branes 
with D4-branes suspended between them and 
in a background space of $x t = (-1)^{N_f+N_f'+N_f''} v^{2N_f+ 2N_f'+4}
(v^2 -m''^2)^{N_f''}$
was found.
Since we are dealing with seven NS-branes, the magnetic M5-brane 
configuration in Figure 5 with equal mass for $q''$  and massless 
for $Q(Q')$ 
can be characterized by the following seventh order equation for $t$ 
as follows:
\bea
& & t^7 + \left[v^{\widetilde{N}_c''} \right] t^6 + \left[ v^{N_c'} \right] t^5
 + \left[ v^{N_c+N_f'} (v -m'')^{N_f''}\right]  t^4 
\nonu \\
&& + \left[ (-1)^{N_c}
v^{2N_c +2N_f'+ 2} 
(v -m'')^{2N_f''} \right] t^3  + 
\left[ (-1)^{N_c+N_c'} v^{N_c' +3N_c+3N_f'+6}  
(v -m'')^{3N_f''} \right] t^2 \nonu \\
&& + \left[(-1)^{N_f''+N_f'+\widetilde{N}_c''}
v^{10+5N_c+4N_f'+\widetilde{N}_c''}  (v -m'')^{4N_f''} 
 (v +m'')^{N_f''}\right] t  \nonu \\
&& + \left[(-1)^{N_c} v^{14+5N_f'+7N_c} (v-m'')^{5N_f''} 
(v+m'')^{2N_f''}\right] = 0.
\nonu
\eea

At nonzero string coupling constant, 
the NS5-branes bend due to their interactions with the D4-branes and
D6-branes.
Then the behavior of the supersymmetric M5-brane curves can be
summarized 
as follows:

1. $v \rightarrow \infty$ limit implies
\bea
w & \rightarrow & 0, \quad y \sim    v^{\widetilde{N}_c''} + \cdots \quad
\mbox{$\overline{NS5}$ 
asymptotic region}, \nonu \\
w & \rightarrow  & 0, \quad y \sim    
v^{N_c+N_f''+N_f'+2} + \cdots \quad
\mbox{$NS5_{M}$ asymptotic region}, 
\nonu \\
w & \rightarrow  & 0, \quad y \sim    
v^{2N_f''+N_f'+2N_c-\widetilde{N}_c''+4} + \cdots \quad
\mbox{NS5 asymptotic region}.   
\nonu
\eea

2.  $w \rightarrow \infty$ limit implies
\bea
v & \rightarrow &   -m'', \quad 
y \sim  w^{N_c-N_c'+N_f''+N_f'}
 +\cdots
\quad \mbox{$\overline{NS5_{L}'}$ asymptotic region}, 
\nonu
\\
v & \rightarrow &  -m'', \quad  
y \sim w^{N_c'-\widetilde{N}_c''}
+\cdots
\quad \mbox{$\overline{NS5_{R}'}$ asymptotic region}, 
\nonu \\
v & \rightarrow &   +m'', \quad 
y \sim  w^{N_c+N_c'+N_f''+N_f'+4}
 +\cdots
\quad \mbox{$NS5_{L}'$ asymptotic region}, 
\nonu
\\
v & \rightarrow &  +m'', \quad  
y \sim w^{\widetilde{N}_c''-N_c'+2N_c+2N_f''+N_f'+4}
+\cdots
\quad \mbox{$NS5_{R}'$ asymptotic region}. 
\nonu
\eea
We denote the mirror branes by writing the bar on the
corresponding NS-brane.
The two $NS5_{L,R}'$-branes 
are moving in the $+v$ direction respectively holding everything 
else fixed instead of moving D6-branes in the $+v$ direction.
The corresponding mirrors of D4-branes are moved appropriately.

\subsection{Magnetic theories 
for the multiple product gauge groups}

Now one can generalize the method for the triple product gauge groups
to the finite $n$-multiple product gauge groups characterized by 
\bea
SU(N_{c,1}) \times SU(N_{c,2}) \cdots
\times SU(N_{c,n})
\nonu
\eea
with the matter, 
the $(n-1)$ bifundametals $({\bf \Box_1, \overline{\Box}_2, 1, \cdots,  1_n})$,
$\cdots$, and $({\bf 1_1, \cdots, 1, \Box_{n-1}, \overline{\Box}_{n}})$, their
complex conjugate $(n-1)$ fields $({\bf \overline{\Box}_1, \Box_2, 1, 
\cdots, 1_n})$,
$\cdots$, 
and $({\bf 1_1, \cdots, 1, \overline{\Box}_{n-1}, \Box_n})$, linking the
gauge groups together,
$n$-fundamentals $({\bf \Box_1, 1, \cdots, 1_n})$, $\cdots$, and 
$({\bf 1_1, \cdots,  1, \Box_n})$, $n$-antifundamentals  
$({\bf \overline{\Box}_1, 1, \cdots, 1_n})$, $\cdots$, and 
$({\bf 1_1, \cdots,  1, \overline{\Box}_n})$, and there exist  
a symmetric tensor $({\bf
  symm, 1_2, \cdots, 1_n})$, and
a conjugate symmetric tensor $({\bf \overline{symm}, 1_2, \cdots, 1_n})$.
Then the mass-deformed superpotential can be written as
$
W_{elec} = \sum_{i=1}^n m_i Q_i \widetilde{Q}_i$. 
The brane configuration can be constructed from Figure 1 by adding
$(n-3)$ NS-branes, $(n-3)$ sets of D6-branes and $(n-3)$
sets of D4-branes  to the right of $NS5_R'$-brane(and its mirrors)
leading to the fact that 
any two neighboring NS-branes should be perpendicular to each other. 

There exist $(2n-2)$ magnetic theories and they can be classified 
as follows.

$\bullet$ When the dual magnetic theory contains $SU(\widetilde{N}_{c,1})$

When the Seiberg dual is taken for the first gauge group factor
 by
assuming that $\Lambda_1 >> \Lambda_i$ where $i=2, \cdots, n$, 
one follows the procedure given in the subsection 2.2.
The gauge group is 
\bea
SU(\widetilde{N}_{c,1} \equiv 2N_{f,1} +2N_{c,2}-N_{c,1}) \times 
SU(N_{c,2}) \times \cdots \times SU(N_{c,n})
\nonu
\eea
and the matter contents are given by 
the dual quarks $q_1$ $({\bf \Box_1, 1, \cdots,
1_n})$ 
and $\widetilde{q}_1$ in the 
representation $({\bf \overline{\Box}_1, 1, \cdots, 1_n})$ 
as well as $(n-1)$ 
quarks $Q_i$ and
$\widetilde{Q}_i$ where $i=2, \cdots, n$, 
the bifundamentals $f_1$ in the representation 
 $({\bf \Box_1, \overline{\Box}_2, 1, \cdots,  1_n})$ under the dual 
gauge group, and $\widetilde{f}_1$ in the representation
$({\bf \overline{\Box}_1, \Box_2, 1, \cdots, 1_n})$ under the dual gauge
group  in
addition to $(n-2)$ bifundamentals $G_i$ and $\widetilde{G}_i$, 
a symmetric tensor $({\bf
  symm, 1_2, \cdots, 1_n})$, and
a conjugate symmetric tensor $({\bf \overline{symm}, 1_2, \cdots, 1_n})$,
and
various gauge singlets $X_2, \widetilde{X}_2, M_1$ and $\Phi_2$.
The corresponding brane configuration can be 
obtained similarly and 
the extra $(n-3)$ NS-branes, $(n-3)$ sets of D6-branes and $(n-3)$
sets of D4-branes  
are present at the right hand side of the $NS5_R'$-brane
of Figure 2(and their mirrors).
The magnetic superpotential can be written as
\bea
W_{dual} = \left(M_1 q_1 \widetilde{s} s 
\widetilde{q}_1 + f_1 \widetilde{X}_2 \widetilde{q}_1 + 
\widetilde{f}_1 q_1
X_2 + \Phi_2 f_1 \widetilde{f}_1 \right) + m_1 M_1.
\nonu
\eea
By computing the contribution for the one loop as in the subsection
2.2, 
the vacua are stable and the asymptotic behavior of $(2n+1)$ NS-branes
can be obtained also.  

$\bullet$ When the dual magnetic theory contains $SU(\widetilde{N}_{c,2})$

When the Seiberg dual is taken for the second gauge group factor
by
assuming that $\Lambda_2 >> \Lambda_j$ where $j=1, 3, \cdots,
\cdots, n$, 
one follows the procedure given in the subsection 2.3.
The gauge group is given by
\bea
SU(N_{c,1}) \times 
SU(\widetilde{N}_{c,2} \equiv N_{f,2}+N_{c,3}+N_{c,1}-N_{c,2}) 
\times \cdots \times SU(N_{c,n}).
\nonu
\eea
The corresponding brane configuration can be 
obtained similarly and 
the extra $(n-3)$ NS-branes, $(n-3)$ sets of D6-branes and $(n-3)$
sets of D4-branes  are present at the right hand side of the $NS5_R'$-brane
of Figure 3(and their mirrors).
The magnetic superpotential can be written as
\bea
W_{dual} = \left(M_2 q_2 \widetilde{q}_2 + 
g_2 \widetilde{X}_3 \widetilde{q}_2 + 
\widetilde{g}_2 q_2
X_3 + \Phi_3 g_2 \widetilde{g}_2 \right) + m_2 M_2.
\nonu
\eea
By computing the contribution for the one loop as in the subsection
2.3, 
the vacua are stable and the asymptotic behavior of $(2n+1)$ NS-branes
can be obtained. 

When the Seiberg dual is taken for the second gauge group factor with
different brane motion
by
assuming that $\Lambda_2 >> \Lambda_j$ where $j=1, 3,
\cdots, n$, 
one follows the procedure given in the subsection 2.4.
The gauge group is given by
\bea
SU(N_{c,1}) \times 
SU(\widetilde{N}_{c,2} \equiv N_{f,2}+N_{c,3}+N_{c,1}-N_{c,2}) 
\times \cdots \times SU(N_{c,n}).
\nonu
\eea
The corresponding brane configuration can be 
obtained similarly and 
the extra $(n-3)$ NS-branes, $(n-3)$ sets of D6-branes and $(n-3)$
sets of D4-branes  are present at the right hand side of the $NS5_R'$-brane
of Figure 4(and their mirrors).
The magnetic superpotential can be written as
\bea
W_{dual} = \left(M_2 q_2 \widetilde{q}_2 + f_1 X_1 q_2 + 
\widetilde{f}_1 \widetilde{q}_2
\widetilde{X}_1 + \Phi_1 f_1 \widetilde{f}_1 \right) + m_2 M_2.
\nonu
\eea
By computing the contribution for the one loop as in the subsection
2.4, 
the vacua are stable and the asymptotic behavior of $(2n+1)$ NS-branes
can be obtained.

$\bullet$ When the dual magnetic theory contains $SU(\widetilde{N}_{c,i})$ where
$ 3 \leq i \leq n-1$

When the Seiberg dual is taken for the middle gauge group factor
by
assuming that $\Lambda_i >> \Lambda_j$ where $j=1,2, \cdots, i-1, i+1,
\cdots, n$, 
one follows the procedure given in the subsection 2.3 of \cite{Ahn07-8}.
The gauge group is given by
\bea
SU(N_{c,1}) \times \cdots  \times 
SU(\widetilde{N}_{c,i} \equiv N_{f,i}+N_{c,i+1}+N_{c,i-1}-N_{c,i}) 
 \times \cdots \times SU(N_{c,n}).
\nonu
\eea
The corresponding brane configuration can be 
obtained similarly and 
the extra $(i-2)$ NS-branes, $(i-2)$ sets of D6-branes and $(i-2)$
sets of D4-branes  
are present between the $NS5_M$-brane and the NS5-brane
and the extra $(n-i-1)$ NS-branes, $(n-i-1)$ sets of D6-branes and $(n-i-1)$
sets of D4-branes  are present at the right hand side of the $NS5_R'$-brane
of Figure 3(and their mirrors).
The magnetic superpotential can be written as
\bea
W_{dual} = \left(M_{i} q_i \widetilde{q}_i + 
g_{i} \widetilde{X}_{i+1} \widetilde{q}_i + 
\widetilde{g}_{i} q_i
X_{i+1} + \Phi_{i+1} g_{i} \widetilde{g}_{i} \right) + m_i M_{i}.
\nonu
\eea
By computing the contribution for the one loop as in the subsection
2.3 of \cite{Ahn07-8}, 
the vacua are stable and the asymptotic behavior of $(2n+1)$ NS-branes
can be obtained. 

When the Seiberg dual is taken for the middle gauge group factor with
different brane motion
by
assuming that $\Lambda_i >> \Lambda_j$ where $j=1,2, \cdots, i-1, i+1,
\cdots, n$, 
one follows the procedure given in the subsection 2.4 of \cite{Ahn07-8}.
The gauge group is given by
\bea
SU(N_{c,1}) \times \cdots  \times 
SU(\widetilde{N}_{c,i} \equiv N_{f,i}+N_{c,i+1}+N_{c,i-1}-N_{c,i}) 
\times  \cdots \times SU(N_{c,n}).
\nonu
\eea
The corresponding brane configuration can be 
obtained similarly and 
the extra $(i-2)$ NS-branes, $(i-2)$ sets of D6-branes and $(i-2)$
sets of D4-branes  
are present between the $NS5_M$-brane and the NS5-brane of Figure 4
and the extra $(n-i-1)$ NS-branes, $(n-i-1)$ sets of D6-branes and $(n-i-1)$
sets of D4-branes  are present at the right hand side of the $NS5_R'$-brane
of Figure 4(and their mirrors).
The magnetic superpotential can be written as
\bea
W_{dual} = \left(M_{i} q_i \widetilde{q}_i + f_{i-1} X_{i-1} q_i + 
\widetilde{f}_{i-1} \widetilde{q}_i
\widetilde{X}_{i-1} + \Phi_{i-1} f_{i-1} \widetilde{f}_{i-1} \right) + m_i M_{i}.
\nonu
\eea
By computing the contribution for the one loop as in the subsection
2.4 of \cite{Ahn07-8}, 
the vacua are stable and the asymptotic behavior of $(2n+1)$ NS-branes
can be obtained.

$\bullet$ When the dual magnetic theory contains $SU(\widetilde{N}_{c,n})$

When the Seiberg dual is taken for the last gauge group factor by
assuming that $\Lambda_n >> \Lambda_i$ where $i=1,2, \cdots, (n-1)$, 
one follows the procedure given in the subsection 2.5.
The gauge group is given by
\bea
SU(N_{c,1}) \times \cdots \times 
SU(N_{c,n-1}) \times SU(\widetilde{N}_{c,n} \equiv N_{f,n} +N_{c,n-1}-N_{c,n}).
\nonu
\eea
The corresponding brane configuration can be 
obtained similarly and 
the extra $(n-3)$ NS-branes, $(n-3)$ sets of D6-branes and $(n-3)$
sets of D4-branes  
are present between the $NS5_M$-brane and  the $NS5_L'$-brane
of Figure 5(and their mirrors).
The magnetic superpotential can be written as
\bea
W_{dual} = \left(M_n q_n \widetilde{q}_n + g_{n-1} X_{n-1} q_n + 
\widetilde{g}_{n-1} \widetilde{q}_n
\widetilde{X}_{n-1} + \Phi_{n-1} g_{n-1} \widetilde{g}_{n-1} \right) + m_n M_n.
\nonu
\eea
By computing the contribution for the one loop as in the subsection
2.5, 
the vacua are stable and the asymptotic behavior of $(2n+1)$ NS-branes
can be obtained.  

\section{Meta-stable brane configurations
of multiple product gauge theories with different matters}


\subsection{Electric theory}

We describe the gauge theory with triple product gauge groups 
$SU(N_c) \times SU(N_c') \times SU(N_c'')$ where the antisymmetric, a
conjugate symmetric tensors and eight fundamentals 
are present in addition to the
fundamentals and bifundamentals.
The matter contents 
are 

$\bullet$
$N_f$-chiral multiplets $Q$ are  in the
representation $({\bf N_c, 1, 1
})$, and 
$N_f$-chiral multiplets $\widetilde{Q}$ are in  
the representation $({\bf \overline{N_c}, 1, 1})$

$\bullet$
$N_f'$-chiral multiplets $Q'$ are  in the
representation $({\bf 1, N_c', 1})$, and 
$N_f'$-chiral multiplets $\widetilde{Q}'$ are in  
the representation $({\bf 1,\overline{N_c'}, 1})$

$\bullet$
$N_f''$-chiral multiplets $Q''$ are  in the
representation $({\bf 1, 1, N_c''
})$, and 
$N_f''$-chiral multiplets $\widetilde{Q}''$ are in  
the representation $({\bf 1, 1, \overline{N_c''}})$

$\bullet$
Eight-chiral multiplets $\hat{Q}$ are  in the
representation $({\bf N_c, 1, 1
})$ 

$\bullet$
The flavor-singlet field $F$ is 
in the bifundamental representation $({\bf N_c, \overline{N_c'}, 1 })$, 
and its conjugate field $\widetilde{F}$
 is 
in the bifundamental representation $({\bf \overline{N_c}, N_c', 1})$

$\bullet$
The flavor-singlet field $G$ is 
in the bifundamental representation $({\bf 1, N_c', \overline{N_c''} })$, 
and its conjugate field $\widetilde{G}$
 is 
in the bifundamental representation $({\bf 1, \overline{N_c'}, N_c''})$

$\bullet$ The flavor-singlet field $A$, which is 
in an antisymmetric tensor representation under the $SU(N_c)$, is in the
representation $({\bf \frac{1}{2} N_c(N_c-1), 1, 1})$, and
conjugate field of symmetric tensor field $\widetilde{S}$ is in the 
representation $({\bf \overline{\frac{1}{2} N_c(N_c+1)}, 1, 1})$

In this case also if 
we ignore the antisymmetric and conjugate symmetric tensors $A$ and
$\widetilde{S}$ and eight-fundamentals $\hat{Q}$, 
this
theory was studied in \cite{BH,Ahn07-8}. If we put to $Q'',
\widetilde{Q}'', G, \widetilde{G}, A, \hat{Q}$ and 
$\widetilde{S}$ zero, then 
this becomes the product gauge group theory with fundamentals and
bifundamentals
\cite{ILS,BIWW,BH,AT97}.
Furthermore, if we put to $Q', \widetilde{Q}', Q'',
\widetilde{Q}'', F, \widetilde{F}, G$, and $\widetilde{G}$ zero, then 
this becomes the single gauge group theory with fundamentals and
bifundamentals, eight-fundamentals, antisymmetric and conjugate
symmetric tensors \cite{LLL1,BHKL,EGKT}.


The coefficient of the beta function of the first gauge group 
is given by
$
b_{SU(N_c)}=3N_c-(N_f+4)-N_c'-\frac{1}{2}(N_c+2)-\frac{1}{2}(N_c-2)
$
by realizing the index of the antisymmetric and symmetric
representations
of $SU(N_c)$ gauge group
and 
the coefficient of the beta function of the second gauge group  
is given by
$
b_{SU(N_c')}=3N_c'-N_f'-N_c-N_c''
$
and finally 
the coefficient of the beta function of the third gauge group
is given by
$
b_{SU(N_c'')}=3N_c''-N_f''-N_c'$.
We'll see how these coefficients change in the magnetic theory.
We denote the
strong coupling scales for $SU(N_c)$ as $\Lambda_1$, for $SU(N_c')$
as $\Lambda_2$ and for $SU(N_c'')$
as $\Lambda_3$ respectively.  

The electric superpotential with mass-deformed terms and an
interaction term between eight-fundamentals and conjugate symmetric
tensor field is 
\bea
W_{elec} & = & 
\left( \mu A_d^2 + \lambda Q A_d \widetilde{Q} + 
A A_d \widetilde{S} + \widetilde{F} A_d F +
\mu' A_d'^2 + \lambda' Q' A_d' \widetilde{Q}' + \widetilde{F} A_d' F +
\widetilde{G} A_d' G \right. \nonu \\
&+& \left.
 \mu'' A''^2 + \lambda'' Q'' A'' \widetilde{Q}'' + \widetilde{G} A'' G
\right)
+ \hat{Q} \widetilde{S} \hat{Q} +
m Q \widetilde{Q} + m' Q' \widetilde{Q}' + m'' Q'' \widetilde{Q}''.
\nonu
\eea
After integrating the adjoint fields 
$A_d$ for $SU(N_c)$, $A_d'$ for $SU(N_c')$ and $A_d''$ for $SU(N_c'')$ 
and taking $\mu, \mu'$ and $\mu''$ to infinity limit which is
equivalent to take any two NS-branes be perpendicular to each other,
the mass-deformed electric superpotential becomes 
$
W_{elec}  =  \hat{Q} \widetilde{S} \hat{Q} +
m Q \widetilde{Q} + m' Q' \widetilde{Q}' + m'' Q'' \widetilde{Q}''$.

The type IIA brane configuration for this mass-deformed theory 
can be described by as follows. 
The $N_c$-color 
D4-branes (01236) are suspended between the $NS5_M'$-brane (012389)
located at $x^6=0$ 
and the $NS5_L$-brane (012345) along positive $x^6$
direction,
together with $N_f$ D6-branes (0123789) 
which are parallel to $NS5_M'$-brane and have nonzero $v$ direction.
The NS5'-brane 
is located at the right hand side of
the $NS5_L$-brane along the positive $x^6$ direction and 
there exist $N_c'$-color D4-branes
suspended 
between them, with  $N_f'$ D6-branes which have nonzero $v$ direction. 
Moreover, 
the $NS5_R$-brane 
is located at the right hand side of
the NS5'-brane along the positive $x^6$ direction and there 
exist $N_c''$-color D4-branes
suspended 
between them, with  $N_f''$ D6-branes which have nonzero $v$ direction.
There exist two types of  orientifold 6-plane (0123789) at the origin $x^6=0$
and they act as $(x^4, x^5, x^6) \rightarrow (-x^4, -x^5, -x^6)$. 
Then the mirrors of above branes appear in 
the negative $x^6$ region and are denoted by bar on the corresponding branes.
From the left to the right, there are $\overline{NS5_R}$-,
$\overline{NS5'}$-, 
$\overline{NS5_L}$-, $NS5_M'$-,
$NS5_L$-, $NS5'$-, and $NS5_R$-branes.

We summarize the ${\cal N}=1$ supersymmetric electric brane
configuration in type IIA string theory as follows:

$\bullet$ Four
NS5-branes in (012345) directions. 

$\bullet$ Three
NS5'-branes in (012389) directions.

$\bullet$ Two sets of
$N_c(N_c')[N_c'']$-color D4-branes in (01236) directions. 
  
$\bullet$ Two sets of
$N_f(N_f')[N_f'']$ D6-branes in (0123789) directions. 

$\bullet$ Eight
half D6-branes in (0123789) directions. 

$\bullet$ Two
$O6^{\pm}$-planes in (0123789) directions with $x^6=0$

Now we draw this electric brane configuration in Figure 6 and we put
the coincident $N_f(N_f')[N_f'']$ D6-branes with positive $x^6$ in 
the nonzero $v$ direction in general. 
This brane configuration can be obtained from the brane configuration
of \cite{Ahn07-4} by adding the two outer NS5-branes(i.e., 
$\overline{NS5_R}$-brane and $NS5_R$-brane), two sets of $N_c''$ D4-branes
and two sets of $N_f''$ D6-branes or from the one of \cite{BH}
with the gauge theory of triple product gauge groups
by adding O6-planes and the extra NS-branes, D4-branes and D6-branes.
Then the mirrors with 
negative $x^6$ can be constructed by using the action of O6-plane and
are located at the positions by changing (456) directions for original
branes with minus signs.

\begin{figure}[ht]
   \epsfxsize=3.0in 
\centerline{\epsffile{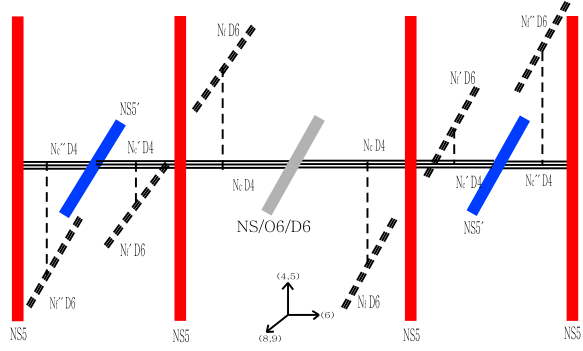}}
   \caption[FIG. \arabic{figure}.]{ 
The ${\cal N}=1$ supersymmetric electric brane configuration with
$SU(N_c) \times SU(N_c') \times SU(N_c'')$ gauge group with
fundamentals $Q(Q')[Q'']$ and
$\widetilde{Q}(\widetilde{Q}')[\widetilde{Q}'']$ 
for each gauge group and bifundamentals $F(G)$,  
$\widetilde{F}(\widetilde{G})$, an antisymmetric tensor $A$, a conjugate
symmetric tensor $\widetilde{S}$ and eight-fundamentals.
The $O6^{\pm}$-planes are located at the
origin $x^6=0$.
The  two NS5-branes with positive $x^6$ can be denoted by $NS5_{L,R}$-branes.
At the
origin of $(x^6, v,w)$ coordinates, there exist NS5'-brane,
$O6^{+}$-plane, $O6^{-}$-plane and eight half-D6-branes.  
One denotes this combination as $NS5/O6/D6$-branes here.
 }
\end{figure}

\subsection{Magnetic theory with dual for first gauge group}

By moving the mirror of  $NS5_L$-brane to the right all the way past
O6-plane(and the $NS5_L$-brane to the left) we arrive at the Figure
7A. Then the linking number
of $NS5_L$-brane from Figure 7A becomes 
$L_5 =\frac{N_f}{2}+\frac{1}{2}(4) -\widetilde{N}_c+N_c' +  N_f$. 
Originally, the linking number was $L_5=-
\frac{N_f}{2}-\frac{1}{2}(4) +N_c-N_c'$ from Figure 6.
This implies that the number of D4-branes in magnetic theory,
$\widetilde{N}_c$, 
becomes \cite{Ahn07-1}
\bea
\widetilde{N}_c = 2(N_f + N_c')-N_c +4.
\nonu
\eea

Let us draw this magnetic brane configuration in Figure 7A and recall
that we put
the coincident $N_f$ D6-branes in the nonzero $v$-direction in the
electric theory and consider massless flavors for $Q'$ and $Q''$ by
putting $N_f'$ and $N_f''$ D6-branes at $v=0$.
Because $N_c'$ or $N_c''$ D4-branes are suspending between two equal
$NS5_{L,R}$-branes located at different $x^6$ coordinate, these D4-branes
can slide along the $v$-direction.
If we ignore the NS5'-brane, $N_c'$ D4-branes, $N_f'$ 
D6-branes, the $NS5_R$-brane, $N_c''$ D4-branes and $N_f''$ 
D6-branes(detaching these
branes from Figure 7A), 
then this brane configuration 
leads to the ${\cal N}=1$ magnetic theory with gauge group 
$SU(\widetilde{N}_c=2N_f-N_c+4)$ with
$N_f$ massive fundamental 
flavors plus antisymmetric, conjugate symmetric flavors, eight-fundamentals 
and gauge singlets \cite{FGU,Ahn07-1}.
On the other hand, when 
we ignore the $NS5_R$-brane, $N_c''$ D4-branes and $N_f''$ 
D6-branes(detaching these
branes from Figure 7A), 
then this brane configuration 
leads to the ${\cal N}=1$ magnetic theory with gauge group 
$SU(\widetilde{N}_c=2N_f+2N_c'-N_c+4) \times SU(N_c')$ with
fundamental 
flavors, bifundamentals, antisymmetric, conjugate 
symmetric flavors, eight-fundamentals and 
gauge singlets in Figure 6 of  \cite{Ahn07-4}.

Now let us recombine $\widetilde{N}_c$ flavor D4-branes among $N_f$
flavor 
D4-branes(connecting between D6-branes and $NS5_L$-brane) with the same number of 
color D4-branes(connecting between $NS5_M'$-brane and $NS5_L$-brane) and push
them in $+v$ direction from Figure 7A. We assume that $N_c \geq N_f+2N_c'+4$. 
After this procedure, there are no color D4-branes between 
$NS5_M'$-brane and $NS5_L$-brane.
For the flavor D4-branes, we are left with only 
$(N_f-\widetilde{N}_c)=N_c-N_f-2N_c'-4$ flavor D4-branes
connecting between D6-branes and $NS5_M'$-brane.  

\begin{figure}[ht]
   \epsfxsize=4.0in 
\centerline{\epsffile{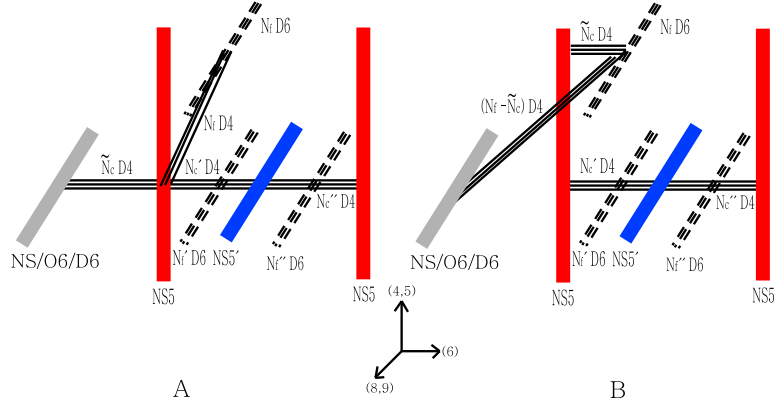}}
   \caption[FIG. \arabic{figure}.]{ 
The ${\cal N}=1$ supersymmetric magnetic brane configuration with
$SU(\widetilde{N}_c =2N_f + 2N_c'-N_c +4) \times SU(N_c') 
\times SU(N_c'')$ gauge group
with fundamentals $q(Q')[Q'']$ and
$\widetilde{q}(\widetilde{Q}')[\widetilde{Q}'']$ 
for each gauge group, $\hat{q}$,  and bifundamentals $f(G)$ and
$\widetilde{f}(\widetilde{G})$, $a$ and $\widetilde{s}$, and 
gauge singlets in Figure 7A. In
Figure 7B, the nonsupersymmetric minimal energy brane configuration
with the same gauge group and matter contents above 
for massless  $Q'(Q'')$ and
$\widetilde{Q}'(\widetilde{Q}'')$ is given. 
}
\end{figure}

The additional $N_f$-$SU(N_c')$ fundamentals $X'$ and $N_f$-$SU(N_c')$
antifundamentals $\widetilde{X}'$ 
are originating from 
the $SU(N_c)$ chiral mesons $\widetilde{F} Q$ and $F
\widetilde{Q}$ 
respectively. Therefore, 
there are free indices for a single color and a single flavor. 
Then the strings stretching between the $N_f$ $D6$-branes and $N_c'$
D4-branes will give rise to these additional $N_f$-$SU(N_c')$
fundamentals and  $N_f$-$SU(N_c')$
antifundamentals.
The gauge singlet $M$ corresponds to the $SU(N_c)$ 
chiral meson $Q \widetilde{Q}$
where the color indices are contracted.
Because the
$N_f$ D6-branes are parallel to the $NS5'_M$-brane from Figure 7B, 
the newly created $N_f$-flavor D4-branes can slide along the plane
consisting of these $N_f$ $D6$-branes and  $NS5'_M$-brane
freely.
The fluctuations of the gauge-singlet $M$ correspond to the motion of $N_f$
flavor D4-branes along (789) directions in Figure 7B.
For the nonsupersymmetric brane configuration,  
a misalignment for the $N_f$-flavor D4-branes arises and some of the
vacuum expectation value of $M$ is fixed and the remaining
components are arbitrary.
The $\Phi'$ corresponds to the $SU(N_c)$ chiral meson $F \widetilde{F}$
where the color indices for the first gauge group 
are contracted each other.
The fluctuations of the singlet $\Phi'$ correspond to the motion of
$N_c'$ D4-branes suspended two $NS5_{L,R}$-branes along the $v$
direction in Figure 7B.  
Although the gauge singlets 
$N, \hat{M}, P$ and $\widetilde{P}$
appear in the dual magnetic
superpotential for the general rotation angles of NS-branes and D6-branes,
the case we are considering 
does not contain these gauge singlets, as observed in \cite{Ahn07-1}.

Then the gauge group and matter contents we consider 
are summarized as follows:
\bea
 & \mbox{gauge group}:& \;\;\;\;\;   SU(\widetilde{N}_c) \times SU(N_c') \times
 SU(N_c'')  \nonu
\\
\mbox{matter}: 
 &q_f \oplus \widetilde{q}_{\widetilde{f}}& \;\;\;\;\;\;\;\;\; 
\;\;\; {(\bf \Box, 1, 1) \oplus (\overline{\Box}, 1, 1)}
\;\;\;\;\; (f, \widetilde{f}=1,  \cdots, N_f) 
\nonu \\
 &Q'_{f'} \oplus \widetilde{Q}'_{\widetilde{f}'}& \;\;\;\;\;\;\;\;\;\;
\;\; {(\bf 1, \Box, 1) \oplus ( 1, \overline{\Box}, 1)}
\;\;\;\;\; (f', \widetilde{f}' =1,  \cdots, N_f') 
\nonu \\
 &Q''_{f''} \oplus \widetilde{Q}''_{\widetilde{f}''}& \;\;\;\;\;\;\;\;\;\; 
\;\; {(\bf 1, 1, \Box) \oplus ( 1, 1, \overline{\Box})} 
\;\;\;\;\; (f'', \widetilde{f}'' =1,  \cdots, N_f'')
\nonu \\
 &\hat{q}_f & \;\;\;\;\;\;\;\;\; 
\;\;\;\;\;\;\;\;\;\;\; {(\bf \Box, 1, 1) }
\;\;\;\;\;\;\;\;\;\;\;\;\;\; (f=1,  \cdots, 8) 
\nonu \\ 
&f \oplus \widetilde{f}& \;\;\;\;\;\;\;\;\; 
\;\; {(\bf \Box, \overline{\Box},1) \oplus (\overline{\Box}, \Box, 1)} 
\nonu \\
 &G \oplus \widetilde{G}& \;\;\;\;\;\;\;\;\; 
\;\; {(\bf 1, \Box, \overline{\Box}) \oplus ( 1, \overline{\Box}, \Box)} 
\nonu \\
 &a \oplus \widetilde{s}& \;\;\;\;\;\;\;\;\; 
\;\; {(\bf asymm, 1, 1) \oplus (\overline{symm}, 1, 1)} 
\nonu \\ 
& (X_{n}' \equiv) \widetilde{F} Q \oplus F \widetilde{Q} (\equiv 
\widetilde{X}_{\widetilde{n}}') & 
 \;\;\;\;\;\;\;\;\;\;
\;\; {(\bf 1, \Box, 1) \oplus ( 1, \overline{\Box}, 1)}
\;\; (n, \widetilde{n} =1,  \cdots, N_f) 
\nonu \\
&  (M_{f,\widetilde{g}} \equiv) Q \widetilde{Q} & 
 \;\;\;\;\;\;\;\;\;\;\;\;\;\;\;\;\;\;\;
\;\; {(\bf 1, 1, 1)} \;\;\;\;\;\;\;\;\;
\;\; (f, \widetilde{g} =1,  \cdots, N_f) \nonu \\
&  (\hat{M}_{f,\widetilde{g}} \equiv) \hat{Q} \widetilde{Q} & 
 \;\;\;\;\;\;\;\;\;\;\;\;\;\;\;\;\;\;\;
\;\; {(\bf 1, 1, 1)} \; (f=1, \cdots, 8, \widetilde{g} =1,  \cdots, N_f) 
\nonu \\
& (\Phi' \equiv) F \widetilde{F} & 
 \;\;\;\;\;\;\;\;\;\;
\;\; {(\bf 1, adj, 1) \oplus ( 1, 1, 1)}
\nonu \\
& (P_{f, g} \equiv)  Q \widetilde{S} Q \oplus \widetilde{Q} A \widetilde{Q} (\equiv 
\widetilde{P}_{\widetilde{f}, \widetilde{g}}) & 
 \;\;\;\;\;\;\;\;\;\;
\;\; {(\bf 1, 1, 1) \oplus ( 1, 1, 1)}
\;\; (f, \widetilde{f}, g, \widetilde{g} =1,  \cdots, N_f) 
\nonu \\
& (N_{f, \widetilde{g}} \equiv) Q \widetilde{S} A \widetilde{Q} & 
 \;\;\;\;\;\;\;\;\;\;\;\;\;\;\;\;\;\;
\;\; {(\bf 1, 1, 1)}  \;\;
(f, \widetilde{g} =1,  \cdots, N_f) 
\nonu
\eea

The coefficient of the beta function of the first dual gauge group factor 
is given by
$
b_{SU(\widetilde{N}_c)}^{mag}=3\widetilde{N}_c-(N_f+4)-N_c'-
\frac{1}{2}(\widetilde{N}_c+2)-\frac{1}{2}(\widetilde{N}_c-2)
$
and 
the coefficient of the beta function of the second gauge group factor 
is given by
$
b_{SU(N_c')}^{mag}=3N_c'-N_f'-\widetilde{N}_c-N_c''-N_f-N_c'
$
and 
the coefficient of the beta function of the third gauge group factor
is given by 
$
b_{SU(N_c'')}^{mag}=3N_c''-N_f''-N_c'=b_{SU(N_c'')}$.

From the superpotential obtained from \cite{LLL1,BHKL,EGKT} partly
\bea
W_{dual}= \left(M q \widetilde{s} a \widetilde{q} +  m M  +
\hat{q} \widetilde{s} \hat{q} + \hat{M} \hat{q} \widetilde{q} \right)+ 
\widetilde{f} X' q +
f \widetilde{q} \widetilde{X}' + \Phi' f \widetilde{f}, 
\nonu
\eea
then, $q \widetilde{s} a \widetilde{q}$ has rank 
$\widetilde{N}_c$ while $m$ has a
rank $N_f$.  Therefore, the derivative of the 
superpotential $W_{dual}$ with respect to $M$, cannot be satisfied 
if the rank $N_f$ exceeds $\widetilde{N}_c$ and the supersymmetry is broken. 
The classical moduli space of vacua can be obtained from F-term
equations.
Then, it is easy to see that 
$
a \widetilde{q} M =0= M q \widetilde{s}, 
 q \widetilde{s} a \widetilde{q} +  m  =  0$.
Then the solutions can be written as
\bea
<q \widetilde{s}>  & = &  \left(
\begin{array}{c}
\sqrt{m} e^{\phi} {\bf 1}_{\widetilde{N}_c}  \\
0
\end{array}
\right),  
<a \widetilde{q}> =
 \left(
\begin{array}{cc}
\sqrt{m} e^{-\phi}  {\bf 1}_{\widetilde{N}_c}   &
0
\end{array}
\right), 
<M>  =
 \left(
\begin{array}{cc}
0  & 0 
 \\
0 & M_0  {\bf 1}_{N_f-\widetilde{N}_c} 
\end{array}
\right),
\nonu \\
<f> & = & <\widetilde{f}> = <X'> = <\widetilde{X}'>=<\hat{q}> = 
<\hat{M}>= 0.
\nonu
\eea
Let us expand around on a point on the vacua, as done in
\cite{ISS}. 
Then the remaining relevant terms of superpotential are given by
$
W_{dual}^{rel}  =   M_0 \left( \delta \hat{\varphi}  
\; \delta \hat{\widetilde{\varphi}} + m \right) +
  \delta Z \; \delta \hat{\varphi} \; a_0 \; \widetilde{q}_0 
+ \delta \widetilde{Z} \; q_0 \; \widetilde{s}_0 \;
\delta \hat{\widetilde{\varphi}}
$
by following the fluctuations for the various fields in \cite{Ahn07-1}.
Note that there exist five kinds of terms, 
the vacuum  $<q>$ multiplied by 
$\delta \widetilde{f} \delta X'$,  
the vacuum  $<\widetilde{q}>$ multiplied by $\delta \widetilde{X}' 
\delta f$, 
the vacuum  $<\Phi'>$ multiplied by $\delta f 
\delta \widetilde{f}$, the vacuum $<\widetilde{s}>$ multiplied by 
$\delta \hat{q} \delta \hat{q}$, and the vacuum $\widetilde{q}$
multiplied by $\delta \hat{M} \delta \hat{q}$.
By redefining these as before, they do not enter the 
contributions for the one loop result, up to quadratic order. 
As done in \cite{Ahn07}, the defining function ${\cal F}(v^2)$ can be
computed
and using the equation (2.14) of \cite{Shih} 
of $m_{M_0}^2$ and ${\cal F}(v^2)$, one gets 
that $m_{M_0}^2$ will contain $(\log 4 -1) > 0$ implying that these
are stable.

The nonsupersymmetric minimal energy brane configuration  Figure 7B
with a replacement $N_f''$ D6-branes by 
the NS5'-brane(neglecting  the
$NS5_R$-brane, $N_f''$ D6-branes and $N_c''$ D4-branes 
and $N_f'$ D6-branes)
leads to 
the Figure 7B of \cite{Ahn07-7} with a rotation of NS5-brane by
$\frac{\pi}{2}$
angle.

At nonzero string coupling constant, 
the NS5-branes bend due to their interactions with the D4-branes and
D6-branes.
Now the asymptotic regions of various NS-branes 
can be determined by reading off the first two terms of the seventh order
curve above giving the
$\overline{NS5_R}$-brane
asymptotic region, next two terms giving 
the $\overline{NS5'}$-brane asymptotic region, next two terms
giving the $\overline{NS5_L}$-brane asymptotic region, next two terms 
giving $NS5_M'$-brane asymptotic region, 
 next two terms 
giving $NS5_L$-brane asymptotic region, 
 next two terms 
giving NS5'-brane asymptotic region, 
and 
final two terms giving $NS5_R$-brane asymptotic region.
Then the behavior of the supersymmetric M5-brane curves can be
summarized 
as follows:

1. $v \rightarrow \infty$ limit implies
\bea
w & \rightarrow & 0, \quad y \sim    v^{N_c''} + \cdots \quad
\mbox{$\overline{NS5_R}$ 
asymptotic region}, \nonu \\
w & \rightarrow  & 0, \quad y \sim    
v^{-\widetilde{N}_c+N_c'+N_f''+N_f+N_f'+4} + \cdots \quad
\mbox{$NS5_{L}$ asymptotic region}, 
\nonu \\
w & \rightarrow  & 0, \quad y \sim    
v^{\widetilde{N}_c-N_c'+N_f''+N_f+N_f'} + \cdots \quad
\mbox{$\overline{NS5_L}$ asymptotic region},
\nonu \\
w & \rightarrow &   0, \quad 
y \sim  v^{-N_c''+2N_f''+2N_f+2N_f'+4}
 +\cdots
\quad \mbox{$NS5_{R}$ asymptotic region}. 
\nonu
\eea

2.  $w \rightarrow \infty$ limit implies
\bea
v & \rightarrow &  -m, \quad  
y \sim w^{N_c'+N_f''-N_c''}
+\cdots
\quad \mbox{$\overline{NS5'}$ asymptotic region}, 
\nonu \\
v & \rightarrow &   +m, \quad 
y \sim  w^{N_f''+N_f+N_f'+2}
 +\cdots
\quad \mbox{$NS5_{M}'$ asymptotic region}, 
\nonu
\\
v & \rightarrow &  +m, \quad  
y \sim w^{N_f''+2N_f+2N_f'-N_c'+N_c''+4}
+\cdots
\quad \mbox{NS5' asymptotic region}. 
\nonu
\eea

\subsection{Magnetic theory with dual for second gauge group}

By moving the NS5'-brane in Figure 6 with massive $N_f'$ D6-branes
to the left all the way past the  
$NS5_L$-brane, one arrives at the Figure 8A.
The linking number of NS5'-brane from Figure 8A
is 
$
L_5 = -\frac{N_f'}{2} +\widetilde{N}_c'-N_c
$ 
while
the linking number of NS5'-brane from Figure 6
is
$
L_5 = \frac{N_f'}{2} + N_c'' -N_c'$. 
From these two relations, one obtains
the number of colors of dual magnetic theory
\bea
\widetilde{N}_c' = N_f' + N_c''+ N_c-N_c'.
\nonu
\eea

Let us draw this magnetic brane configuration in Figure 8A and recall
that we put
the coincident $N_f'$ D6-branes in the nonzero $v$-direction in the
electric theory and consider massless flavors for $Q$ and $Q''$ by
putting $N_f$ and $N_f''$ D6-branes at $v=0$.
If we ignore the mirror branes corresponding to 
$NS5_{L,R}'$-branes, NS5-brane, $N_f(N_f')[N_f'']$ D6-branes
and $N_c(N_c')[N_c'']$ D4-branes, 
O6-planes and half D6-branes(detaching these
branes from Figure 8A), 
then this brane configuration 
looks similar to the standard ${\cal N}=1$ magnetic gauge theory 
$SU(N_c) \times 
SU(\widetilde{N}_c'=N_f'+N_c+N_c''-N_c') \times SU(N_c'')$ with fundamentals, 
bifundamentals, and singlets in Figure 4 of \cite{Ahn07-8}.

Now let us recombine $\widetilde{N}_c'$ flavor D4-branes among $N_f'$
flavor 
D4-branes(connecting between D6-branes and $NS5_L$-brane) with the same number of 
color D4-branes(connecting between NS5'-brane and $NS5_L$-brane) and push
them in $+v$ direction from Figure 8A. 
For the flavor D4-branes, we are left with only 
$(N_f'-\widetilde{N}_c')=N_c'-N_c''-N_c$ flavor D4-branes
connecting between D6-branes and NS5'-brane.

\begin{figure}[ht]
   \epsfxsize=4.0in 
\centerline{\epsffile{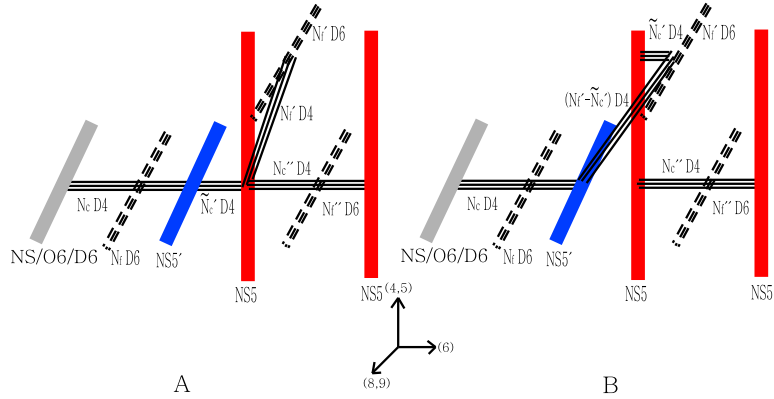}}
   \caption[FIG. \arabic{figure}.]{ 
The ${\cal N}=1$ supersymmetric magnetic brane configuration with
$SU(N_c) \times SU(\widetilde{N}_c'= N_f' + N_c''+ N_c-N_c') 
\times SU(N_c'')$ gauge group
with fundamentals $Q(q')[Q'']$ and
$\widetilde{Q}(\widetilde{q}')[\widetilde{Q}'']$ 
for each gauge group and bifundamentals $F(g)$, $\hat{Q}$, 
$\widetilde{F}(\widetilde{g})$, $A$ and $\widetilde{S}$,  
and gauge singlets in Figure 8A. In
Figure 8B, the nonsupersymmetric minimal energy brane configuration
with the same gauge group and matter contents above 
for massless  $Q(Q'')$ and
$\widetilde{Q}(\widetilde{Q}'')$ is given. 
}
\end{figure}

The additional $N_f'$-$SU(N_c'')$ fundamentals $X''$ and $N_f'$-$SU(N_c'')$
antifundamentals $\widetilde{X}''$ 
are originating from 
the $SU(N_c')$ chiral mesons $\widetilde{G} Q'$ and $G
\widetilde{Q}'$ 
respectively. 
The gauge singlet $M'$ corresponds to the $SU(N_c')$ 
chiral meson $Q' \widetilde{Q}'$
where the color indices are contracted.
The fluctuations of the gauge-singlet $M'$ correspond to the motion of $N_f'$
flavor D4-branes along (789) directions in Figure 8B.
The $\Phi''$ corresponds to the $SU(N_c')$ chiral meson $G \widetilde{G}$
where the color indices for the second gauge group 
are contracted each other.

The coefficient of the beta function of the first gauge group factor 
is given by
$
b_{SU(N_c)}^{mag}=3N_c-(N_f+4)-\widetilde{N}_c'-
\frac{1}{2}(N_c+2)-\frac{1}{2}(N_c-2)
$
as before and 
the coefficient of the beta function of the second gauge group factor 
is given by
$
b_{SU(\widetilde{N}_c')}^{mag}=3\widetilde{N}_c'-N_f'-N_c-N_c''$.
The coefficient of the beta function of the third gauge group factor 
is given by
$
b_{SU(N_c'')}^{mag}=3N_c''-N_f''-\widetilde{N}_c'-N_f'-N_c''$.

Then the gauge group and matter contents we consider 
are summarized as follows:
\bea
 & \mbox{gauge group}:& \;\;\;\;\;   SU(N_c) \times SU(\widetilde{N}_c') \times
 SU(N_c'')  \nonu
\\
\mbox{matter}: 
 &Q_f \oplus \widetilde{Q}_{\widetilde{f}}& \;\;\;\;\;\;\;\;\; 
\;\;\; {(\bf \Box, 1, 1) \oplus (\overline{\Box}, 1, 1)}
\;\;\;\;\; (f, \widetilde{f}=1,  \cdots, N_f) 
\nonu \\
 &q'_{f'} \oplus \widetilde{q}'_{\widetilde{f}'}& \;\;\;\;\;\;\;\;\;\;
\;\; {(\bf 1, \Box, 1) \oplus ( 1, \overline{\Box}, 1)}
\;\;\;\;\; (f', \widetilde{f}' =1,  \cdots, N_f') 
\nonu \\
 &Q''_{f''} \oplus \widetilde{Q}''_{\widetilde{f}''}& \;\;\;\;\;\;\;\;\;\; 
\;\; {(\bf 1, 1, \Box) \oplus ( 1, 1, \overline{\Box})} 
\;\;\;\;\; (f'', \widetilde{f}'' =1,  \cdots, N_f'')
\nonu \\
 &\hat{Q}_f & \;\;\;\;\;\;\;\;\; 
\;\;\;\;\;\;\;\;\;\;\; {(\bf \Box, 1, 1) }
\;\;\;\;\;\;\;\;\;\;\;\;\;\; (f=1,  \cdots, 8) 
\nonu \\ 
 &F \oplus \widetilde{F}& \;\;\;\;\;\;\;\;\; 
\;\; {(\bf \Box, \overline{\Box},1) \oplus (\overline{\Box}, \Box, 1)} 
\nonu \\
 &g \oplus \widetilde{g}& \;\;\;\;\;\;\;\;\; 
\;\; {(\bf 1, \Box, \overline{\Box}) \oplus ( 1, \overline{\Box}, \Box)} 
\nonu \\
 &A \oplus \widetilde{S}& \;\;\;\;\;\;\;\;\; 
\;\; {(\bf asymm, 1, 1) \oplus (\overline{symm}, 1, 1)} 
\nonu \\ 
& (X_{n'}'' \equiv) \widetilde{G} Q' \oplus G \widetilde{Q}' (\equiv 
\widetilde{X}_{\widetilde{n}'}'') & 
 \;\;\;\;\;\;\;\;\;\;
\;\; {(\bf 1, 1, \Box) \oplus ( 1, 1, \overline{\Box})}
\;\; (n', \widetilde{n}' =1,  \cdots, N_f') 
\nonu \\
&  (M_{f',\widetilde{g}'}' \equiv) Q' \widetilde{Q}' & 
 \;\;\;\;\;\;\;\;\;\;\;\;\;\;\;\;\;\;\;
\;\; {(\bf 1, 1, 1)} \;\;\;\;\;\;\;\;\;
\;\; (f', \widetilde{g}' =1,  \cdots, N_f') 
\nonu \\
& (\Phi'' \equiv) G \widetilde{G} & 
 \;\;\;\;\;\;\;\;\;\;
\;\; {(\bf 1, 1, adj) \oplus ( 1, 1, 1)}
\nonu 
\eea

From the superpotential
\bea
W_{dual}= \left( M' q' \widetilde{q}'  + m' M' \right) + 
X'' \widetilde{g} q' +
\widetilde{X}'' \widetilde{q}' g + \Phi'' g \widetilde{g} 
\nonu
\eea
one sees that $q' \widetilde{q}'$ has rank 
$\widetilde{N}_c'$ while $m'$ has a
rank $N_f'$.  If the rank $N_f'$ exceeds $\widetilde{N}_c'$, then 
the supersymmetry is broken. 
The classical moduli space of vacua can be obtained from F-term
equations.
Then, it is easy to see that 
$
\widetilde{q'} M' =0= M' q', 
 q'  \widetilde{q'} +  m'  =  0$.
Then the solutions can be written as
\bea
<q' >  & = &  \left(
\begin{array}{c}
\sqrt{m'} e^{\phi} {\bf 1}_{\widetilde{N}_c'}  \\
0
\end{array}
\right),  
< \widetilde{q'}> =
 \left(
\begin{array}{cc}
\sqrt{m'} e^{-\phi}  {\bf 1}_{\widetilde{N}_c'}   &
0
\end{array}
\right), 
<M'>  =
 \left(
\begin{array}{cc}
0  & 0 
 \\
0 & M_0'  {\bf 1}_{N_f'-\widetilde{N}_c'} 
\end{array}
\right),
\nonu \\
<g> & = & <\widetilde{g}> = <X''> = <\widetilde{X}''>= 0.
\nonu
\eea
By expanding the fields around the vacua and it 
turns out that states are stable by realizing the mass of 
$m_{M_0'}^2$ positive.

The Figure 8B
with a replacement $N_f'$ D6-branes by 
the NS5'-brane(neglecting  the
$NS5_R$-brane, $N_f''$ D6-branes and $N_c''$ D4-branes 
and $N_f$ D6-branes)
leads to 
the Figure 8B of \cite{Ahn07-7}.

At nonzero string coupling constant, 
the NS5-branes bend due to their interactions with the D4-branes and
D6-branes.
Then the behavior of the supersymmetric M5-brane curves can be
summarized 
as follows:

1. $v \rightarrow \infty$ limit implies
\bea
w & \rightarrow & 0, \quad y \sim    v^{N_c''} + \cdots \quad
\mbox{$\overline{NS5_R}$ 
asymptotic region}, \nonu \\
w & \rightarrow  & 0, \quad y \sim    
v^{-\widetilde{N}_c'+3N_c+N_c''+N_f''+N_f'+6} + \cdots \quad
\mbox{$NS5_{L}$ asymptotic region}, 
\nonu \\
w & \rightarrow  & 0, \quad y \sim    
v^{\widetilde{N}_c'-N_c''+N_f''+N_f'} + \cdots \quad
\mbox{$\overline{NS5_L}$ asymptotic region},
\nonu \\
w & \rightarrow &   0, \quad 
y \sim  v^{-N_c''+2N_c+N_f''+2N_f'+4}
 +\cdots
\quad \mbox{$NS5_{R}$ asymptotic region}. 
\nonu
\eea

2.  $w \rightarrow \infty$ limit implies
\bea
v & \rightarrow &  -m', \quad  
y \sim w^{N_c+N_f''-\widetilde{N}_c'+N_f'}
+\cdots
\quad \mbox{$\overline{NS5'}$ asymptotic region}, 
\nonu \\
v & \rightarrow &   +m', \quad 
y \sim  w^{N_c+N_f''+N_f'+2}
 +\cdots
\quad \mbox{$NS5_{M}'$ asymptotic region}, 
\nonu
\\
v & \rightarrow &  +m', \quad  
y \sim w^{N_f''+N_f'+\widetilde{N}_c'+2}
+\cdots
\quad \mbox{NS5' asymptotic region}. 
\nonu
\eea

\subsection{Magnetic theory with dual for second gauge group}

By moving the $NS5_L$-brane in Figure 6
with massive $N_f'$ D6-branes to the right all the way past the  
NS5'-brane, one arrives at the Figure 9A.
The linking number of $NS5_L$-brane from Figure 9A
is 
$
L_5 = \frac{N_f'}{2} -\widetilde{N}_c'+N_c''$ and
the linking number of $NS5_L$-brane from the  
Figure 6
is
$
L_5 = -\frac{N_f'}{2} + N_c' -N_c$. 
From these two relations, one obtains
the number of colors of dual magnetic theory
\bea
\widetilde{N}_c' = N_f' + N_c''+ N_c-N_c'.
\nonu
\eea

Let us draw this magnetic brane configuration in Figure 9A and recall
that we put
the coincident $N_f'$ D6-branes in the nonzero $v$-direction in the
electric theory and consider massless flavors for $Q$ and $Q''$ by
putting $N_f$ and $N_f''$ D6-branes at $v=0$.
If we ignore  
$NS5_R$-brane, $N_f''$ D6-branes
and $N_c''$ D4-branes(detaching these
branes from Figure 9A), 
then this brane configuration 
leads to the standard ${\cal N}=1$ magnetic gauge theory 
$SU(N_c) \times SU(\widetilde{N}_c'=N_f'+N_c-N_c')$ with fundamentals, 
bifundamentals, eight-fundamentals, antisymmetric, a conjugate
symmetric flavors, and singlets in Figure 7 of \cite{Ahn07-4}.

Now let us recombine $\widetilde{N}_c'$ flavor D4-branes among $N_f'$
flavor 
D4-branes(connecting between D6-branes and NS5'-brane) with the same number of 
color D4-branes(connecting between NS5'-brane and $NS5_L$-brane) and push
them in $+v$ direction from Figure 9A. 
For the flavor D4-branes, we are left with only 
$(N_f'-\widetilde{N}_c')=N_c'-N_c''-N_c$ flavor D4-branes
connecting between D6-branes and NS5'-brane.  

\begin{figure}[ht]
   \epsfxsize=4.0in 
\centerline{\epsffile{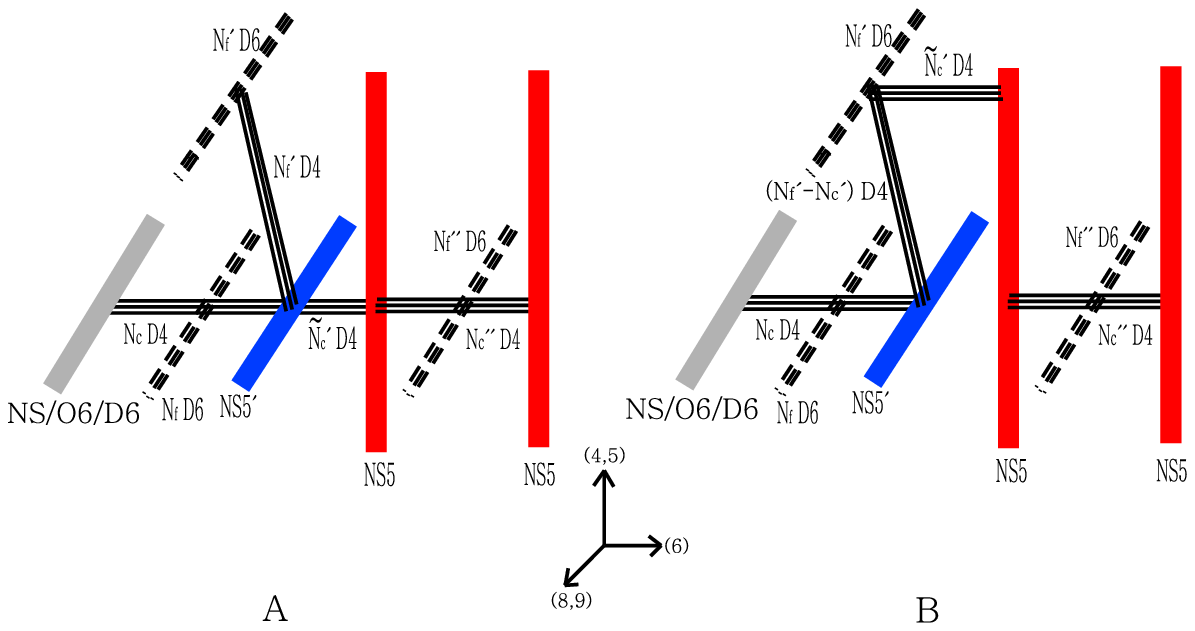}}
   \caption[FIG. \arabic{figure}.]{ 
The ${\cal N}=1$ supersymmetric magnetic brane configuration with
$SU(N_c) \times SU(\widetilde{N}_c'= 
N_f' + N_c''+ N_c-N_c') \times SU(N_c'')$ gauge group
with fundamentals $Q(q')[Q'']$ and
$\widetilde{Q}(\widetilde{q}')[\widetilde{Q}'']$ 
for each gauge group and bifundamentals $f(G)$ and
$\widetilde{f}(\widetilde{G})$, $A$ and $\widetilde{S}$, $\hat{Q}$, 
and gauge singlets in Figure 9A. In
Figure 9B, the nonsupersymmetric minimal energy brane configuration
with the same gauge group and matter contents above 
for massless  $Q(Q'')$ and
$\widetilde{Q}(\widetilde{Q}'')$ is given. 
}
\end{figure}

The additional $N_f'$-$SU(N_c)$ fundamentals $X$ and $N_f'$-$SU(N_c)$
antifundamentals $\widetilde{X}$ 
are originating from 
the $SU(N_c')$ chiral mesons $F Q'$ and $\widetilde{F}
\widetilde{Q}'$ 
respectively. 
The gauge singlet $M'$ corresponds to the $SU(N_c')$ 
chiral meson $Q' \widetilde{Q}'$
where the color indices are contracted.
The fluctuations of the gauge-singlet $M'$ correspond to the motion of $N_f'$
flavor D4-branes along (789) directions in Figure 9B.
The $\Phi$ corresponds to the $SU(N_c')$ chiral meson $F \widetilde{F}$
where the color indices for the second gauge group 
are contracted each other.

The coefficient of the beta function of the first gauge group factor 
is given by
$
b_{SU(N_c)}^{mag}=3N_c-(N_f+4)-\widetilde{N}_c'-
\frac{1}{2}(N_c+2)-\frac{1}{2}(N_c-2)-N_f'-N_c
$
and 
the coefficient of the beta function of the second gauge group factor 
is given by
$
b_{SU(\widetilde{N}_c')}^{mag}=3\widetilde{N}_c'-N_f'-N_c-N_c''$.
The coefficient of the beta function of the third gauge group factor 
is given by
$
b_{SU(N_c'')}^{mag}=3N_c''-N_f''-\widetilde{N}_c'$.

From the superpotential
\bea
W_{dual}= \left( M' q' \widetilde{q}'  + m' M' \right) + 
X f q' +
\widetilde{X} \widetilde{q}' \widetilde{f} + \Phi f \widetilde{f} 
\nonu
\eea
one sees that $q' \widetilde{q}'$ has rank 
$\widetilde{N}_c'$ while $m'$ has a
rank $N_f'$.  If the rank $N_f'$ exceeds $\widetilde{N}_c'$, then 
the supersymmetry is broken. 
The classical moduli space of vacua can be obtained from F-term
equations.
Then, it is easy to see that 
$
\widetilde{q}' M' =0= M' q', 
 q'  \widetilde{q}' +  m'  =  0$.
Then the solutions can be written as
\bea
<q' >  & = &  \left(
\begin{array}{c}
\sqrt{m'} e^{\phi} {\bf 1}_{\widetilde{N}_c'}  \\
0
\end{array}
\right),  
< \widetilde{q}'> =
 \left(
\begin{array}{cc}
\sqrt{m'} e^{-\phi}  {\bf 1}_{\widetilde{N}_c'}   &
0
\end{array}
\right), 
<M'>  =
 \left(
\begin{array}{cc}
0  & 0 
 \\
0 & M_0'  {\bf 1}_{N_f'-\widetilde{N}_c'} 
\end{array}
\right),
\nonu \\
<f> & = & <\widetilde{f}> = <X> = <\widetilde{X}>= 0.
\nonu
\eea
By expanding the fields around the vacua and it 
turns out that states are stable by realizing the mass of 
$m_{M_0'}^2$ positive.

Then the gauge group and matter contents we consider 
are summarized as follows:
\bea
 & \mbox{gauge group}:& \;\;\;\;\;   SU(N_c) \times SU(\widetilde{N}_c') \times
 SU(N_c'')  \nonu
\\
\mbox{matter}: 
 &Q_f \oplus \widetilde{Q}_{\widetilde{f}}& \;\;\;\;\;\;\;\;\; 
\;\;\; {(\bf \Box, 1, 1) \oplus (\overline{\Box}, 1, 1)}
\;\;\;\;\; (f, \widetilde{f}=1,  \cdots, N_f) 
\nonu \\
 &q'_{f'} \oplus \widetilde{q}'_{\widetilde{f}'}& \;\;\;\;\;\;\;\;\;\;
\;\; {(\bf 1, \Box, 1) \oplus ( 1, \overline{\Box}, 1)}
\;\;\;\;\; (f', \widetilde{f}' =1,  \cdots, N_f') 
\nonu \\
 &Q''_{f''} \oplus \widetilde{Q}''_{\widetilde{f}''}& \;\;\;\;\;\;\;\;\;\; 
\;\; {(\bf 1, 1, \Box) \oplus ( 1, 1, \overline{\Box})} 
\;\;\;\;\; (f'', \widetilde{f}'' =1,  \cdots, N_f'')
\nonu \\
 &\hat{Q}_f & \;\;\;\;\;\;\;\;\; 
\;\;\;\;\;\;\;\;\;\;\; {(\bf \Box, 1, 1) }
\;\;\;\;\;\;\;\;\;\;\;\;\;\; (f=1,  \cdots, 8) 
\nonu \\ 
 &f \oplus \widetilde{f}& \;\;\;\;\;\;\;\;\; 
\;\; {(\bf \Box, \overline{\Box},1) \oplus (\overline{\Box}, \Box, 1)} 
\nonu \\
 &G \oplus \widetilde{G}& \;\;\;\;\;\;\;\;\; 
\;\; {(\bf 1, \Box, \overline{\Box}) \oplus ( 1, \overline{\Box}, \Box)} 
\nonu \\
 &A \oplus \widetilde{S}& \;\;\;\;\;\;\;\;\; 
\;\; {(\bf asymm, 1, 1) \oplus (\overline{symm}, 1, 1)} 
\nonu \\ 
& (X_{n'} \equiv) F Q' \oplus \widetilde{F} \widetilde{Q}' (\equiv 
\widetilde{X}_{\widetilde{n}'}) & 
 \;\;\;\;\;\;\;\;\;\;
\;\; {(\bf \Box, 1, 1) \oplus ( \overline{\Box}, 1, 1)}
\;\; (n', \widetilde{n}' =1,  \cdots, N_f') 
\nonu \\
&  (M_{f',\widetilde{g}'}' \equiv) Q' \widetilde{Q}' & 
 \;\;\;\;\;\;\;\;\;\;\;\;\;\;\;\;\;\;\;
\;\; {(\bf 1, 1, 1)} \;\;\;\;\;\;\;\;\;
\;\; (f', \widetilde{g}' =1,  \cdots, N_f') 
\nonu \\
& (\Phi \equiv) F \widetilde{F} & 
 \;\;\;\;\;\;\;\;\;\;
\;\; {(\bf adj, 1, 1) \oplus ( 1, 1, 1)}
\nonu 
\eea

The nonsupersymmetric minimal energy brane configuration Figure 9B
(neglecting  the O6-planes, half D6-branes, and the mirrors)
looks similar to 
the Figure 3B of \cite{Ahn07-8} and leads to a reflection of 
Figure 3B of \cite{Ahn07-8} with respect to NS5-brane.

At nonzero string coupling constant, 
the NS5-branes bend due to their interactions with the D4-branes and
D6-branes.
Then the behavior of the supersymmetric M5-brane curves can be
summarized 
as follows:

1. $v \rightarrow \infty$ limit implies
\bea
w & \rightarrow & 0, \quad y \sim    v^{N_c''} + \cdots \quad
\mbox{$\overline{NS5_R}$ 
asymptotic region}, \nonu \\
w & \rightarrow  & 0, \quad y \sim    
v^{-\widetilde{N}_c'+N_c''+N_f''+N_f+2N_f'+4} + \cdots \quad
\mbox{$NS5_{L}$ asymptotic region}, 
\nonu \\
w & \rightarrow  & 0, \quad y \sim    
v^{\widetilde{N}_c'-N_c''+N_f''} + \cdots \quad
\mbox{$\overline{NS5_L}$ asymptotic region},
\nonu \\
w & \rightarrow &   0, \quad 
y \sim  v^{-N_c''+2N_f''+N_f+2N_f'+6}
 +\cdots
\quad \mbox{$NS5_{R}$ asymptotic region}. 
\nonu
\eea

2.  $w \rightarrow \infty$ limit implies
\bea
v & \rightarrow &  -m', \quad  
y \sim w^{N_c+N_f''+2-\widetilde{N}_c'}
+\cdots
\quad \mbox{$\overline{NS5'}$ asymptotic region}, 
\nonu \\
v & \rightarrow &   +m', \quad 
y \sim  w^{N_f''+N_f+N_f'+4}
 +\cdots
\quad \mbox{$NS5_{M}'$ asymptotic region}, 
\nonu
\\
v & \rightarrow &  +m', \quad  
y \sim w^{N_f''+N_f+2N_f'-\widetilde{N}_c'-N_c+4}
+\cdots
\quad \mbox{NS5' asymptotic region}. 
\nonu
\eea

\subsection{Magnetic theory with dual for third gauge group}

By moving the $NS5_R$-brane to the left all the way past the  
NS5'-brane, one arrives at the Figure 10A.
The linking number of $NS5_R$-brane from Figure 10A
is given by 
$
L_5 = \frac{N_f''}{2} -\widetilde{N}_c''$ and
the linking number of $NS5_R$-brane from Figure 6
is
$
L_5 = -\frac{N_f''}{2} + N_c'' -N_c'$. 
From these two relations, one obtains
the number of colors of dual magnetic theory
\bea
\widetilde{N}_c'' = N_f'' + N_c'-N_c''.
\nonu
\eea

Let us draw this magnetic brane configuration in Figure 10A and recall
that we put
the coincident $N_f''$ D6-branes in the nonzero $v$-direction in the
electric theory and consider massless flavors for $Q$ and $Q'$ by
putting $N_f$ and $N_f'$ D6-branes at $v=0$.
If we ignore the mirror branes corresponding to 
$NS5_{L,R}'$-branes, NS5-brane, $N_f(N_f')[N_f'']$ D6-branes
and $N_c(N_c')[N_c'']$ D4-branes, 
O6-planes and half-D6-branes(detaching these
branes from Figure 10A), 
then this brane configuration 
looks similar to the standard ${\cal N}=1$ magnetic gauge theory with
$SU(N_c) \times SU(N_c') 
\times SU(\widetilde{N}_c''=N_f''+N_c'-N_c'')$ with fundamentals, 
bifundamentals, and singlets in Figure 5 of \cite{Ahn07-8}.

Now let us recombine $\widetilde{N}_c''$ flavor D4-branes among $N_f''$
flavor 
D4-branes(connecting between D6-branes and NS5'-brane) with the same number of 
color D4-branes(connecting between $NS5_R$-brane and NS5'-brane) and push
them in $+v$ direction from Figure 10A. 
For the flavor D4-branes, we are left with only 
$(N_f''-\widetilde{N}_c'')=N_c''-N_c'$ flavor D4-branes
connecting between D6-branes and NS5'-brane. 

\begin{figure}[ht]
   \epsfxsize=4.0in 
\centerline{\epsffile{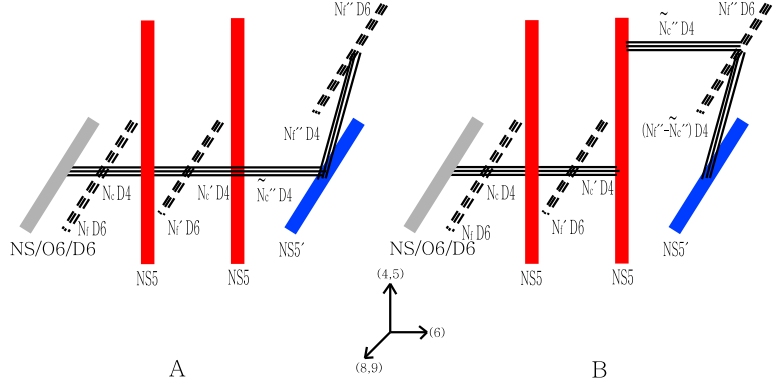}}
   \caption[FIG. \arabic{figure}.]{ 
The ${\cal N}=1$ supersymmetric magnetic brane configuration with
$SU(N_c) \times SU(N_c') \times SU(\widetilde{N}_c''=N_f'' + N_c'-N_c'')$ gauge group
with fundamentals $Q(Q')[q'']$ and 
$\widetilde{Q}(\widetilde{Q}')[\widetilde{q}'']$ 
for each gauge group, bifundamentals $F(g)$ and
$\widetilde{F}(\widetilde{g})$, $A$ and $\widetilde{S}$, $\hat{Q}$, 
and gauge singlets in Figure 10A. In
Figure 10B, the nonsupersymmetric minimal energy brane configuration
with the same gauge group and matter contents above 
for massless  $Q(Q')$ and
$\widetilde{Q}(\widetilde{Q}')$ is given. 
}
\end{figure}

The additional $N_f''$-$SU(N_c')$ fundamentals $X'$ and $N_f''$-$SU(N_c')$
antifundamentals $\widetilde{X}'$ 
are originating from 
the $SU(N_c'')$ chiral mesons $G Q''$ and $\widetilde{G}
\widetilde{Q}''$ 
respectively. 
The gauge singlet $M''$ corresponds to the $SU(N_c'')$ 
chiral meson $Q'' \widetilde{Q}''$
where the color indices are contracted.
The fluctuations of the gauge-singlet $M''$ correspond to the motion of $N_f''$
flavor D4-branes along (789) directions in Figure 10A.
The $\Phi'$ corresponds to the $SU(N_c'')$ chiral meson $G \widetilde{G}$
where the color indices for the third gauge group 
are contracted each other.

The coefficient of the beta function of the first gauge group factor 
is given by
$
b_{SU(N_c)}^{mag}=3N_c-(N_f+4)-N_c'-
\frac{1}{2}(N_c+2)-\frac{1}{2}(N_c-2)=b_{SU(N_c)}
$
and 
the coefficient of the beta function of the second gauge group factor 
is given by
$
b_{SU(N_c')}^{mag}=3N_c'-N_f'-N_c-\widetilde{N}_c''-N_f''-N_c'$.
The coefficient of the beta function of the third gauge group factor 
is given by
$
b_{SU(\widetilde{N}_c'')}^{mag}=3\widetilde{N}_c''-N_f''-N_c'$.
Since $b_{SU(N_c')}-b_{SU(N_c')}^{mag} > 0$, $SU(N_c')$ is more
asymptotically free than $SU(N_c')^{mag}$.

Then the gauge group and matter contents we consider 
are summarized as follows:
\bea
 & \mbox{gauge group}:& \;\;\;\;\;   SU(N_c) \times SU(N_c') \times
 SU(\widetilde{N}_c'')  \nonu
\\
\mbox{matter}:  
&Q_f \oplus \widetilde{Q}_{\widetilde{f}}& \;\;\;\;\;\;\;\;\; 
\;\;\; {(\bf \Box, 1, 1) \oplus (\overline{\Box}, 1, 1)}
\;\; (f, \widetilde{f}=1,  \cdots, N_f) 
\nonu \\
 &Q'_{f'} \oplus \widetilde{Q}'_{\widetilde{f}'}& \;\;\;\;\;\;\;\;\;\;
\;\; {(\bf 1, \Box, 1) \oplus ( 1, \overline{\Box}, 1)}
\;\; (f', \widetilde{f}' =1,  \cdots, N_f') 
\nonu \\
 &q''_{f''} \oplus \widetilde{q}''_{\widetilde{f}''}& \;\;\;\;\;\;\;\;\;\; 
\;\; {(\bf 1, 1, \Box) \oplus ( 1, 1, \overline{\Box})} 
\;\; (f'', \widetilde{f}'' =1,  \cdots, N_f'')
\nonu \\
 &\hat{Q}_f & \;\;\;\;\;\;\;\;\; 
\;\;\;\;\;\;\;\;\;\;\; {(\bf \Box, 1, 1) }
\;\;\;\;\;\;\;\;\;\;\;\;\;\; (f=1,  \cdots, 8) 
\nonu \\ 
&F \oplus \widetilde{F}& \;\;\;\;\;\;\;\;\; 
\;\; {(\bf \Box, \overline{\Box},1) \oplus (\overline{\Box}, \Box, 1)} 
\nonu \\
 &g \oplus \widetilde{g}& \;\;\;\;\;\;\;\;\; 
\;\; {(\bf 1, \Box, \overline{\Box}) \oplus ( 1, \overline{\Box}, \Box)} 
\nonu \\
 &A \oplus \widetilde{S}& \;\;\;\;\;\;\;\;\; 
\;\; {(\bf asymm, 1, 1) \oplus (\overline{symm}, 1, 1)} 
\nonu \\ 
& (X_{n''}' \equiv) G Q'' \oplus \widetilde{G} \widetilde{Q}'' (\equiv 
\widetilde{X}_{\widetilde{n}''}') & 
 \;\;\;\;\;\;\;\;\;\;
\;\; {(\bf 1, \Box, 1) \oplus ( 1, \overline{\Box}, 1)}
\;\; (n'', \widetilde{n}'' =1,  \cdots, N_f'') 
\nonu \\
&  (M''_{f'',\widetilde{g}''} \equiv) Q'' \widetilde{Q}'' & 
 \;\;\;\;\;\;\;\;\;\;\;\;\;\;\;\;\;\;\;
\;\; {(\bf 1, 1, 1)} \;\;\;\;\;\;\;\;\;
\;\; (f'', \widetilde{g}'' =1,  \cdots, N_f'') 
\nonu \\
& (\Phi' \equiv) G \widetilde{G} & 
 \;\;\;\;\;\;\;\;\;\;
\;\; {(\bf 1, adj, 1) \oplus ( 1, 1, 1)}
\nonu
\eea

The superpotential is
\bea
W_{dual}= \left( M'' q'' \widetilde{q}''  + m'' M'' \right) + 
X' g q'' +
\widetilde{X}' \widetilde{q}'' \widetilde{g} + \Phi' g \widetilde{g}. 
\nonu
\eea
Then, $q'' \widetilde{q}''$ has rank 
$\widetilde{N}_c''$ while $m''$ has a
rank $N_f''$.  The derivative of the 
superpotential $W_{dual}$ with respect to $M''$, cannot be satisfied 
if the rank $N_f''$ exceeds $\widetilde{N}_c''$ and the supersymmetry is broken. 
The classical moduli space of vacua can be obtained from F-term
equations.
Then, it is easy to see that 
$
\widetilde{q}'' M'' =0= M'' q'', 
 q''  \widetilde{q}'' +  m''  =  0$.
Then the solutions can be written as
\bea
<q'' >  & = &  \left(
\begin{array}{c}
\sqrt{m''} e^{\phi} {\bf 1}_{\widetilde{N}_c''}  \\
0
\end{array}
\right),  
< \widetilde{q}''> =
 \left(
\begin{array}{cc}
\sqrt{m''} e^{-\phi}  {\bf 1}_{\widetilde{N}_c''}   &
0
\end{array}
\right), 
<M''>  =
 \left(
\begin{array}{cc}
0  & 0 
 \\
0 & M_0''  {\bf 1}_{N_f''-\widetilde{N}_c''} 
\end{array}
\right),
\nonu \\
<g> & = & <\widetilde{g}> = <X'> = <\widetilde{X}'>= 0.
\nonu
\eea
One can check 
that states are stable by realizing the mass of 
$m_{M_0''}^2$ positive, by expanding the fields around the vacua.

The nonsupersymmetric minimal energy brane configuration Figure 10B
(neglecting  the O6-plane, half D6-branes, and the mirrors)
looks similar to 
the Figure 5B of \cite{Ahn07-8} and leads to a reflection of 
Figure 5B of \cite{Ahn07-8} with respect to NS5-brane. 

At nonzero string coupling constant, 
the NS5-branes bend due to their interactions with the D4-branes and
D6-branes.
Then the behavior of the supersymmetric M5-brane curves can be
summarized 
as follows:

1. $v \rightarrow \infty$ limit implies
\bea
w & \rightarrow & 0, \quad y \sim    v^{N_c'-\widetilde{N}_c''} + \cdots \quad
\mbox{$\overline{NS5_R}$ 
asymptotic region}, \nonu \\
w & \rightarrow  & 0, \quad y \sim    
v^{N_c+N_c'+N_f''+N_f'+4} + \cdots \quad
\mbox{$NS5_{L}$ asymptotic region}, 
\nonu \\
w & \rightarrow  & 0, \quad y \sim    
v^{N_c-N_c'+N_f''+N_f'} + \cdots \quad
\mbox{$\overline{NS5_L}$ asymptotic region},
\nonu \\
w & \rightarrow &   0, \quad 
y \sim  v^{\widetilde{N}_c''-N_c'+2N_c+2N_f''+N_f'+4}
 +\cdots
\quad \mbox{$NS5_{R}$ asymptotic region}. 
\nonu
\eea

2.  $w \rightarrow \infty$ limit implies
\bea
v & \rightarrow &  -m'', \quad  
y \sim w^{\widetilde{N}_c''}
+\cdots
\quad \mbox{$\overline{NS5'}$ asymptotic region}, 
\nonu \\
v & \rightarrow &   +m'', \quad 
y \sim  w^{N_c+N_f''+N_f'+2}
 +\cdots
\quad \mbox{$NS5_{M}'$ asymptotic region}, 
\nonu
\\
v & \rightarrow &  +m'', \quad  
y \sim w^{2N_f''+N_f'+2N_c-\widetilde{N}_c''+4}
+\cdots
\quad \mbox{NS5' asymptotic region}. 
\nonu
\eea

\subsection{Magnetic theories 
for the multiple product gauge groups}

Now one can generalize the method for the triple product gauge groups
to the finite $n$-multiple product gauge groups characterized by 
\bea
SU(N_{c,1}) \times SU(N_{c,2}) \times \cdots
\times SU(N_{c,n})
\nonu
\eea
with the matter, 
the $(n-1)$ bifundametals $({\bf \Box_1, \overline{\Box}_2, 1, 
\cdots,  1_n})$,
$\cdots$, and $({\bf 1_1, \cdots, 1, \Box_{n-1}, \overline{\Box}_{n}})$, their
complex conjugate $(n-1)$ fields $({\bf \overline{\Box}_1, \Box_2, 1, 
\cdots, 1_n})$,
$\cdots$, 
and $({\bf 1_1, \cdots, 1, \overline{\Box}_{n-1}, \Box_n})$, linking the
gauge groups together,
$n$-fundamentals $({\bf \Box_1, 1, \cdots, 1_n})$, $\cdots$, and 
$({\bf 1_1, \cdots,  1, \Box_n})$, and $n$-antifundamentals  
$({\bf \overline{\Box}_1, 1, \cdots, 1_n})$, $\cdots$,  
$({\bf 1_1, \cdots,  1, \overline{\Box}_n})$, eight-fundamentals, an
antisymmetric tensor, and a conjugate symmetric tensor.
Then the mass-deformed superpotential can be written as
$
W_{elec} = \sum_{i=1}^n m_i Q_i \widetilde{Q}_i$. 
The brane configuration can be constructed from Figure 6 by adding
$(n-3)$ NS-branes, $(n-3)$ sets of D6-branes and $(n-3)$
sets of D4-branes  to the right of $NS5_R$-brane(and its mirrors)
leading to the fact that 
any two neighboring NS-branes should be perpendicular to each other. 

There exist $(2n-2)$ magnetic theories and they can be classified 
as follows.

$\bullet$ When the dual magnetic theory contains $SU(\widetilde{N}_{c,1})$

When the Seiberg dual is taken for the first gauge group factor
 by
assuming that $\Lambda_1 >> \Lambda_i$ where $i=2, \cdots, n$, 
one follows the procedure given in the subsection 3.2.
The gauge group is 
\bea
SU(\widetilde{N}_{c,1} \equiv 2N_{f,1} +2N_{c,2}-N_{c,1}+4) \times 
SU(N_{c,2}) \times \cdots \times SU(N_{c,n})
\nonu
\eea
and the matter contents are given by 
the dual quarks $q_1$ $({\bf \Box_1, 1, \cdots,
1_n})$ 
and $\widetilde{q}_1$ in the 
representation $({\bf \overline{\Box}_1, 1, \cdots, 1_n})$ 
as well as $(n-1)$ 
quarks $Q_i$ and
$\widetilde{Q}_i$ where $i=2, \cdots, n$, 
the bifundamentals $f_1$ in the representation 
 $({\bf \Box_1, \overline{\Box}_2, 1, \cdots,  1_n})$ under the dual gauge group,
and $\widetilde{f}_1$ in the representation
$({\bf \overline{\Box}_1, \Box_2, 1, \cdots, 1_n})$ under the dual gauge
group  in
addition to $(n-2)$ bifundamentals $G_i$ and $\widetilde{G}_i$, 
 eight-fundamentals, an
antisymmetric tensor, a conjugate symmetric tensor,
and
various gauge singlets $X_2, \widetilde{X}_2, M_1, \hat{M}$ and
 $\Phi_2$ in addition to $P, \widetilde{P}$ and $N$.
The corresponding brane configuration can be 
obtained similarly and 
the extra $(n-3)$ NS-branes, $(n-3)$ sets of D6-branes and $(n-3)$
sets of D4-branes  
are present at the right hand side of the $NS5_R$-brane
of Figure 7(and their mirrors).
The magnetic superpotential can be written as
\bea
W_{dual} = \left(M_1 q_1 \widetilde{s} a \widetilde{q}_1 + \hat{q}
 \widetilde{s} \hat{q} + \hat{M} \hat{q} \widetilde{q} +
f_1 \widetilde{X}_2 \widetilde{q}_1 + 
\widetilde{f}_1 q_1
X_2 + \Phi_2 f_1 \widetilde{f}_1 \right) + m_1 M_1.
\nonu
\eea
By computing the contribution for the one loop as in the subsection
3.2, 
the vacua are stable and the asymptotic behavior of $(2n+1)$ NS-branes
can be obtained also.  

$\bullet$ When the dual magnetic theory contains $SU(\widetilde{N}_{c,2})$

When the Seiberg dual is taken for the second gauge group factor
by
assuming that $\Lambda_2 >> \Lambda_j$ where $j=1, 3,
\cdots, n$, 
one follows the procedure given in the subsection 3.3.
The gauge group is given by
\bea
SU(N_{c,1}) \times 
SU(\widetilde{N}_{c,2} \equiv N_{f,2}+N_{c,3}+N_{c,1}-N_{c,2}) 
\times SU(N_{c,3}) \times \cdots \times SU(N_{c,n}).
\nonu
\eea
The corresponding brane configuration can be 
obtained similarly and 
the extra $(n-3)$ NS-branes, $(n-3)$ sets of D6-branes and $(n-3)$
sets of D4-branes  are present at the right hand side of the $NS5_R$-brane
of Figure 8.
The magnetic superpotential can be written as
\bea
W_{dual} = \left(M_2 q_2 \widetilde{q}_2 + 
g_2 \widetilde{X}_3 \widetilde{q}_2 + 
\widetilde{g}_2 q_2
X_3 + \Phi_3 g_2 \widetilde{g}_2 \right) + m_2 M_2.
\nonu
\eea

When the Seiberg dual is taken for the second gauge group factor with
different brane motion
by
assuming that $\Lambda_2 >> \Lambda_j$ where $j=1, 3,
\cdots, n$, 
one follows the procedure given in the subsection 3.4.
The gauge group is given by
\bea
SU(N_{c,1}) \times 
SU(\widetilde{N}_{c,2} \equiv N_{f,2}+N_{c,3}+N_{c,1}-N_{c,2}) 
\times SU(N_{c,3}) \times \cdots \times SU(N_{c,n}).
\nonu
\eea
The corresponding brane configuration can be 
obtained similarly and 
the extra $(n-3)$ NS-branes, $(n-3)$ sets of D6-branes and $(n-3)$
sets of D4-branes  are present at the right hand side of the $NS5_R$-brane
of Figure 9.
The magnetic superpotential can be written as
\bea
W_{dual} = \left(M_2 q_2 \widetilde{q}_2 + f_1 X_1 q_2 + 
\widetilde{f}_1 \widetilde{q}_2
\widetilde{X}_1 + \Phi_1 f_1 \widetilde{f}_1 \right) + m_2 M_2.
\nonu
\eea

$\bullet$ When the dual magnetic theory contains $SU(\widetilde{N}_{c,i})$ where
$ 3 \leq i \leq n-1$

When the Seiberg dual is taken for the middle gauge group factor
by
assuming that $\Lambda_i >> \Lambda_j$ where $j=1,2, \cdots, i-1, i+1,
\cdots, n$, 
one follows the procedure given in the subsection 2.3 of \cite{Ahn07-8}.
The gauge group is given by
\bea
SU(N_{c,1}) \times \cdots \times 
SU(\widetilde{N}_{c,i} \equiv N_{f,i}+N_{c,i+1}+N_{c,i-1}-N_{c,i}) 
\times  \cdots \times SU(N_{c,n}).
\nonu
\eea
The corresponding brane configuration can be 
obtained similarly and 
the extra $(i-2)$ NS-branes, $(i-2)$ sets of D6-branes and $(i-2)$
sets of D4-branes  
are present between the $NS5_M'$-brane and the  NS5'-brane
and the extra $(n-i-1)$ NS-branes, $(n-i-1)$ sets of D6-branes and $(n-i-1)$
sets of D4-branes  are present at the right hand side of the $NS5_R$-brane
of Figure 8.
The magnetic superpotential can be written as
\bea
W_{dual} = \left(M_{i} q_i \widetilde{q}_i + 
g_{i} \widetilde{X}_{i+1} \widetilde{q}_i + 
\widetilde{g}_{i} q_i
X_{i+1} + \Phi_{i+1} g_{i} \widetilde{g}_{i} \right) + m_i M_{i}.
\nonu
\eea

When the Seiberg dual is taken for the middle gauge group factor with
different brane motion
by
assuming that $\Lambda_i >> \Lambda_j$ where $j=1,2, \cdots, i-1, i+1,
\cdots, n$, 
one follows the procedure given in the subsection 2.4 of \cite{Ahn07-8}.
The gauge group is given by
\bea
SU(N_{c,1}) \times \cdots  \times 
SU(\widetilde{N}_{c,i} \equiv N_{f,i}+N_{c,i+1}+N_{c,i-1}-N_{c,i}) 
\times \cdots \times SU(N_{c,n}).
\nonu
\eea
The corresponding brane configuration can be 
obtained similarly and 
the extra $(i-2)$ NS-branes, $(i-2)$ sets of D6-branes and $(i-2)$
sets of D4-branes  
are present between the $NS5_M'$-brane and the NS5'-brane
and the extra $(n-i-1)$ NS-branes, $(n-i-1)$ sets of D6-branes and $(n-i-1)$
sets of D4-branes  are present at the right hand side of the $NS5_R$-brane
of Figure 9.
The magnetic superpotential can be written as
\bea
W_{dual} = \left(M_{i} q_i \widetilde{q}_i + f_{i-1} X_{i-1} q_i + 
\widetilde{f}_{i-1} \widetilde{q}_i
\widetilde{X}_{i-1} + \Phi_{i-1} f_{i-1} \widetilde{f}_{i-1} \right) + m_i M_{i}.
\nonu
\eea

$\bullet$ When the dual magnetic theory contains $SU(\widetilde{N}_{c,n})$

When the Seiberg dual is taken for the last gauge group factor by
assuming that $\Lambda_n >> \Lambda_i$ where $i=1,2, \cdots, (n-1)$, 
one follows the procedure given in the subsection 3.5.
The gauge group is given by
\bea
SU(N_{c,1}) \times \cdots \times 
SU(N_{c,n-1}) \times SU(\widetilde{N}_{c,n} \equiv N_{f,n} +N_{c,n-1}-N_{c,n}).
\nonu
\eea
The corresponding brane configuration can be 
obtained similarly and 
the extra $(n-3)$ NS-branes, $(n-3)$ sets of D6-branes and $(n-3)$
sets of D4-branes  
are present between the $NS5_M'$-brane and the $NS5_L$-brane
of Figure 10.
The magnetic superpotential can be written as
\bea
W_{dual} = \left(M_n q_n \widetilde{q}_n + g_{n-1} X_{n-1} q_n + 
\widetilde{g}_{n-1} \widetilde{q}_n
\widetilde{X}_{n-1} + \Phi_{n-1} g_{n-1} \widetilde{g}_{n-1} \right) + m_n M_n.
\nonu
\eea

\section{Meta-stable brane configurations
of other multiple product gauge theories }

After we describe the electric brane configuration as we did previously, 
we present the three magnetic brane configurations, and then the
nonsupersymmetric meta-stable brane configurations are  found.
The generalization to multiple product gauge groups is discussed.

\subsection{Electric theory}

Let us describe the gauge theory with triple product gauge groups 
$SO(N_c) \times SU(N_c') \times SU(N_c'')$ and
the matter contents 
are 

$\bullet$
$2N_f$-chiral multiplets $Q$ are  in the
representation $({\bf N_c, 1, 1
})$

$\bullet$
$N_f'$-chiral multiplets $Q'$ are  in the
representation $({\bf 1, N_c', 1})$, and 
$N_f'$-chiral multiplets $\widetilde{Q}'$ are in  
the representation $({\bf 1,\overline{N_c'}, 1})$

$\bullet$
$N_f''$-chiral multiplets $Q''$ are  in the
representation $({\bf 1, 1, N_c''
})$, and 
$N_f''$-chiral multiplets $\widetilde{Q}''$ are in  
the representation $({\bf 1, 1, \overline{N_c''}})$

$\bullet$
The flavor-singlet field $F$ is 
in the bifundamental representation $({\bf N_c, \overline{N_c'}, 1 })$, 
and its conjugate field $\widetilde{F}$
 is 
in the bifundamental representation $({\bf N_c, N_c', 1})$

$\bullet$
The flavor-singlet field $G$ is 
in the bifundamental representation $({\bf 1, N_c', \overline{N_c''} })$, 
and its conjugate field $\widetilde{G}$
 is 
in the bifundamental representation $({\bf 1, \overline{N_c'}, N_c''})$

If we put to $Q'',
\widetilde{Q}'', G$, and $\widetilde{G}$ zero, then 
this becomes the product gauge group theory with fundamentals and
bifundamentals
\cite{LO,Ahn07-3}. On the other hand, if we ignore $Q, F$, and
$\widetilde{F}$, then this theory is given by \cite{BH}.


In the electric 
theory, 
the coefficient of the beta function
of the first gauge group factor is
$
b_{SO(N_c)} = 3(N_c-2)-2N_f-2N_c'
$
and similarly
the coefficient of the beta
function
of the second gauge group factor
is
$
b_{SU(N_c')} = 3N_c' -N_f'-N_c-N_c''
$
and 
the coefficient of the beta function
of the third gauge group factor is
$
b_{SU(N_c'')} = 3N_c'' -N_f''-N_c'$.
We'll see how these coefficients change in the magnetic theory.
We denote the
strong coupling scales for $SO(N_c)$ as $\Lambda_1$, for $SU(N_c')$
as $\Lambda_2$ and for $SU(N_c'')$
as $\Lambda_3$ respectively. 

From the electric superpotential 
\bea
W_{elec} & = & 
\left( \mu A^2 +  Q A \widetilde{Q} +  \widetilde{F} A F + A_s^2 + 
Q A_s \widetilde{Q} + \widetilde{F} A_s F    
 +
\mu' A'^2 +  Q' A' \widetilde{Q}' + \widetilde{F} A' F \right. \nonu \\
&+& \left.  
\widetilde{G} A' G +
 \mu'' A''^2 +  Q'' A'' \widetilde{Q}'' + \widetilde{G} A'' G \right)
+ m Q Q + m' Q' \widetilde{Q}' + m'' Q'' \widetilde{Q}''
\nonu
\eea
one  integrates out  the adjoint fields 
$A$ for $SO(N_c)$,  $A'$ for $SU(N_c')$ and $A''$ for $SU(N_c'')$ 
and the symmetric field $A_s$ for $SO(N_c)$ 
and taking $\mu, \mu'$ and $\mu''$ to infinity limit which is
equivalent to take any two NS-branes be perpendicular to each other,
the mass-deformed electric superpotential becomes 
$
W_{elec}  = 
 m Q Q + m' Q' \widetilde{Q}' + m'' Q'' \widetilde{Q}''$.

The type IIA brane configuration for this mass-deformed theory 
can be described by as follows. 
The $N_c$-color 
D4-branes (01236) are suspended between the $NS5_L$-brane (012345) and
its mirror $\overline{NS5_L}$-brane 
together with $N_f$ D6-branes (0123789) 
which have nonzero $v$ direction.
The NS5'-brane 
is located at the right hand side of
the $NS5_L$-brane along the positive $x^6$ direction and 
there exist $N_c'$-color D4-branes
suspended 
between them, with  $N_f'$ D6-branes which have nonzero $v$ direction. 
Moreover, 
the $NS5_R$-brane 
is located at the right hand side of
the NS5'-brane along the positive $x^6$ direction and there 
exist $N_c''$-color D4-branes
suspended 
between them, with  $N_f''$ D6-branes which have nonzero $v$ direction.
There exists an orientifold 6-plane (0123789) at the origin $x^6=0$
and it acts as $(x^4, x^5, x^6) \rightarrow (-x^4, -x^5, -x^6)$. 
Then the mirrors of above branes appear in 
the negative $x^6$ region and are denoted by bar on the corresponding branes.
From the left to the right, there are $\overline{NS5_R}$-,
$\overline{NS5'}$-, 
$\overline{NS5_L}$-,
$NS5_L$-, $NS5'$-, and $NS5_R$-branes.

We summarize the ${\cal N}=1$ supersymmetric electric brane
configuration in type IIA string theory as follows:

$\bullet$ Four 
NS5-branes in (012345) directions. 

$\bullet$ Two
NS5'-branes in (012389) directions.

$\bullet$ Two sets of
$N_c(N_c')[N_c'']$-color D4-branes in (01236) directions. 
  
$\bullet$ Two sets of
$N_f(N_f')[N_f'']$ D6-branes in (0123789) directions. 

$\bullet$
$O6^{+}$-plane in (0123789) directions with $x^6=0$

Now we draw this electric brane configuration in Figure 11 and we put
the coincident $N_f(N_f')[N_f'']$ D6-branes with positive $x^6$ in 
the nonzero $v$ direction in general. 
This brane configuration can be obtained from the brane configuration
of \cite{LO,Ahn07-3} by adding the two outer NS5-branes(i.e., 
$\overline{NS5_R}$-brane and $NS5_R$-brane), two sets of $N_c''$ D4-branes
and two sets of $N_f''$ D6-branes or from the one of \cite{BH}
with the gauge theory of triple product gauge groups
by adding $O6^{+}$-plane and the extra NS-branes, D4-branes and D6-branes.
The brane configuration for the single gauge group $SO(N_c)$
was studied in \cite{CSST}.
Then the mirrors with 
negative $x^6$ can be constructed by using the action of O6-plane and
are located at the positions by changing (456) directions for the
original branes with minus signs.

\begin{figure}[ht]
   \epsfxsize=3.0in 
\centerline{\epsffile{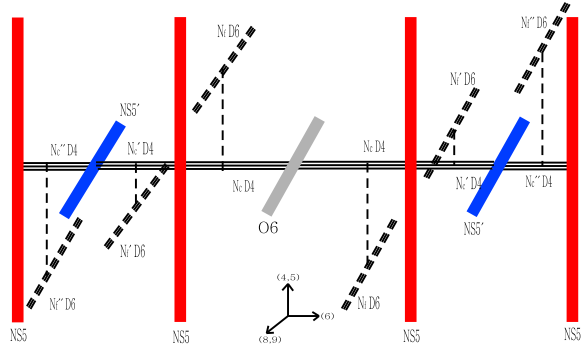}}
   \caption[FIG. \arabic{figure}.]{ 
The ${\cal N}=1$ supersymmetric electric brane configuration with
$SO(N_c) \times SU(N_c') \times SU(N_c'')$ gauge group with
flavors $Q(Q')[Q'']$ and
$(\widetilde{Q}')[\widetilde{Q}'']$ 
for each gauge group and bifundamentals $F(G)$,  
$\widetilde{F}(\widetilde{G})$.
 The $O6^{+}$-plane is located at the
origin $x^6=0$.
The  two NS5-branes with positive $x^6$ coordinates
are denoted by $NS5_{L,R}$-branes. 
 }
\end{figure}

\subsection{Magnetic theory with dual for first gauge group}

In this case, there is no extra NS-brane which should be present in
order to construct the recombination of flavor D4-branes and splitting
procedure for meta-stable brane configuration. 
Although the magnetic dual theory is present, there is no
nonsupersymmetric meta-stable
brane configuration.

\subsection{Magnetic theory with dual for second gauge group}

By moving the NS5'-brane in Figure 11 with massive $N_f'$ flavors and massless 
$N_f$ and $N_f''$ flavors
to the left all the way past the  
$NS5_L$-brane, one arrives at the Figure 12A.
The linking number of NS5'-brane from Figure 12A
is 
$
L_5 = -\frac{N_f'}{2} +\widetilde{N}_c'-N_c
$ 
while
the linking number of NS5'-brane from Figure 11
is
$
L_5 = \frac{N_f'}{2} + N_c'' -N_c'$. 
From these two relations, one obtains
the number of colors of dual magnetic theory
\bea
\widetilde{N}_c' = N_f' + N_c''+ N_c-N_c'.
\nonu
\eea

Let us draw this magnetic brane configuration in Figure 12A and recall
that we put
the coincident $N_f'$ D6-branes in the nonzero $v$-direction in the
electric theory and consider massless flavors for $Q$ and $Q''$ by
putting $N_f$ and $N_f''$ D6-branes at $v=0$.
If we ignore  
$NS5_R$-brane, $N_f''$ D6-branes
and $N_c''$ D4-branes(detaching these
branes from Figure 12A), 
then this brane configuration 
looks similar  to the Figure 6 of \cite{Ahn07-3} for
the standard ${\cal N}=1$ magnetic gauge theory 
$SO(N_c) \times SU(\widetilde{N}_c'=N_f'+N_c-N_c')$ with fundamentals, 
bifundamentals, and singlets.

Now let us recombine $\widetilde{N}_c'$ flavor D4-branes among $N_f'$
flavor 
D4-branes(connecting between D6-branes and $NS5_L$-brane) with the
same 
number of 
color D4-branes(connecting between NS5'-brane and $NS5_L$-brane) and push
them in $+v$ direction from Figure 12A. 
For the flavor D4-branes, we are left with only 
$(N_f'-\widetilde{N}_c')=N_c'-N_c''-N_c$ flavor D4-branes
connecting between D6-branes and NS5'-brane. 

\begin{figure}[ht]
   \epsfxsize=4.0in 
\centerline{\epsffile{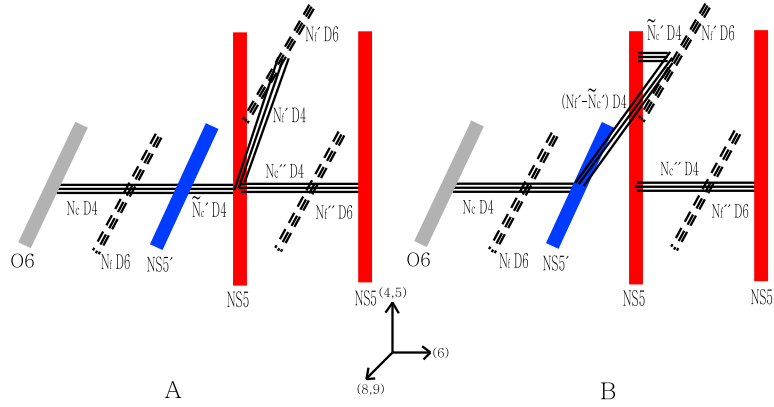}}
   \caption[FIG. \arabic{figure}.]{ 
The ${\cal N}=1$ supersymmetric magnetic brane configuration with
$SO(N_c) \times SU(\widetilde{N}_c'= N_f' +N_c +N_c'' -N_c') 
\times SU(N_c'')$ gauge group
with fundamentals $Q(q')[Q'']$ and
$(\widetilde{q}')[\widetilde{Q}'']$ 
for each gauge group and bifundamentals $F(g)$ and
$\widetilde{F}(\widetilde{g})$,   
and gauge singlets in Figure 12A. In
Figure 12B, the nonsupersymmetric minimal energy brane configuration
with the same gauge group and matter contents above 
for massless  $Q(Q'')$ and
$(\widetilde{Q}'')$ is given. Compared with the Figure 12A, there
exists a misalignment of the flavor D4-branes.
}
\end{figure}

In the dual theory, 
the coefficient of the beta function of the first gauge group factor
is
$
b_{SO(N_c)}^{mag} 
= 3(N_c-2)-2N_f-2\widetilde{N}_c' 
$
and 
the coefficient of the beta function 
of the second gauge group factor is
$
b_{SU(\widetilde{N}_c')}^{mag}
= 3\widetilde{N}_c'-N_f'-N_c-N_c''
$
and 
the coefficient of the beta function 
of the third gauge group factor is
$
b_{SU(N_c'')}^{mag}
= 3N_c''-N_f''-\widetilde{N}_c'-N_f'-N_c''$.

The superpotential from \cite{Ahn07-3} is 
\bea
W_{dual}= \left( M' q' \widetilde{q}'  + m' M' \right) + 
X'' \widetilde{g} q' +
\widetilde{X}'' \widetilde{q}' g + \Phi'' g \widetilde{g} 
\nonu
\eea
and one sees that $q' \widetilde{q}'$ has rank 
$\widetilde{N}_c'$ while $m'$ has a
rank $N_f'$.  If the rank $N_f'$ exceeds $\widetilde{N}_c'$, then 
the supersymmetry is broken. 
The classical moduli space of vacua can be obtained from F-term
equations.
Then the solutions can be written as
\bea
<q' >  & = &  \left(
\begin{array}{c}
\sqrt{m'} e^{\phi} {\bf 1}_{\widetilde{N}_c'}  \\
0
\end{array}
\right),  
< \widetilde{q}'> =
 \left(
\begin{array}{cc}
\sqrt{m'} e^{-\phi}  {\bf 1}_{\widetilde{N}_c'}   &
0
\end{array}
\right), 
<M'>  =
 \left(
\begin{array}{cc}
0  & 0 
 \\
0 & M_0'  {\bf 1}_{N_f'-\widetilde{N}_c'} 
\end{array}
\right),
\nonu \\
<g> & = & <\widetilde{g}> = <X''> = <\widetilde{X}''> = 0.
\nonu
\eea
By expanding the fields around the vacua and it 
turns out that states are stable by realizing the mass of 
$m_{M_0'}^2$ positive.

Then the gauge group and matter contents we consider 
are summarized as follows:
\bea
 & \mbox{gauge group}:& \;\;\;\;\;   SO(N_c) \times SU(\widetilde{N}_c') \times
 SU(N_c'')  \nonu
\\
\mbox{matter}: 
 &Q_f & \;\;\;\;\;\;\;\;\; 
\;\;\;\;\;\;\;\;\;\;\;\;\; {(\bf \Box, 1, 1)}
\;\;\;\;\;\;\;\;\;\;\;\; (f=1,  \cdots, 2N_f) 
\nonu \\
 &q'_{f'} \oplus \widetilde{q}'_{\widetilde{f}'}& \;\;\;\;\;\;\;\;\;\;
\;\; {(\bf 1, \Box, 1) \oplus ( 1, \overline{\Box}, 1)}
\;\;\;\;\; (f', \widetilde{f}' =1,  \cdots, N_f') 
\nonu \\
 &Q''_{f''} \oplus \widetilde{Q}''_{\widetilde{f}''}& \;\;\;\;\;\;\;\;\;\; 
\;\; {(\bf 1, 1, \Box) \oplus ( 1, 1, \overline{\Box})} 
\;\;\;\;\; (f'', \widetilde{f}'' =1,  \cdots, N_f'')
\nonu \\
 &F \oplus \widetilde{F}& \;\;\;\;\;\;\;\;\; 
\;\; {(\bf \Box, \overline{\Box}, 1) \oplus (\Box, \Box, 1)} 
\nonu \\
 &g \oplus \widetilde{g}& \;\;\;\;\;\;\;\;\; 
\;\; {(\bf 1, \Box, \overline{\Box}) \oplus ( 1, \overline{\Box}, \Box)} 
\nonu \\
& (X_{n'}'' \equiv) \widetilde{G} Q' \oplus G \widetilde{Q}' (\equiv 
\widetilde{X}_{\widetilde{n}'}'') & 
 \;\;\;\;\;\;\;\;\;\;
\;\; {(\bf 1, 1, \Box) \oplus ( 1, 1, \overline{\Box})}
\;\; (n', \widetilde{n}' =1,  \cdots, N_f') 
\nonu \\
&  (M_{f',\widetilde{g}'}' \equiv) Q' \widetilde{Q}' & 
 \;\;\;\;\;\;\;\;\;\;\;\;\;\;\;\;\;\;\;
\;\; {(\bf 1, 1, 1)} \;\;\;\;\;\;\;\;\;
\;\; (f', \widetilde{g}' =1,  \cdots, N_f') 
\nonu \\
& (\Phi'' \equiv) G \widetilde{G} & 
 \;\;\;\;\;\;\;\;\;\;
\;\; {(\bf 1, 1, adj) \oplus ( 1, 1, 1)}
\nonu 
\eea

The nonsupersymmetric minimal energy brane configuration Figure 12B
with a replacement $N_f'$ D6-branes by 
the NS5'-brane(neglecting  the
$NS5_R$-brane, $N_f''$ D6-branes, $N_f$ D6-branes, and $N_c''$ D4-branes)
leads to 
the Figure 15B of \cite{Ahn07-6}.

In \cite{LL}, the Riemann surface 
describing a set of NS5-branes 
with D4-branes suspended between them and 
in a background space of 
$x t = (-1)^{N_f+N_f'+N_f''} v^{2N_f+2N_f''+4}
(v^2 - m'^2)^{N_f'}$
was found.
Since we are dealing with six NS5-branes, the magnetic M5-brane 
configuration in Figure 12 with equal mass for $q'$ and $\widetilde{q}'$ 
and massless 
for $Q$ and $Q''(\widetilde{Q}'')$ 
can be characterized by the following sixth order equation for $t$ 
as follows:
\bea
& & t^6 + \left[v^{N_c''} \right] t^5 + \left[
  v^{\widetilde{N}_c'+N_f''}(v+m')^{N_f'} 
\right] t^4
 + \left[ v^{N_c+2N_f''}(v+m')^{2N_f'} \right]  t^3 
\nonu \\
&& + \left[ (-1)^{\widetilde{N}_c'+N_f}
v^{\widetilde{N}_c' +3N_f''+2N_f+ 4} 
(v +m')^{3N_f'} \right] t^2  + 
\left[ (-1)^{N_c''} v^{N_c'' +4N_f''+4N_f+8}  
(v +m')^{4N_f'} \right] t \nonu \\
&& + \left[(-1)^{N_f+N_f'+N_f''}
v^{12+6N_f+6N_f''}  (v +m')^{5N_f'} 
 (v-m')^{N_f'}\right] =0.  
\nonu
\eea

At nonzero string coupling constant, 
the NS-branes bend due to their interactions with the D4-branes and
D6-branes.
Then the behavior of the supersymmetric M5-brane curves can be
summarized 
as follows:

1. $v \rightarrow \infty$ limit implies
\bea
w & \rightarrow & 0, \quad y \sim    v^{N_c''} + \cdots \quad
\mbox{$\overline{NS5_R}$ 
asymptotic region}, \nonu \\
w & \rightarrow  & 0, \quad y \sim    
v^{N_f''+N_f'+\widetilde{N}_c'-N_c''} + \cdots \quad
\mbox{$\overline{NS5_L}$ asymptotic region},
\nonu \\
w & \rightarrow & 0, \quad y \sim
v^{-\widetilde{N}_c'+N_f''+2N_f+N_f'+
N_c''+4} + \cdots \quad
\mbox{$NS5_L$ 
asymptotic region}, \nonu \\
w & \rightarrow  & 0, \quad y \sim    
v^{2N_f''+2N_f+2N_f'-N_c''+4} + \cdots \quad
\mbox{$NS5_R$ asymptotic region}.   
\nonu
\eea

2.  $w \rightarrow \infty$ limit implies
\bea
v & \rightarrow &   -m', \quad 
y \sim  w^{-\widetilde{N}_c'+N_c+N_f''+N_f'}
 +\cdots
\quad \mbox{$\overline{NS5'}$ asymptotic region}, 
\nonu
\\
v & \rightarrow &  +m', \quad  
y \sim w^{\widetilde{N}_c'-N_c+N_f''+2N_f+N_f'+4}
+\cdots
\quad \mbox{NS5' asymptotic region}. 
\nonu
\eea


\subsection{Magnetic theory with dual for second gauge group}

By moving the $NS5_L$-brane in Figure 11 with massive $N_f'$ D6-branes
to the right all the way past the  
NS5'-brane, one arrives at the Figure 13A.
The linking number of $NS5_L$-brane from Figure 13A
is 
$
L_5 = \frac{N_f'}{2} -\widetilde{N}_c'+N_c''$ and
the linking number of $NS5_L$-brane from the  
Figure 11
is
$
L_5 = -\frac{N_f'}{2} + N_c' -N_c$. 
From these two relations, one obtains
the number of colors of dual magnetic theory
\bea
\widetilde{N}_c' = N_f' + N_c''+ N_c-N_c'.
\nonu
\eea

Let us draw this magnetic brane configuration in Figure 13A and recall
that we put
the coincident $N_f'$ D6-branes in the nonzero $v$-direction in the
electric theory and consider massless flavors for $Q$ and $Q''$ by
putting $N_f$ and $N_f''$ D6-branes at $v=0$.
If we ignore  
$NS5_R$-brane, $N_f''$ D6-branes
and $N_c''$ D4-branes(detaching these
branes from Figure 13A), 
then this brane configuration 
leads to the Figure 6 of \cite{Ahn07-3} for
the standard ${\cal N}=1$ magnetic gauge theory 
$SO(N_c) \times SU(\widetilde{N}_c'=N_f'+N_c-N_c')$ with fundamentals, 
bifundamentals, and singlets.

Now let us recombine $\widetilde{N}_c'$ flavor D4-branes among $N_f'$
flavor 
D4-branes(connecting between D6-branes and NS5'-brane) with the same number of 
color D4-branes(connecting between NS5'-brane and $NS5_L$-brane) and push
them in $+v$ direction from Figure 13A. 
For the flavor D4-branes, we are left with only 
$(N_f'-\widetilde{N}_c')=N_c'-N_c''-N_c$ flavor D4-branes
connecting between D6-branes and NS5'-brane.  

\begin{figure}[ht]
   \epsfxsize=4.0in 
\centerline{\epsffile{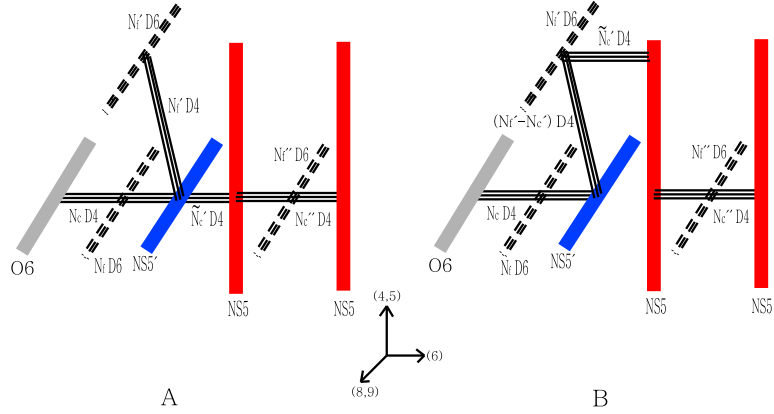}}
   \caption[FIG. \arabic{figure}.]{ 
The ${\cal N}=1$ supersymmetric magnetic brane configuration with
$SO(N_c) \times SU(\widetilde{N}_c'=N_f' +N_c +N_c''-N_c') 
\times SU(N_c'')$ gauge group
with fundamentals $Q(q')[Q'']$ and
$(\widetilde{q}')[\widetilde{Q}'']$ 
for each gauge group and bifundamentals $f(G)$ and
$\widetilde{f}(\widetilde{G})$,  
and gauge singlets in Figure 13A. In
Figure 13B, the nonsupersymmetric minimal energy brane configuration
with the same gauge group and matter contents above 
for massless  $Q(Q'')$ and
$(\widetilde{Q}'')$ is given. 
}
\end{figure}

In the dual theory, 
the coefficient of the beta function of the first gauge group factor
 is
$
b_{SO(N_c)}^{mag} 
= 3(N_c-2)-2N_f-2\widetilde{N}_c'-2N_f'-2N_c 
$
and 
the coefficient of the beta function 
of the second gauge group factor is
$
b_{SU(\widetilde{N}_c')}^{mag}
= 3\widetilde{N}_c'-N_f'-N_c-N_c''
$
and 
the coefficient of the beta function 
of the third gauge group factor is
$
b_{SU(N_c'')}^{mag}
= 3N_c''-N_f''-\widetilde{N}_c'$.

Then the gauge group and matter contents we consider 
are summarized as follows:
\bea
 & \mbox{gauge group}:& \;\;\;\;\;   SO(N_c) \times SU(\widetilde{N}_c') \times
 SU(N_c'')  \nonu
\\
\mbox{matter}: 
 &Q_f & \;\;\;\;\;\;\;\;\; 
\;\;\;\;\;\;\;\;\;\;\;\;\; {(\bf \Box, 1, 1)}
\;\;\;\;\;\;\;\;\;\;\;\; (f=1,  \cdots, 2N_f) 
\nonu \\
 &q'_{f'} \oplus \widetilde{q}'_{\widetilde{f}'}& \;\;\;\;\;\;\;\;\;\;
\;\; {(\bf 1, \Box, 1) \oplus ( 1, \overline{\Box}, 1)}
\;\;\;\;\; (f', \widetilde{f}' =1,  \cdots, N_f') 
\nonu \\
 &Q''_{f''} \oplus \widetilde{Q}''_{\widetilde{f}''}& \;\;\;\;\;\;\;\;\;\; 
\;\; {(\bf 1, 1, \Box) \oplus ( 1, 1, \overline{\Box})} 
\;\;\;\;\; (f'', \widetilde{f}'' =1,  \cdots, N_f'')
\nonu \\
 &f \oplus \widetilde{f}& \;\;\;\;\;\;\;\;\; 
\;\; {(\bf \Box, \overline{\Box}, 1) \oplus (\Box, \Box, 1)} 
\nonu \\
 &G \oplus \widetilde{G}& \;\;\;\;\;\;\;\;\; 
\;\; {(\bf 1, \Box, \overline{\Box}) \oplus ( 1, \overline{\Box}, \Box)} 
\nonu \\
& (X_{\widetilde{n}'} \equiv) F Q' 
\oplus \widetilde{F} \widetilde{Q}' (\equiv 
\widetilde{X}_{n'}) & 
 \;\;\;\;\;\;\;\;\;\;
\;\; {(\bf \Box, 1, 1) \oplus ( \Box, 1, 1)}
\;\; (n', \widetilde{n}' =1,  \cdots, N_f') 
\nonu \\
&  (M_{f',\widetilde{g}'}' \equiv) Q' \widetilde{Q}' & 
 \;\;\;\;\;\;\;\;\;\;\;\;\;\;\;\;\;\;\;
\;\; {(\bf 1, 1, 1)} \;\;\;\;\;\;\;\;\;\;\; (f', \widetilde{g}' =1,  \cdots, N_f') 
\nonu \\
& (\Phi \equiv) F \widetilde{F} & 
 \;\;\;\;\;\;\;\;\;\;
\;\; {(\bf adj, 1, 1) \oplus ( symm, 1, 1)}
\nonu 
\eea

From the superpotential from \cite{Ahn07-3}
\bea
W_{dual}= \left( M' q' \widetilde{q}'  + m' M' \right) + 
X f q' +
\widetilde{X} \widetilde{q}' \widetilde{f} + \Phi f \widetilde{f} + Q
\Phi Q
\nonu
\eea
one sees that $q' \widetilde{q}'$ has rank 
$\widetilde{N}_c'$ while $m'$ has a
rank $N_f'$.  If the rank $N_f'$ exceeds $\widetilde{N}_c'$, then 
the supersymmetry is broken. 
The classical moduli space of vacua can be obtained from F-term
equations.
Then the solutions can be written as
\bea
<q' >  & = &  \left(
\begin{array}{c}
\sqrt{m'} e^{\phi} {\bf 1}_{\widetilde{N}_c'}  \\
0
\end{array}
\right),  
< \widetilde{q}'> =
 \left(
\begin{array}{cc}
\sqrt{m'} e^{-\phi}  {\bf 1}_{\widetilde{N}_c'}   &
0
\end{array}
\right), 
<M'>  =
 \left(
\begin{array}{cc}
0  & 0 
 \\
0 & M_0'  {\bf 1}_{N_f'-\widetilde{N}_c'} 
\end{array}
\right),
\nonu \\
<f> & = & <\widetilde{f}> = <X> = <\widetilde{X}> = <Q>= 0.
\nonu
\eea
Let us expand around a point on the vacua, as done in
\cite{ISS}. 
Then the remaining relevant terms of superpotential are given by
$
W_{dual}^{rel}  =   M_0' \left( \delta \varphi  
\; \delta \widetilde{\varphi} + m' \right) +
  \delta Z \; \delta \varphi  \; \widetilde{q}_0' 
+ \delta \widetilde{Z} \; q_0'  
\delta \widetilde{\varphi}
$
by following the similar fluctuations 
for the various fields as in \cite{Ahn07}.
Note that there exist also four kinds of terms, 
the vacuum  $<q'>$ multiplied by 
$\delta f \delta X$,  
the vacuum  $<\widetilde{q}'>$ multiplied by $\delta \widetilde{X} 
\delta \widetilde{f}$, 
the vacuum  $<\Phi>$ multiplied by $\delta f 
\delta \widetilde{f}$, and 
the vacuum  $<\Phi>$ multiplied by $\delta Q 
\delta Q$. However,
by redefining these, they do not enter the 
contributions for the one loop result, up to quadratic order. 
As done in \cite{Shih}, one gets 
that $m_{M_0'}^2$ will contain $(\log 4 -1) > 0$ implying that these
are stable.

The nonsupersymmetric minimal energy brane configuration Figure 13B
(neglecting  the
$NS5_R$-brane, $N_f''$ D6-branes and $N_c''$ D4-branes)
looks similar to 
the Figure 6 of \cite{Ahn07-3}.

The Riemann surface 
describing a set of NS5-branes 
with D4-branes suspended between them and 
in a background space of $x t = (-1)^{N_f+N_f'+N_f''}
v^{2N_f+2N_f''+4} 
(v^2 - m'^2)^{N_f'}$
was found.
Since we are dealing with six NS-branes, the magnetic M5-brane 
configuration in Figure 13 with equal mass for $q$ and $\widetilde{q}'$ and massless 
for $Q(Q'')$ and $\widetilde{Q}''$ 
can be characterized by the following sixth order equation for $t$ 
as follows:
\bea
& & t^6 + \left[v^{N_c''} \right] t^5 + \left[ v^{\widetilde{N}_c'+N_f''} \right] t^4
 + \left[ v^{N_c+2N_f''}\right]  t^3 
\nonu \\
&& + \left[ (-1)^{\widetilde{N}_c'+N_f+N_f'}
v^{\widetilde{N}_c' +3N_f''+2N_f+ 4} 
(v^2 -m'^2)^{N_f'} \right] t^2  + 
\left[ (-1)^{N_c''} v^{N_c'' +4N_f''+4N_f+8}  
(v^2 -m'^2)^{2N_f'} \right] t \nonu \\
&& + \left[(-1)^{N_f+N_f'+N_f''}
v^{12+6N_f+6N_f''}  
 (v^2 -m'^2)^{3N_f'}\right] =0.  
\nonu
\eea

At nonzero string coupling constant, 
the NS-branes bend due to their interactions with the D4-branes and
D6-branes.
Then the behavior of the supersymmetric M5-brane curves can be
summarized 
as follows:

1. $v \rightarrow \infty$ limit implies
\bea
w & \rightarrow & 0, \quad y \sim    v^{N_c''} + \cdots \quad
\mbox{$\overline{NS5_R}$ 
asymptotic region}, \nonu \\
w & \rightarrow  & 0, \quad y \sim    
v^{N_f''+\widetilde{N}_c'-N_c''} + \cdots \quad
\mbox{$\overline{NS5_L}$ asymptotic region},
\nonu \\
w & \rightarrow & 0, \quad y \sim    
v^{-\widetilde{N}_c'+N_f''+N_c''+2N_f'+2N_f+4} + \cdots \quad
\mbox{$NS5_L$ 
asymptotic region}, \nonu \\
w & \rightarrow  & 0, \quad y \sim    
v^{2N_f''+2N_f+2N_f'-N_c''+4} + \cdots \quad
\mbox{$NS5_R$ asymptotic region}.   
\nonu
\eea

2.  $w \rightarrow \infty$ limit implies
\bea
v & \rightarrow &   -m', \quad 
y \sim  w^{-\widetilde{N}_c'+N_c+N_f''}
 +\cdots
\quad \mbox{$\overline{NS5'}$ asymptotic region}, 
\nonu
\\
v & \rightarrow &  +m', \quad  
y \sim w^{\widetilde{N}_c'-N_c+N_f''+2N_f+2N_f'+4}
+\cdots
\quad \mbox{NS5' asymptotic region}. 
\nonu
\eea


\subsection{Magnetic theory with dual for third gauge group}

By moving the $NS5_R$-brane with massive $N_f''$ D6-branes 
to the left all the way past the  
NS5'-brane, one arrives at the Figure 14A.
The linking number of $NS5_R$-brane from Figure 14A
is given by 
$
L_5 = \frac{N_f''}{2} -\widetilde{N}_c''$ and
the linking number of $NS5_R$-brane from Figure 11
is
$
L_5 = -\frac{N_f''}{2} + N_c'' -N_c'$. 
From these two relations, one obtains
the number of colors of dual magnetic theory
\bea
\widetilde{N}_c'' = N_f'' + N_c'-N_c''.
\nonu
\eea

Let us draw this magnetic brane configuration in Figure 14A and recall
that we put
the coincident $N_f''$ D6-branes in the nonzero $v$-direction in the
electric theory and consider massless flavors for $Q$ and $Q'$ by
putting $N_f$ and $N_f'$ D6-branes at $v=0$.

Now let us recombine $\widetilde{N}_c''$ flavor D4-branes among $N_f''$
flavor 
D4-branes(connecting between D6-branes and NS5'-brane) with the same number of 
color D4-branes(connecting between $NS5_R$-brane and NS5'-brane) and push
them in $+v$ direction from Figure 14A. 
For the flavor D4-branes, we are left with only 
$(N_f''-\widetilde{N}_c'')=N_c''-N_c'$ flavor D4-branes
connecting between D6-branes and NS5'-brane.  

\begin{figure}[ht]
   \epsfxsize=4.0in 
\centerline{\epsffile{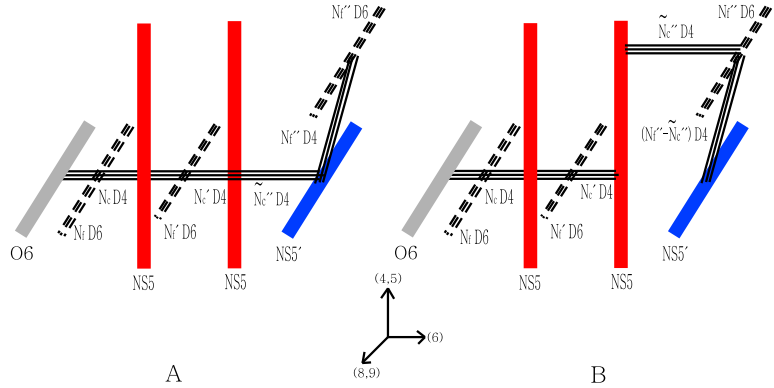}}
   \caption[FIG. \arabic{figure}.]{ 
The ${\cal N}=1$ supersymmetric magnetic brane configuration with
$SO(N_c) \times SU(N_c') \times SU(\widetilde{N}_c''= 
N_f'' + N_c'-N_c'')$ gauge group
with fundamentals $Q(Q')[q'']$ and 
$(\widetilde{Q}')[\widetilde{q}'']$ 
for each gauge group, bifundamentals $F(g)$ and
$\widetilde{F}(\widetilde{g})$, 
and gauge singlets in Figure 14A. In
Figure 14B, the nonsupersymmetric minimal energy brane configuration
with the same gauge group and matter contents above 
for massless  $Q(Q')$ and
$(\widetilde{Q}')$ is given. 
}
\end{figure}

The coefficient of the beta function
of first gauge group factor is
$
b_{SO(N_c)}^{mag} 
= 3(N_c-2)-2N_f-2N_c'=b_{SO(N_c)} 
$
and
the coefficient of the beta function
of second gauge group factor is 
$
b_{SU(N_c')}^{mag}
= 3N_c'-N_f'-N_c-\widetilde{N}_c''-N_f''-N_c'
$
and 
the coefficient of the beta function
of third gauge group factor is 
$
b_{SU(\widetilde{N}_c'')}^{mag}
= 3\widetilde{N}_c''-N_f''-N_c'$.
Since $b_{SU(N_c')}-b_{SU(N_c')}^{mag} > 0$, $SU(N_c')$ is more
asymptotically free than $SU(N_c')^{mag}$.

Then the gauge group and matter contents we consider 
are summarized as follows:
\bea
 & \mbox{gauge group}:& \;\;\;\;\;   SO(N_c) \times SU(N_c') \times
 SU(\widetilde{N}_c'')  \nonu
\\
\mbox{matter}:  
 &Q_f & \;\;\;\;\;\;\;\;\; 
\;\;\;\;\;\;\;\;\;\;\;\;\; {(\bf \Box, 1, 1)}
\;\;\;\;\;\;\;\;\;\;\;\; (f=1,  \cdots, 2N_f) 
\nonu \\
 &Q'_{f'} \oplus \widetilde{Q}'_{\widetilde{f}'}& \;\;\;\;\;\;\;\;\;\;
\;\; {(\bf 1, \Box, 1) \oplus ( 1, \overline{\Box}, 1)}
\;\; (f', \widetilde{f}' =1,  \cdots, N_f') 
\nonu \\
 &q''_{f''} \oplus \widetilde{q}''_{\widetilde{f}''}& \;\;\;\;\;\;\;\;\;\; 
\;\; {(\bf 1, 1, \Box) \oplus ( 1, 1, \overline{\Box})} 
\;\; (f'', \widetilde{f}'' =1,  \cdots, N_f'')
\nonu \\
&F \oplus \widetilde{F}& \;\;\;\;\;\;\;\;\; 
\;\; {(\bf \Box, \overline{\Box}, 1) \oplus (\Box, \Box, 1)} 
\nonu \\
 &g \oplus \widetilde{g}& \;\;\;\;\;\;\;\;\; 
\;\; {(\bf 1, \Box, \overline{\Box}) \oplus ( 1, \overline{\Box}, \Box)} 
\nonu \\
& (X_{n''}' \equiv) G Q'' \oplus \widetilde{G} \widetilde{Q}'' (\equiv 
\widetilde{X}_{\widetilde{n}''}') & 
 \;\;\;\;\;\;\;\;\;\;
\;\; {(\bf 1, \Box, 1) \oplus ( 1, \overline{\Box}, 1)}
\;\; (n'', \widetilde{n}'' =1,  \cdots, N_f'') 
\nonu \\
&  (M''_{f'',\widetilde{g}''} \equiv) Q'' \widetilde{Q}'' & 
 \;\;\;\;\;\;\;\;\;\;\;\;\;\;\;\;\;\;\;
\;\; {(\bf 1, 1, 1)} \;\;\;\;\;\;\;\;\;
\;\; (f'', \widetilde{g}'' =1,  \cdots, N_f'') 
\nonu \\
& (\Phi' \equiv) G \widetilde{G} & 
 \;\;\;\;\;\;\;\;\;\;
\;\; {(\bf 1, adj, 1) \oplus ( 1, 1, 1)}
\nonu
\eea

From the superpotential
\bea
W_{dual}= \left( M'' q'' \widetilde{q}''  + m'' M'' \right) + 
X' g q'' +
\widetilde{X}' \widetilde{q}'' \widetilde{g} + \Phi' g \widetilde{g} 
\nonu
\eea
$q'' \widetilde{q}''$ has rank 
$\widetilde{N}_c''$ while $m''$ has a
rank $N_f''$.  The derivative of the 
superpotential $W_{dual}$ with respect to $M''$ cannot be satisfied 
if the rank $N_f''$ exceeds $\widetilde{N}_c''$ and the supersymmetry is broken.
The classical moduli space of vacua can be obtained from F-term
equations.
Then the solutions can be written as
\bea
<q'' >  & = &  \left(
\begin{array}{c}
\sqrt{m''} e^{\phi} {\bf 1}_{\widetilde{N}_c''}  \\
0
\end{array}
\right),  
< \widetilde{q}''> =
 \left(
\begin{array}{cc}
\sqrt{m''} e^{-\phi}  {\bf 1}_{\widetilde{N}_c''}   &
0
\end{array}
\right), 
<M''>  =
 \left(
\begin{array}{cc}
0  & 0 
 \\
0 & M_0''  {\bf 1}_{N_f''-\widetilde{N}_c''} 
\end{array}
\right),
\nonu \\
<g> & = & <\widetilde{g}> = <X'> = <\widetilde{X}'> = 0.
\nonu
\eea
It 
turns out that states are stable by realizing the mass of 
$m_{M_0''}^2$ positive, by expanding the fields around the vacua.

The nonsupersymmetric minimal energy brane configuration Figure 14B
with a replacement $N_f''$ D6-branes by 
the NS5'-brane(neglecting  the
$NS5_L$-brane, $N_f$ D6-branes, $N_f'$ D6-branes, and $N_c$ D4-branes)
looks similar to 
the Figure 16B of \cite{Ahn07-6}. The position of NS5'-brane
is different from each other.

The Riemann surface 
describing a set of NS5-branes 
with D4-branes suspended between them and 
in a background space of $x t = (-1)^{N_f+N_f'+N_f''} v^{2N_f+2N_f'+4}
(v^2 - m''^2)^{N_f''}$
was found.
Since we are dealing with six NS-branes, the magnetic M5-brane 
configuration in Figure 14 with equal mass for 
$q''$ and $\widetilde{q}''$ and massless 
for $Q$ and $Q'(\widetilde{Q}')$ 
can be characterized by the following sixth order equation for $t$ 
as follows:
\bea
& & t^6 + \left[v^{\widetilde{N}_c''} \right] t^5 + 
\left[ v^{N_c'}(v+m'')^{N_f''} \right] t^4
 + \left[ v^{N_c+N_f'}(v+m'')^{2N_f''}\right]  t^3 
\nonu \\
&& + \left[ (-1)^{N_c'+N_f}
v^{N_c' +2N_f'+2N_f+ 4} 
(v+m'')^{3N_f''} \right] t^2  + 
\left[ (-1)^{\widetilde{N}_c''+N_f'} v^{\widetilde{N}_c'' +4N_f'+4N_f+8}  
(v+m'')^{4N_f''} \right] t \nonu \\
&& + \left[(-1)^{N_f+N_f''}
v^{12+6N_f'+6N_f}  (v -m'')^{N_f''} 
 (v+m'')^{5N_f''}\right] =0.  
\nonu
\eea

At nonzero string coupling constant, 
the NS5-branes bend due to their interactions with the D4-branes and
D6-branes.
Then the behavior of the supersymmetric M5-brane curves can be
summarized 
as follows:

1. $v \rightarrow \infty$ limit implies
\bea
w & \rightarrow & 0, \quad y \sim    v^{N_c'+N_f''-\widetilde{N}_c''} + \cdots \quad
\mbox{$\overline{NS5_R}$ 
asymptotic region}, \nonu \\
w & \rightarrow  & 0, \quad y \sim    
v^{N_f''+N_f'+N_c-N_c'} + \cdots \quad
\mbox{$\overline{NS5_L}$ asymptotic region},
\nonu \\
w & \rightarrow & 0, \quad y \sim    v^{N_c'+N_f''+2N_f+N_f'-N_c+4} + \cdots \quad
\mbox{$NS5_L$ 
asymptotic region}, \nonu \\
w & \rightarrow  & 0, \quad y \sim    
v^{N_f''+2N_f+2N_f'-N_c'+\widetilde{N}_c''+4} + \cdots \quad
\mbox{$NS5_R$ asymptotic region}.   
\nonu
\eea

2.  $w \rightarrow \infty$ limit implies
\bea
v & \rightarrow &   -m'', \quad 
y \sim  w^{\widetilde{N}_c''}
 +\cdots
\quad \mbox{$\overline{NS5'}$ asymptotic region}, 
\nonu
\\
v & \rightarrow &  +m'', \quad  
y \sim w^{-\widetilde{N}_c''+2N_f''+2N_f'+2N_f+4}
+\cdots
\quad \mbox{NS5' asymptotic region}. 
\nonu
\eea


\subsection{Magnetic theories 
for the multiple product gauge groups}

Now one can generalize the method for the triple product gauge groups
to the finite $n$-multiple product gauge groups characterized by 
\bea
SO(N_{c,1}) \times SU(N_{c,2}) \cdots
\times SU(N_{c,n})
\nonu
\eea
with the matter, 
the $(n-1)$ bifundametals $({\bf \Box_1, \overline{\Box}_2, 1, \cdots,  1_n})$,
$\cdots$, and $({\bf 1_1, \cdots, 1, \Box_{n-1}, \overline{\Box}_{n}})$, their
complex conjugate $(n-1)$ fields $({\bf \Box_1, 
\Box_2, 1, \cdots, 1_n})$,
$\cdots$, 
and $({\bf 1_1, \cdots, 1, \overline{\Box}_{n-1}, \Box_n})$, linking the
gauge groups together,
$(n-1)$-fundamentals $({\bf 1_1, \Box_2, \cdots, 1_n})$, $\cdots$, and 
$({\bf 1_1, \cdots,  1, \Box_n})$, and $(n-1)$-antifundamentals  
$({\bf 1_1, \overline{\Box}_2, \cdots, 1_n})$, $\cdots$, and 
$({\bf 1_1, \cdots,  1, \overline{\Box}_n})$ and one-vector in the
representation 
$({\bf \Box_1, \cdots, 1_n})$.
Then the mass-deformed superpotential can be written as
$
W_{elec} = m_1 Q_1 Q_1 + \sum_{i=2}^n m_i Q_i \widetilde{Q}_i$. 
The brane configuration can be constructed from Figure 11 by adding
$(n-3)$ NS-branes, $(n-3)$ sets of D6-branes and $(n-3)$
sets of D4-branes  to the right of $NS5_R$-brane(and its mirrors)
leading to the fact that 
any two neighboring NS-branes should be perpendicular to each other. 

There exist $(2n-3)$ magnetic theories and they can be classified as follows.

$\bullet$ When the dual magnetic gauge group is $SO(\widetilde{N}_{c,1})$

There is no nonsupersymmetric meta-stable brane configuration.

$\bullet$ When the dual magnetic gauge group is $SU(\widetilde{N}_{c,2})$

When the Seiberg dual is taken for the second gauge group factor
by
assuming that $\Lambda_2 >> \Lambda_j$ where $j=1, 3, \cdots, n$, 
one follows the procedure given in the subsection 4.3.
The gauge group is given by
\bea
SO(N_{c,1}) \times 
SU(\widetilde{N}_{c,2} \equiv N_{f,2}+N_{c,1}+N_{c,3}-N_{c,2}) 
\times SU(N_{c,3}) \times \cdots \times SU(N_{c,n}).
\nonu
\eea
The corresponding brane configuration can be 
obtained similarly and 
the extra $(n-3)$ NS-branes, $(n-3)$ sets of D6-branes and $(n-3)$
sets of D4-branes  are present at the right hand side of the $NS5_R$-brane
of Figure 12.
The magnetic superpotential can be written as
\bea
W_{dual} = \left(M_2 q_2 \widetilde{q}_2 + 
g_2 \widetilde{X}_3 \widetilde{q}_2 + 
\widetilde{g}_2 q_2
X_3 + \Phi_3 g_2 \widetilde{g}_2 \right) + m_2 M_2.
\nonu
\eea
By computing the contribution for the one loop as in the subsection
4.3, 
the vacua are stable and the asymptotic behavior of $2n$ NS-branes
can be obtained. 

When the Seiberg dual is taken for the second gauge group factor with
different brane motion
by
assuming that $\Lambda_2 >> \Lambda_j$ where $j=1, 3, \cdots, n$, 
one follows the procedure given in the subsection 4.4.
The gauge group is given by
\bea
SO(N_{c,1}) \times 
SU(\widetilde{N}_{c,2} \equiv N_{f,2}+N_{c,3}+N_{c,1}-N_{c,2}) 
\times SU(N_{c,3}) \times \cdots \times SU(N_{c,n}).
\nonu
\eea
The corresponding brane configuration can be 
obtained similarly and 
the extra $(n-3)$ NS-branes, $(n-3)$ sets of D6-branes and $(n-3)$
sets of D4-branes  are present at the right hand side of the $NS5_R$-brane
of Figure 13.
The magnetic superpotential can be written as
\bea
W_{dual} = \left(M_2 q_2 \widetilde{q}_2 + f_1 X_1 q_2 + 
\widetilde{f}_1 \widetilde{q}_2
\widetilde{X}_1 + \Phi_1 f_1 \widetilde{f}_1 + Q \Phi_1 Q\right) + m_2 M_2.
\nonu
\eea

$\bullet$ When the dual magnetic gauge group is $SU(\widetilde{N}_{c,i})$ where
$ 3 \leq i \leq n-1$

When the Seiberg dual is taken for the middle gauge group factor
by
assuming that $\Lambda_i >> \Lambda_j$ where $j=1,2, \cdots, i-1, i+1,
\cdots, n$, 
one follows the procedure given in the subsection 2.3 of \cite{Ahn07-8}.
The gauge group is given by
\bea
SO(N_{c,1}) \times \cdots  \times 
SU(\widetilde{N}_{c,i} \equiv N_{f,i}+N_{c,i+1}+N_{c,i-1}-N_{c,i}) 
 \times \cdots \times SU(N_{c,n}).
\nonu
\eea
The corresponding brane configuration can be 
obtained similarly and 
the extra $(i-2)$ NS-branes, $(i-2)$ sets of D6-branes and $(i-2)$
sets of D4-branes  
are present between O6-plane and the NS5'-brane in Figure 12
and the extra $(n-i-1)$ NS-branes, $(n-i-1)$ sets of D6-branes and $(n-i-1)$
sets of D4-branes  are present at the right hand side of the $NS5_R$-brane
of Figure 12.
The magnetic superpotential can be written as
\bea
W_{dual} = \left(M_{i} q_i \widetilde{q}_i + 
g_{i} \widetilde{X}_{i+1} \widetilde{q}_i + 
\widetilde{g}_{i} q_i
X_{i+1} + \Phi_{i+1} g_{i} \widetilde{g}_{i} \right) + m_i M_{i}.
\nonu
\eea

When the Seiberg dual is taken for the middle gauge group factor with
different brane motion
by
assuming that $\Lambda_i >> \Lambda_j$ where $j=1,2, \cdots, i-1, i+1,
\cdots, n$, 
one follows the procedure given in the subsection 2.4 of \cite{Ahn07-8}.
The gauge group is given by
\bea
SO(N_{c,1}) \times \cdots \times 
SU(\widetilde{N}_{c,i} \equiv N_{f,i}+N_{c,i+1}+N_{c,i-1}-N_{c,i}) 
\times \cdots \times SU(N_{c,n}).
\nonu
\eea
The corresponding brane configuration can be 
obtained similarly and 
the extra $(i-2)$ NS-branes, $(i-2)$ sets of D6-branes and $(i-2)$
sets of D4-branes  
are present between O6-plane and the NS5'-brane in Figure 13
and the extra $(n-i-1)$ NS-branes, $(n-i-1)$ sets of D6-branes and $(n-i-1)$
sets of D4-branes  are present at the right hand side of the $NS5_R$-brane
of Figure 13.
The magnetic superpotential can be written as
\bea
W_{dual} = \left(M_{i} q_i \widetilde{q}_i + f_{i-1} X_{i-1} q_i + 
\widetilde{f}_{i-1} \widetilde{q}_i
\widetilde{X}_{i-1} + \Phi_{i-1} f_{i-1} \widetilde{f}_{i-1} \right) + m_i M_{i}.
\nonu
\eea

$\bullet$ When the dual magnetic gauge group is $SU(\widetilde{N}_{c,n})$

When the Seiberg dual is taken for the last gauge group factor by
assuming that $\Lambda_n >> \Lambda_i$ where $i=1,2, \cdots, (n-1)$, 
one follows the procedure given in the subsection 4.5.
The gauge group is given by
\bea
SO(N_{c,1}) \times \cdots \times 
SU(N_{c,n-1}) \times SU(\widetilde{N}_{c,n} \equiv N_{f,n} +N_{c,n-1}-N_{c,n}).
\nonu
\eea
The corresponding brane configuration can be 
obtained similarly and 
the extra $(n-3)$ NS-branes, $(n-3)$ sets of D6-branes and $(n-3)$
sets of D4-branes  
are present between O6-plane and the $NS5_L$-brane
of Figure 14.
The magnetic superpotential can be written as
\bea
W_{dual} = \left(M_n q_n \widetilde{q}_n + g_{n-1} X_{n-1} q_n + 
\widetilde{g}_{n-1} \widetilde{q}_n
\widetilde{X}_{n-1} + \Phi_{n-1} g_{n-1} \widetilde{g}_{n-1} \right) + m_n M_n.
\nonu
\eea

\section{More meta-stable brane configurations
of other multiple
product gauge theories }

After we describe the electric brane configuration, 
we present the three magnetic brane configurations, and then the
nonsupersymmetric meta-stable brane configurations are  found.
The case of multiple product gauge groups is also discussed.

\subsection{Electric theory}

Let us describe the gauge theory with triple product gauge groups 
$Sp(N_c) \times SU(N_c') \times SU(N_c'')$.
The matter contents 
are 

$\bullet$
$2N_f$-chiral multiplets $Q$ are  in the
representation $({\bf 2N_c, 1, 1
})$

$\bullet$
$N_f'$-chiral multiplets $Q'$ are  in the
representation $({\bf 1, N_c', 1})$, and 
$N_f'$-chiral multiplets $\widetilde{Q}'$ are in  
the representation $({\bf 1,\overline{N_c'}, 1})$

$\bullet$
$N_f''$-chiral multiplets $Q''$ are  in the
representation $({\bf 1, 1, N_c''
})$, and 
$N_f''$-chiral multiplets $\widetilde{Q}''$ are in  
the representation $({\bf 1, 1, \overline{N_c''}})$

$\bullet$
The flavor-singlet field $F$ is 
in the bifundamental representation $({\bf 2N_c, \overline{N_c'}, 1 })$, 
and its conjugate field $\widetilde{F}$
 is 
in the bifundamental representation $({\bf 2N_c, N_c', 1})$

$\bullet$
The flavor-singlet field $G$ is 
in the bifundamental representation $({\bf 1, N_c', \overline{N_c''} })$, 
and its conjugate field $\widetilde{G}$
 is 
in the bifundamental representation $({\bf 1, \overline{N_c'}, N_c''})$

If we put to $Q'',
\widetilde{Q}'', G$, and $\widetilde{G}$ zero, then 
this becomes the product gauge group theory with fundamentals and
bifundamentals
\cite{LO,Ahn07-3}. On the other hand, if we ignore $Q, F$, and
$\widetilde{F}$, then this theory is given by \cite{BH}.


The coefficient of the beta function
of the first gauge group factor is
$
b_{Sp(N_c)} = 3(N_c+2)-2N_f-2N_c'
$
and similarly 
the coefficient of the beta function
of the second gauge group factor is
$
b_{SU(N_c')} = 3N_c' -N_f'-N_c-N_c''
$
and finally
the coefficient of the beta function
of the third gauge group factor is
$
b_{SU(N_c'')} = 3N_c'' -N_f''-N_c'$.

From the electric superpotential
\bea
W_{elec} & = & 
\left( \mu A^2 + Q A \widetilde{Q} +  \widetilde{F} A F + A_a^2 + 
Q A_a \widetilde{Q} + \widetilde{F} A_a F    
 +
\mu' A'^2 +  Q' A' \widetilde{Q}' + \widetilde{F} A' F \right. \nonu \\
&+& \left.  
\widetilde{G} A' G +
 \mu'' A''^2 +  Q'' A'' \widetilde{Q}'' + \widetilde{G} A'' G \right)
+ m Q Q + m' Q' \widetilde{Q}' + m'' Q'' \widetilde{Q}''
\nonu
\eea
one  integrates out  the adjoint fields 
$A$ for $Sp(N_c)$,  $A'$ for $SU(N_c')$ and $A''$ for $SU(N_c'')$ 
and the antisymmetric field $A_a$ for $Sp(N_c)$ 
and taking $\mu, \mu'$ and $\mu''$ to infinity limit which is
equivalent to take any two NS-branes be perpendicular to each other,
the mass-deformed electric superpotential becomes 
$
W_{elec}  = 
m Q Q + m' Q' \widetilde{Q}' + m'' Q'' \widetilde{Q}''$.

The type IIA brane configuration for this mass-deformed theory 
can be described by as follows. 
The $2N_c$-color 
D4-branes (01236) are suspended between the $NS5_L'$-brane (012389) and
its mirror $\overline{NS5_L'}$-brane 
together with $N_f$ D6-branes (0123789) 
which have nonzero $v$ direction.
The NS5-brane 
is located at the right hand side of
the $NS5_L'$-brane along the positive $x^6$ direction and 
there exist $N_c'$-color D4-branes
suspended 
between them, with  $N_f'$ D6-branes which have nonzero $v$ direction. 
Moreover, 
the $NS5_R'$-brane 
is located at the right hand side of
the NS5-brane along the positive $x^6$ direction and there 
exist $N_c''$-color D4-branes
suspended 
between them, with  $N_f''$ D6-branes which have nonzero $v$ direction.
There exists an orientifold 6-plane (0123789) at the origin $x^6=0$
and it acts as $(x^4, x^5, x^6) \rightarrow (-x^4, -x^5, -x^6)$. 
Then the mirrors of above branes appear in 
the negative $x^6$ region and are denoted by bar on the corresponding branes.
From the left to the right, there are $\overline{NS5_R'}$-,
$\overline{NS5}$-, 
$\overline{NS5_L'}$-,
$NS5_L'$-, $NS5$-, and $NS5_R'$-branes.

We summarize the ${\cal N}=1$ supersymmetric electric brane
configuration in type IIA string theory as follows:

$\bullet$ Two 
NS5-branes in (012345) directions. 

$\bullet$ Four
NS5'-branes in (012389) directions.

$\bullet$ Two sets of
$N_c(N_c')[N_c'']$-color D4-branes in (01236) directions. 
  
$\bullet$ Two sets of
$N_f(N_f')[N_f'']$ D6-branes in (0123789) directions. 

$\bullet$
$O6^{-}$-plane in (0123789) directions with $x^6=0$

Now we draw this electric brane configuration in Figure 15 and we put
the coincident $N_f(N_f')[N_f'']$ D6-branes with positive $x^6$ in 
the nonzero $v$ direction in general. 
This brane configuration can be obtained from the brane configuration
of \cite{LO,Ahn07-3} by adding the two outer NS5'-branes(i.e., 
$\overline{NS5_R'}$-brane and $NS5_R'$-brane), two sets of $N_c''$ D4-branes
and two sets of $N_f''$ D6-branes or from the one of \cite{BH}
with the gauge theory of triple product gauge groups
by adding O6-plane and the extra NS-branes, D4-branes and D6-branes.
The brane configuration for single gauge $Sp(N_c)$ theory was
presented in \cite{CSST}. 
Then the mirrors with 
negative $x^6$ can be constructed by using the action of O6-plane and
are located at the positions by changing (456) directions of original
branes with minus signs.

\begin{figure}[ht]
   \epsfxsize=3.0in 
\centerline{\epsffile{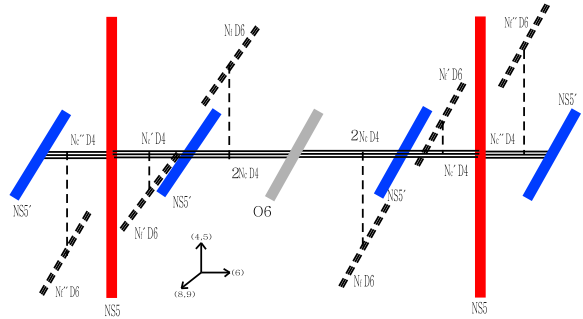}}
   \caption[FIG. \arabic{figure}.]{ 
The ${\cal N}=1$ supersymmetric electric brane configuration with
$Sp(N_c) \times SU(N_c') \times SU(N_c'')$ gauge group with
fundamentals $Q(Q')[Q'']$ and
$(\widetilde{Q}')[\widetilde{Q}'']$ 
for each gauge group and bifundamentals $F(G)$,  
$\widetilde{F}(\widetilde{G})$.
 The $O6^{-}$-plane is located at the
origin $x^6=0$.
The  two NS5'-branes with positive $x^6$ coordinates
are denoted by $NS5_{L,R}'$-branes. 
}
\end{figure}

\subsection{Magnetic theory with dual for first gauge group}

In this case, there is no extra NS-brane which should be present in
order to construct the recombination of flavor D4-branes and splitting
procedure for meta-stable brane configuration. 
Although the magnetic dual theory is present, there is no
nonsupersymmetric meta-stable
brane configuration.

\subsection{Magnetic theory with dual for second gauge group}

By moving the NS5-brane in Figure 15 with massive $N_f'$ D6-branes
to the left all the way past the  
$NS5'_L$-brane, one arrives at the Figure 16A.
The linking number of NS5-brane from Figure 16A
is 
$
L_5 = -\frac{N_f'}{2} +\widetilde{N}_c'-2N_c
$ 
while
the linking number of NS5-brane from Figure 15
is
$
L_5 = \frac{N_f'}{2} + N_c'' -N_c'$. 
From these two relations, one obtains
the number of colors of dual magnetic theory
$
\widetilde{N}_c' = N_f' + N_c''+ 2N_c-N_c'$.

Let us draw this magnetic brane configuration in Figure 16A and recall
that we put
the coincident $N_f'$ D6-branes in the nonzero $v$-direction in the
electric theory and consider massless flavors for $Q$ and $Q''$ by
putting $N_f$ and $N_f''$ D6-branes at $v=0$.
If we ignore  
$NS5_R'$-brane, $N_f''$ D6-branes
and $N_c''$ D4-branes(detaching these
branes from Figure 16A), 
then this brane configuration 
looks similar  to the Figure 7 of \cite{Ahn07-3} for
the standard ${\cal N}=1$ magnetic gauge theory 
$Sp(N_c) \times SU(\widetilde{N}_c'=N_f'+2N_c-N_c')$ with fundamentals, 
bifundamentals, and singlets.

Now let us recombine $\widetilde{N}_c'$ flavor D4-branes among $N_f'$
flavor 
D4-branes(connecting between D6-branes and $NS5_L'$-brane) with the same number of 
color D4-branes(connecting between NS5-brane and $NS5_L'$-brane) and push
them in $+v$ direction from Figure 16A. 
For the flavor D4-branes, we are left with only 
$(N_f'-\widetilde{N}_c')=N_c'-N_c''-N_c$ flavor D4-branes
connecting between D6-branes and $NS5_L'$-brane. 

\begin{figure}[ht]
   \epsfxsize=4.0in 
\centerline{\epsffile{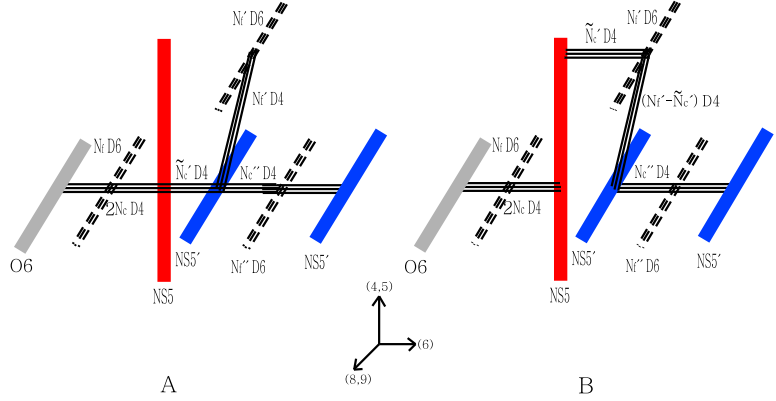}}
   \caption[FIG. \arabic{figure}.]{ 
The ${\cal N}=1$ supersymmetric magnetic brane configuration with
$Sp(N_c) \times SU(\widetilde{N}_c'=N_f' +2N_c +N_c'' -N_c') 
\times SU(N_c'')$ gauge group
with fundamentals $Q(q')[Q'']$ and
$(\widetilde{q}')[\widetilde{Q}'']$ 
for each gauge group and bifundamentals $F(g)$ and
$\widetilde{F}(\widetilde{g})$,   
and gauge singlets in Figure 16A. In
Figure 16B, the nonsupersymmetric minimal energy brane configuration
with the same gauge group and matter contents above 
for massless  $Q(Q'')$ and
$(\widetilde{Q}'')$ is given. 
}
\end{figure}

The coefficient of the beta function
of first gauge group factor  is
$
b_{Sp(N_c)}^{mag} 
= 3(N_c+2)-2N_f-2\widetilde{N}_c' 
$
and 
the coefficient of the beta function
of second gauge group factor is
$
b_{SU(\widetilde{N}_c')}^{mag}
= 3\widetilde{N}_c'-N_f'-N_c-N_c''
$
and 
the coefficient of the beta function
of third gauge group factor is
$
b_{SU(N_c'')}^{mag}
= 3N_c''-N_f''-\widetilde{N}_c'-N_f'-N_c''$.

From the superpotential \cite{Ahn07-3}
$
W_{dual}= \left( M' q' \widetilde{q}'  + m' M' \right) + 
X'' \widetilde{g} q' +
\widetilde{X}'' \widetilde{q}' g + \Phi'' g \widetilde{g}$
one sees that $q' \widetilde{q}'$ has rank 
$\widetilde{N}_c'$ while $m'$ has a
rank $N_f'$.  If the rank $N_f'$ exceeds $\widetilde{N}_c'$, then 
the supersymmetry is broken. 
The classical moduli space of vacua can be obtained from F-term
equations.
Then the solutions can be written as
\bea
<q' >  & = &  \left(
\begin{array}{c}
\sqrt{m'} e^{\phi} {\bf 1}_{\widetilde{N}_c'}  \\
0
\end{array}
\right),  
< \widetilde{q}'> =
 \left(
\begin{array}{cc}
\sqrt{m'} e^{-\phi}  {\bf 1}_{\widetilde{N}_c'}   &
0
\end{array}
\right), 
<M'>  =
 \left(
\begin{array}{cc}
0  & 0 
 \\
0 & M_0'  {\bf 1}_{N_f'-\widetilde{N}_c'} 
\end{array}
\right),
\nonu \\
<g> & = & <\widetilde{g}> = <X''> = <\widetilde{X}''> = 0.
\nonu
\eea
By expanding the fields around the vacua and it 
turns out that states are stable by realizing the mass of 
$m_{M_0'}^2$ positive.

Then the gauge group and matter contents we consider 
are summarized as follows:
\bea
 & \mbox{gauge group}:& \;\;\;\;\;   Sp(N_c) \times SU(\widetilde{N}_c') \times
 SU(N_c'')  \nonu
\\
\mbox{matter}: 
 &Q_f & \;\;\;\;\;\;\;\;\; 
\;\;\;\;\;\;\;\;\;\;\;\;\; {(\bf \Box, 1, 1)}
\;\;\;\;\;\;\;\;\;\;\;\; (f=1,  \cdots, 2N_f) 
\nonu \\
 &q'_{f'} \oplus \widetilde{q}'_{\widetilde{f}'}& \;\;\;\;\;\;\;\;\;\;
\;\; {(\bf 1, \Box, 1) \oplus ( 1, \overline{\Box}, 1)}
\;\;\;\;\; (f', \widetilde{f}' =1,  \cdots, N_f') 
\nonu \\
 &Q''_{f''} \oplus \widetilde{Q}''_{\widetilde{f}''}& \;\;\;\;\;\;\;\;\;\; 
\;\; {(\bf 1, 1, \Box) \oplus ( 1, 1, \overline{\Box})} 
\;\;\;\;\; (f'', \widetilde{f}'' =1,  \cdots, N_f'')
\nonu \\
 &F \oplus \widetilde{F}& \;\;\;\;\;\;\;\;\; 
\;\; {(\bf \Box, \overline{\Box}, 1) \oplus (\Box, \Box, 1)} 
\nonu \\
 &g \oplus \widetilde{g}& \;\;\;\;\;\;\;\;\; 
\;\; {(\bf 1, \Box, \overline{\Box}) \oplus ( 1, \overline{\Box}, \Box)} 
\nonu \\
& (X_{n'}'' \equiv) \widetilde{G} Q' \oplus G \widetilde{Q}' (\equiv 
\widetilde{X}_{\widetilde{n}'}'') & 
 \;\;\;\;\;\;\;\;\;\;
\;\; {(\bf 1, 1, \Box) \oplus ( 1, 1, \overline{\Box})}
\;\; (n', \widetilde{n}' =1,  \cdots, N_f') 
\nonu \\
&  (M_{f',\widetilde{g}'}' \equiv) Q' \widetilde{Q}' & 
 \;\;\;\;\;\;\;\;\;\;\;\;\;\;\;\;\;\;\;
\;\; {(\bf 1, 1, 1)} \;\;\;\;\;\;\;\;\;
\;\; (f', \widetilde{g}' =1,  \cdots, N_f') 
\nonu \\
& (\Phi'' \equiv) G \widetilde{G} & 
 \;\;\;\;\;\;\;\;\;\;
\;\; {(\bf 1, 1, adj) \oplus ( 1, 1, 1)}
\nonu 
\eea

The nonsupersymmetric minimal energy brane configuration Figure 16B
with a replacement $N_f'$ D6-branes by 
the NS5'-brane(neglecting  the
$NS5_R'$-brane, $N_f''$ D6-branes, $N_c''$ D4-branes and $N_f$ D6-branes)
leads to 
the Figure 12B of \cite{Ahn07-6}.

At nonzero string coupling constant, 
the NS5-branes bend due to their interactions with the D4-branes and
D6-branes.
Then the behavior of the supersymmetric M5-brane curves can be
summarized 
as follows:

1. $v \rightarrow \infty$ limit implies
\bea
w & \rightarrow & 0, \quad y \sim
v^{-\widetilde{N}_c'+2N_c+N_f''+N_f'} 
+ \cdots \quad
\mbox{$\overline{NS5}$ 
asymptotic region}, \nonu \\
w & \rightarrow  & 0, \quad y \sim    
v^{\widetilde{N}_c'-2N_c+N_f''+2N_f+N_f'-4} + \cdots \quad
\mbox{NS5 asymptotic region}.
\nonu 
\eea

2.  $w \rightarrow \infty$ limit implies
\bea
v & \rightarrow & +m', \quad y \sim    w^{-\widetilde{N}_c'+N_f''+2N_f+N_f'+
N_c''-4} + \cdots \quad
\mbox{$NS5_L'$ 
asymptotic region}, \nonu \\
v & \rightarrow  & +m', \quad y \sim    
w^{2N_f''+2N_f+2N_f'-N_c''-4} + \cdots \quad
\mbox{$NS5_R'$ asymptotic region},   
\nonu \\
v & \rightarrow &   -m', \quad 
y \sim  w^{N_f''+N_f'+\widetilde{N}_c'-N_c''}
 +\cdots
\quad \mbox{$\overline{NS5_L'}$ asymptotic region}. 
\nonu
\\
v & \rightarrow &  -m', \quad  
y \sim w^{N_c''}
+\cdots
\quad \mbox{$\overline{NS5_R'}$ asymptotic region}.
\nonu
\eea

\subsection{Magnetic theory with dual for second gauge group}

By moving the $NS5_L'$-brane in Figure 15 with massive $N_f'$ D6-branes
to the right all the way past the  
NS5-brane, one arrives at the Figure 17A.
The linking number of $NS5'_L$-brane from Figure 17A
is 
$
L_5 = \frac{N_f'}{2} -\widetilde{N}_c'+N_c''$ and
the linking number of $NS5'_L$-brane from the  
Figure 15
is
$
L_5 = -\frac{N_f'}{2} + N_c' -2N_c$. 
From these two relations, one obtains
the number of colors of dual magnetic theory
$
\widetilde{N}_c' = N_f' + N_c''+ 2N_c-N_c'$.

Let us draw this magnetic brane configuration in Figure 17A and recall
that we put
the coincident $N_f'$ D6-branes in the nonzero $v$-direction in the
electric theory and consider massless flavors for $Q$ and 
$Q''(\widetilde{Q}'')$ by
putting $N_f$ and $N_f''$ D6-branes at $v=0$.
If we ignore  
$NS5_R'$-brane, $N_f''$ D6-branes
and $N_c''$ D4-branes(detaching these
branes from Figure 17A), 
then this brane configuration 
leads to the Figure 7 of \cite{Ahn07-3} for
the standard ${\cal N}=1$ magnetic gauge theory 
$Sp(N_c) \times SU(\widetilde{N}_c'=N_f'+2N_c-N_c')$ with fundamentals, 
bifundamentals, and singlets.

Now let us recombine $\widetilde{N}_c'$ flavor D4-branes among $N_f'$
flavor 
D4-branes(connecting between D6-branes and NS5-brane) with the same number of 
color D4-branes(connecting between NS5-brane and $NS5_L'$-brane) and push
them in $+v$ direction from Figure 17A. 
For the flavor D4-branes, we are left with only 
$(N_f'-\widetilde{N}_c')=N_c'-N_c''-N_c$ flavor D4-branes
connecting between D6-branes and $NS5_L'$-brane.

\begin{figure}[ht]
   \epsfxsize=4.0in 
\centerline{\epsffile{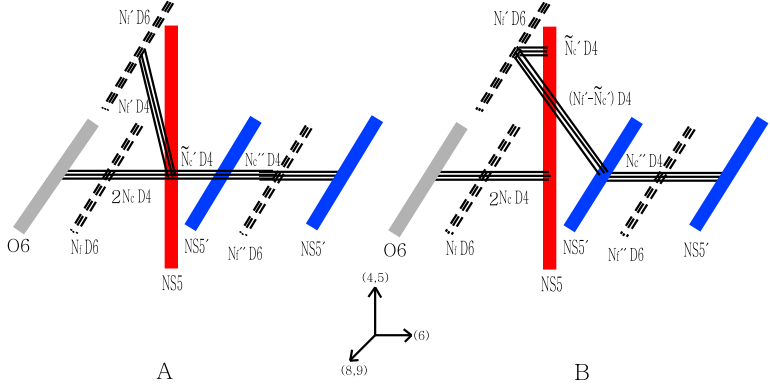}}
   \caption[FIG. \arabic{figure}.]{ 
The ${\cal N}=1$ supersymmetric magnetic brane configuration with
$Sp(N_c) \times SU(\widetilde{N}_c'=N_f' +2N_c+N_c'' -N_c') 
\times SU(N_c'')$ gauge group
with fundamentals $Q(q')[Q'']$ and
$(\widetilde{q}')[\widetilde{Q}'']$ 
for each gauge group and bifundamentals $f(G)$ and
$\widetilde{f}(\widetilde{G})$,  
and gauge singlets in Figure 17A. In
Figure 17B, the nonsupersymmetric minimal energy brane configuration
with the same gauge group and matter contents above 
for massless  $Q(Q'')$ and
$(\widetilde{Q}'')$ is given. 
}
\end{figure}

The coefficient of the beta function
of first gauge group factor is
$
b_{Sp(N_c)}^{mag} 
= 3(N_c+2)-2N_f-2\widetilde{N}_c'-2N_f'-2N_c 
$
and
the coefficient of the beta function
of second gauge group factor is
$
b_{SU(\widetilde{N}_c')}^{mag}
= 3\widetilde{N}_c'-N_f'-N_c-N_c''
$
and 
the coefficient of the beta function
of third gauge group factor is
$
b_{SU(N_c'')}^{mag}
= 3N_c''-N_f''-\widetilde{N}_c'$.

Then the gauge group and matter contents we consider 
are summarized as follows:
\bea
 & \mbox{gauge group}:& \;\;\;\;\;   Sp(N_c) \times SU(\widetilde{N}_c') \times
 SU(N_c'')  \nonu
\\
\mbox{matter}: 
 &Q_f & \;\;\;\;\;\;\;\;\; 
\;\;\;\;\;\;\;\;\;\;\;\;\; {(\bf \Box, 1, 1)}
\;\;\;\;\;\;\;\;\;\;\;\; (f=1,  \cdots, 2N_f) 
\nonu \\
 &q'_{f'} \oplus \widetilde{q}'_{\widetilde{f}'}& \;\;\;\;\;\;\;\;\;\;
\;\; {(\bf 1, \Box, 1) \oplus ( 1, \overline{\Box}, 1)}
\;\;\;\;\; (f', \widetilde{f}' =1,  \cdots, N_f') 
\nonu \\
 &Q''_{f''} \oplus \widetilde{Q}''_{\widetilde{f}''}& \;\;\;\;\;\;\;\;\;\; 
\;\; {(\bf 1, 1, \Box) \oplus ( 1, 1, \overline{\Box})} 
\;\;\;\;\; (f'', \widetilde{f}'' =1,  \cdots, N_f'')
\nonu \\
 &f \oplus \widetilde{f}& \;\;\;\;\;\;\;\;\; 
\;\; {(\bf \Box, \overline{\Box}, 1) \oplus (\Box, \Box, 1)} 
\nonu \\
 &G \oplus \widetilde{G}& \;\;\;\;\;\;\;\;\; 
\;\; {(\bf 1, \Box, \overline{\Box}) \oplus ( 1, \overline{\Box}, \Box)} 
\nonu \\
& (X_{\widetilde{n}'} \equiv) F Q' 
\oplus \widetilde{F} \widetilde{Q}' (\equiv 
\widetilde{X}_{n'}) & 
 \;\;\;\;\;\;\;\;\;\;
\;\; {(\bf \Box, 1, 1) \oplus ( \Box, 1, 1)}
\;\; (n', \widetilde{n}' =1,  \cdots, N_f') 
\nonu \\
&  (M_{f',\widetilde{g}'}' \equiv) Q' \widetilde{Q}' & 
 \;\;\;\;\;\;\;\;\;\;\;\;\;\;\;\;\;\;\;
\;\; {(\bf 1, 1, 1)} \;\;\;\;\;\;\;\;\;
\;\; (f', \widetilde{g}' =1,  \cdots, N_f') 
\nonu \\
& (\Phi \equiv) F \widetilde{F} & 
 \;\;\;\;\;\;\;\;\;\;
\;\; {(\bf adj, 1, 1) \oplus ( asymm, 1, 1)}
\nonu 
\eea

From the superpotential \cite{Ahn07-3}
\bea
W_{dual}= \left( M' q' \widetilde{q}'  + m' M' \right) + 
X f q' +
\widetilde{X} \widetilde{q}' \widetilde{f} + \Phi f \widetilde{f} + Q
\Phi Q
\nonu
\eea
one sees that $q' \widetilde{q}'$ has rank 
$\widetilde{N}_c'$ while $m'$ has a
rank $N_f'$.  If the rank $N_f'$ exceeds $\widetilde{N}_c'$, then 
the supersymmetry is broken. 
The classical moduli space of vacua can be obtained from F-term
equations.
Then the solutions can be written as
\bea
<q' >  & = &  \left(
\begin{array}{c}
\sqrt{m'} e^{\phi} {\bf 1}_{\widetilde{N}_c'}  \\
0
\end{array}
\right),  
< \widetilde{q}'> =
 \left(
\begin{array}{cc}
\sqrt{m'} e^{-\phi}  {\bf 1}_{\widetilde{N}_c'}   &
0
\end{array}
\right), 
<M'>  =
 \left(
\begin{array}{cc}
0  & 0 
 \\
0 & M_0'  {\bf 1}_{N_f'-\widetilde{N}_c'} 
\end{array}
\right),
\nonu \\
<f> & = & <\widetilde{f}> = <X> = <\widetilde{X}> = <Q>= 0.
\nonu
\eea
Let us expand around a point on the vacua, as done in
\cite{ISS}. 
Then the remaining relevant terms of superpotential are given by
$
W_{dual}^{rel}  =   M_0' \left( \delta \varphi  
\; \delta \widetilde{\varphi} + m' \right) +
  \delta Z \; \delta \varphi  \; \widetilde{q}_0' 
+ \delta \widetilde{Z} \; q_0'  
\delta \widetilde{\varphi}
$
by following the similar fluctuations 
for the various fields as in \cite{Ahn07}.
Note that there exist also four kinds of terms, 
the vacuum  $<q'>$ multiplied by 
$\delta f \delta X$,  
the vacuum  $<\widetilde{q}'>$ multiplied by $\delta \widetilde{X} 
\delta \widetilde{f}$, 
the vacuum  $<\Phi>$ multiplied by $\delta f 
\delta \widetilde{f}$, and 
the vacuum  $<\Phi>$ multiplied by $\delta Q 
\delta Q$. However,
by redefining these, they do not enter the 
contributions for the one loop result, up to quadratic order. 
As done in \cite{Shih}, one gets 
that $m_{M_0'}^2$ will contain $(\log 4 -1) > 0$ implying that these
are stable.

The nonsupersymmetric minimal energy brane configuration Figure 17B
(neglecting  the
$NS5_R'$-brane, $N_f''$ D6-branes and $N_c''$ D4-branes)
leads to 
the Figure 7 of \cite{Ahn07-3}.

At nonzero string coupling constant, 
the NS5-branes bend due to their interactions with the D4-branes and
D6-branes.
Then the behavior of the supersymmetric M5-brane curves can be
summarized 
as follows:

1. $v \rightarrow \infty$ limit implies
\bea
w & \rightarrow & 0, \quad y \sim    v^{-\widetilde{N}_c'+2N_c+N_f''} 
+ \cdots \quad
\mbox{$\overline{NS5}$ 
asymptotic region}, \nonu \\
w & \rightarrow  & 0, \quad y \sim    
v^{\widetilde{N}_c'-2N_c+N_f''+2N_f+2N_f'-4} + \cdots \quad
\mbox{NS5 asymptotic region}.
\nonu 
\eea

2.  $w \rightarrow \infty$ limit implies
\bea
v & \rightarrow & +m', \quad y \sim    
w^{-\widetilde{N}_c'+N_f''+N_c''+2N_f'+2N_f-4} + \cdots \quad
\mbox{$NS5_L'$ 
asymptotic region}, \nonu \\
v & \rightarrow  & +m', \quad y \sim    
w^{2N_f''+2N_f+2N_f'-N_c''-4} + \cdots \quad
\mbox{$NS5_R'$ asymptotic region},   
\nonu \\
v & \rightarrow &   -m', \quad 
y \sim  w^{N_f''+\widetilde{N}_c'-N_c''}
 +\cdots
\quad \mbox{$\overline{NS5_L'}$ asymptotic region}, 
\nonu
\\
v & \rightarrow &  -m', \quad  
y \sim w^{N_c''}
+\cdots
\quad \mbox{$\overline{NS5_R'}$ asymptotic region}. 
\nonu
\eea

\subsection{Magnetic theory with dual for third gauge group}

By moving the NS5-brane with massive $N_f''$ D6-branes 
to the right all the way past the  
$NS5'_R$-brane, one arrives at the Figure 18A.
The linking number of NS5-brane from Figure 18A
is given by 
$
L_5 = \frac{N_f''}{2} -\widetilde{N}_c''$ and
the linking number of NS5-brane from Figure 15
is
$
L_5 = -\frac{N_f''}{2} + N_c'' -N_c'$. 
From these two relations, one obtains
the number of colors of dual magnetic theory
$
\widetilde{N}_c'' = N_f'' + N_c'-N_c''$.

Let us draw this magnetic brane configuration in Figure 18A and recall
that we put
the coincident $N_f''$ D6-branes in the nonzero $v$-direction in the
electric theory and consider massless flavors for $Q$ and $Q'$ by
putting $N_f$ and $N_f'$ D6-branes at $v=0$.

Now let us recombine $\widetilde{N}_c''$ flavor D4-branes among $N_f''$
flavor 
D4-branes(connecting between D6-branes and $NS5_R'$-brane) with the same number of 
color D4-branes(connecting between $NS5_R'$-brane and NS5-brane) and push
them in $+v$ direction from Figure 18A. 
For the flavor D4-branes, we are left with only 
$(N_f''-\widetilde{N}_c'')=N_c''-N_c'$ flavor D4-branes
connecting between D6-branes and $NS5_R'$-brane. 

\begin{figure}[ht]
   \epsfxsize=4.0in 
\centerline{\epsffile{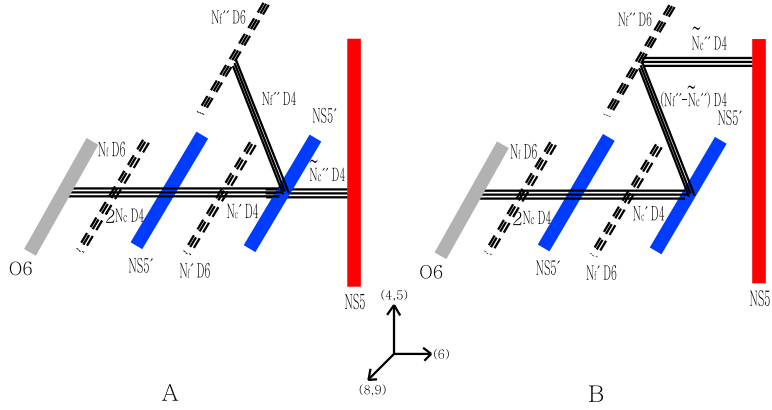}}
   \caption[FIG. \arabic{figure}.]{ 
The ${\cal N}=1$ supersymmetric magnetic brane configuration with
$Sp(N_c) \times SU(N_c') \times SU(\widetilde{N}_c''=N_f'' + 
N_c'-N_c'')$ gauge group
with fundamentals $Q(Q')[q'']$ and 
$(\widetilde{Q}')[\widetilde{q}'']$ 
for each gauge group, bifundamentals $F(g)$ and
$\widetilde{F}(\widetilde{g})$, 
and gauge singlets in Figure 18A. In
Figure 18B, the nonsupersymmetric minimal energy brane configuration
with the same gauge group and matter contents above 
for massless  $Q(Q')$ and
$(\widetilde{Q}')$ is given. 
}
\end{figure}

The coefficient of the beta function
of first gauge group factor is
$
b_{Sp(N_c)}^{mag} 
= 3(N_c+2)-2N_f-2N_c'=b_{Sp(N_c)} 
$
and the coefficient of the beta function
of second gauge group factor is
$
b_{SU(N_c')}^{mag}
= 3N_c'-N_f'-N_c-\widetilde{N}_c''-N_f''-N_c'
$
and 
the coefficient of the beta function
of third gauge group factor is
$
b_{SU(\widetilde{N}_c'')}^{mag}
= 3\widetilde{N}_c''-N_f''-N_c'$.
Since $b_{SU(N_c')}-b_{SU(N_c')}^{mag} > 0$, $SU(N_c')$ is more
asymptotically free than $SU(N_c')^{mag}$.

From the superpotential
$
W_{dual}= \left( M'' q'' \widetilde{q}''  + m'' M'' \right) + 
X' g q'' +
\widetilde{X}' \widetilde{q}'' \widetilde{g} + \Phi' g \widetilde{g}$, 
$q'' \widetilde{q}''$ has rank 
$\widetilde{N}_c''$ while $m''$ has a
rank $N_f''$.  The derivative of the 
superpotential $W_{dual}$ with respect to $M''$ cannot be satisfied 
if the rank $N_f''$ exceeds $\widetilde{N}_c''$ and the supersymmetry is broken. 
The classical moduli space of vacua can be obtained from F-term
equations.
Then the solutions can be written as
\bea
<q'' >  & = &  \left(
\begin{array}{c}
\sqrt{m''} e^{\phi} {\bf 1}_{\widetilde{N}_c''}  \\
0
\end{array}
\right),  
< \widetilde{q}''> =
 \left(
\begin{array}{cc}
\sqrt{m''} e^{-\phi}  {\bf 1}_{\widetilde{N}_c''}   &
0
\end{array}
\right), 
<M''>  =
 \left(
\begin{array}{cc}
0  & 0 
 \\
0 & M_0''  {\bf 1}_{N_f''-\widetilde{N}_c''} 
\end{array}
\right),
\nonu \\
<g> & = & <\widetilde{g}> = <X'> = <\widetilde{X}'> = 0.
\nonu
\eea
It 
turns out that states are stable by realizing the mass of 
$m_{M_0''}^2$ positive, by expanding the fields around the vacua.

Then the gauge group and matter contents we consider 
are summarized as follows:
\bea
 & \mbox{gauge group}:& \;\;\;\;\;   Sp(N_c) \times SU(N_c') \times
 SU(\widetilde{N}_c'')  \nonu
\\
\mbox{matter}:  
 &Q_f & \;\;\;\;\;\;\;\;\; 
\;\;\;\;\;\;\;\;\;\;\;\;\; {(\bf \Box, 1, 1)}
\;\;\;\;\;\;\;\;\;\;\;\; (f=1,  \cdots, 2N_f) 
\nonu \\
 &Q'_{f'} \oplus \widetilde{Q}'_{\widetilde{f}'}& \;\;\;\;\;\;\;\;\;\;
\;\; {(\bf 1, \Box, 1) \oplus ( 1, \overline{\Box}, 1)}
\;\; (f', \widetilde{f}' =1,  \cdots, N_f') 
\nonu \\
 &q''_{f''} \oplus \widetilde{q}''_{\widetilde{f}''}& \;\;\;\;\;\;\;\;\;\; 
\;\; {(\bf 1, 1, \Box) \oplus ( 1, 1, \overline{\Box})} 
\;\; (f'', \widetilde{f}'' =1,  \cdots, N_f'')
\nonu \\
&F \oplus \widetilde{F}& \;\;\;\;\;\;\;\;\; 
\;\; {(\bf \Box, \overline{\Box}, 1) \oplus (\Box, \Box, 1)} 
\nonu \\
 &g \oplus \widetilde{g}& \;\;\;\;\;\;\;\;\; 
\;\; {(\bf 1, \Box, \overline{\Box}) \oplus ( 1, \overline{\Box}, \Box)} 
\nonu \\
& (X_{n''}' \equiv) G Q'' \oplus \widetilde{G} \widetilde{Q}'' (\equiv 
\widetilde{X}_{\widetilde{n}''}') & 
 \;\;\;\;\;\;\;\;\;\;
\;\; {(\bf 1, \Box, 1) \oplus ( 1, \overline{\Box}, 1)}
\;\; (n'', \widetilde{n}'' =1,  \cdots, N_f'') 
\nonu \\
&  (M''_{f'',\widetilde{g}''} \equiv) Q'' \widetilde{Q}'' & 
 \;\;\;\;\;\;\;\;\;\;\;\;\;\;\;\;\;\;\;
\;\; {(\bf 1, 1, 1)} \;\;\;\;\;\;\;\;\;
\;\; (f'', \widetilde{g}'' =1,  \cdots, N_f'') 
\nonu \\
& (\Phi' \equiv) G \widetilde{G} & 
 \;\;\;\;\;\;\;\;\;\;
\;\; {(\bf 1, adj, 1) \oplus ( 1, 1, 1)}
\nonu
\eea

The nonsupersymmetric minimal energy brane configuration Figure 18B
with a replacement of $N_f''$ D6-branes with NS5'-brane(neglecting  the
$NS5_L'$-brane, $N_f$ D6-branes, $N_f'$ D6-branes, and $N_c$ D4-branes)
leads to 
the Figure 14B of \cite{Ahn07-6}. 

At nonzero string coupling constant, 
the NS5-branes bend due to their interactions with the D4-branes and
D6-branes.
Then the behavior of the supersymmetric M5-brane curves can be
summarized 
as follows:

1. $v \rightarrow \infty$ limit implies
\bea
w & \rightarrow & 0, \quad y \sim    v^{\widetilde{N}_c''} + \cdots \quad
\mbox{$\overline{NS5}$ 
asymptotic region}, \nonu \\
w & \rightarrow  & 0, \quad y \sim    
v^{-\widetilde{N}_c''+2N_f''+2N_f'+2N_f-4} + \cdots \quad
\mbox{NS5 asymptotic region}.
\nonu 
\eea

2.  $w \rightarrow \infty$ limit implies
\bea
v & \rightarrow & +m'', \quad y \sim    
w^{N_c'+N_f''+2N_f+N_f'-2N_c-4} + \cdots \quad
\mbox{$NS5_L'$ 
asymptotic region}, \nonu \\
v & \rightarrow  & +m'', \quad y \sim    
w^{N_f''+2N_f+2N_f'-N_c'+\widetilde{N}_c''-4} + \cdots \quad
\mbox{$NS5_R'$ asymptotic region},   
\nonu \\
v & \rightarrow &   -m'', \quad 
y \sim  w^{N_f''+N_f'+2N_c-N_c'}
 +\cdots
\quad \mbox{$\overline{NS5_L'}$ asymptotic region}, 
\nonu
\\
v & \rightarrow &  -m'', \quad  
y \sim w^{N_c'+N_f''-\widetilde{N}_c''}
+\cdots
\quad \mbox{$\overline{NS5_R'}$ asymptotic region}. 
\nonu
\eea

\subsection{Magnetic theories 
for the multiple product gauge groups}

Now one can generalize the method for the triple product gauge groups
to the finite $n$-multiple product gauge groups characterized by 
\bea
Sp(N_{c,1}) \times SU(N_{c,2}) \cdots
\times SU(N_{c,n})
\nonu
\eea
with the matter, 
the $(n-1)$ bifundametals $({\bf \Box_1, \overline{\Box}_2, 1, \cdots,  1_n})$,
$\cdots$, and $({\bf 1_1, \cdots, 1, \Box_{n-1}, \overline{\Box}_{n}})$, their
complex conjugate $(n-1)$ fields $({\bf \Box_1, 
\Box_2, 1, \cdots, 1_n})$,
$\cdots$, 
and $({\bf 1_1, \cdots, 1, \overline{\Box}_{n-1}, \Box_n})$, linking the
gauge groups together,
$(n-1)$-fundamentals $({\bf 1_1, \Box_2, \cdots, 1_n})$, $\cdots$, and 
$({\bf 1_1, \cdots,  1, \Box_n})$, and $(n-1)$-antifundamentals  
$({\bf 1_1, \overline{\Box}_2, \cdots, 1_n})$, $\cdots$, and 
$({\bf 1_1, \cdots,  1, \overline{\Box}_n})$ and one fundamental in the 
representation 
 $({\bf \Box_1, \cdots, 1_n})$.
Then the mass-deformed superpotential can be written as
$
W_{elec} = m_1 Q_1 Q_1 + \sum_{i=2}^n m_i Q_i \widetilde{Q}_i$. 
The brane configuration can be constructed from Figure 15 by adding
$(n-3)$ NS-branes, $(n-3)$ sets of D6-branes and $(n-3)$
sets of D4-branes  to the right of $NS5_R'$-brane(and its mirrors)
leading to the fact that 
any two neighboring NS-branes should be perpendicular to each other. 

There exist $(2n-3)$ magnetic theories and they can be classified as follows.

$\bullet$ When the dual magnetic gauge group is $Sp(\widetilde{N}_{c,1})$

There is no nonsupersymmetric meta-stable brane configuration.

$\bullet$ When the dual magnetic gauge group is $SU(\widetilde{N}_{c,2})$

When the Seiberg dual is taken for the second gauge group factor
by
assuming that $\Lambda_2 >> \Lambda_j$ where $j=1, 3, \cdots, n$, 
one follows the procedure given in the subsection 5.3.
The gauge group is given by
\bea
Sp(N_{c,1}) \times 
SU(\widetilde{N}_{c,2} \equiv N_{f,2}+2N_{c,1}+N_{c,3}-N_{c,2}) 
\times SU(N_{c,3}) \times \cdots \times SU(N_{c,n}).
\nonu
\eea
The corresponding brane configuration can be 
obtained similarly and 
the extra $(n-3)$ NS-branes, $(n-3)$ sets of D6-branes and $(n-3)$
sets of D4-branes  are present at the right hand side of the $NS5_R'$-brane
of Figure 16.
The magnetic superpotential can be written as
\bea
W_{dual} = \left(M_2 q_2 \widetilde{q}_2 + 
g_2 \widetilde{X}_3 \widetilde{q}_2 + 
\widetilde{g}_2 q_2
X_3 + \Phi_3 g_2 \widetilde{g}_2 \right) + m_2 M_2.
\nonu
\eea

When the Seiberg dual is taken for the second gauge group factor with
different brane motion
by
assuming that $\Lambda_2 >> \Lambda_j$ where $j=1, 3, \cdots, n$, 
one follows the procedure given in the subsection 5.4.
The gauge group is given by
\bea
Sp(N_{c,1}) \times 
SU(\widetilde{N}_{c,2} \equiv N_{f,2}+N_{c,3}+2N_{c,1}-N_{c,2}) 
\times SU(N_{c,3}) \times \cdots \times SU(N_{c,n}).
\nonu
\eea
The corresponding brane configuration can be 
obtained similarly and 
the extra $(n-i-1)$ NS-branes, $(n-i-1)$ sets of D6-branes and $(n-i-1)$
sets of D4-branes  are present at the right hand side of the $NS5_R'$-brane
of Figure 17.
The magnetic superpotential can be written as
\bea
W_{dual} = \left(M_2 q_2 \widetilde{q}_2 + f_1 X_1 q_2 + 
\widetilde{f}_1 \widetilde{q}_2
\widetilde{X}_1 + \Phi_1 f_1 \widetilde{f}_1 + Q_1 \Phi_1 Q_1
\right) + m_2 M_2.
\nonu
\eea

$\bullet$ When the dual magnetic gauge group is $SU(\widetilde{N}_{c,i})$ where
$ 3 \leq i \leq n-1$

When the Seiberg dual is taken for the middle gauge group factor
by
assuming that $\Lambda_i >> \Lambda_j$ where $j=1,2, \cdots, i-1, i+1,
\cdots, n$, 
one follows the procedure given in the subsection 2.3 of \cite{Ahn07-8}.
The gauge group is given by
\bea
Sp(N_{c,1}) \times \cdots \times 
SU(\widetilde{N}_{c,i} \equiv N_{f,i}+N_{c,i+1}+N_{c,i-1}-N_{c,i}) 
\times \cdots \times SU(N_{c,n}).
\nonu
\eea
The corresponding brane configuration can be 
obtained similarly and 
the extra $(i-2)$ NS-branes, $(i-2)$ sets of D6-branes and $(i-2)$
sets of D4-branes  
are present between O6-plane and the NS5-brane in Figure 16
and the extra $(n-i-1)$ NS-branes, $(n-i-1)$ sets of D6-branes and $(n-i-1)$
sets of D4-branes  are present at the right hand side of the $NS5_R'$-brane
of Figure 16.
The magnetic superpotential can be written as
\bea
W_{dual} = \left(M_{i} q_i \widetilde{q}_i + 
g_{i} \widetilde{X}_{i+1} \widetilde{q}_i + 
\widetilde{g}_{i} q_i
X_{i+1} + \Phi_{i+1} g_{i} \widetilde{g}_{i}  \right) + m_i M_{i}.
\nonu
\eea

When the Seiberg dual is taken for the middle gauge group factor with
different brane motion
by
assuming that $\Lambda_i >> \Lambda_j$ where $j=1,2, \cdots, i-1, i+1,
\cdots, n$, 
one follows the procedure given in the subsection 2.4 of \cite{Ahn07-8}.
The gauge group is given by
\bea
Sp(N_{c,1}) \times \cdots \times 
SU(\widetilde{N}_{c,i} \equiv N_{f,i}+N_{c,i+1}+N_{c,i-1}-N_{c,i}) 
\times \cdots \times SU(N_{c,n}).
\nonu
\eea
The corresponding brane configuration can be 
obtained similarly and 
the extra $(i-2)$ NS-branes, $(i-2)$ sets of D6-branes and $(i-2)$
sets of D4-branes  
are present between O6-plane and the NS5-brane in Figure 17
and the extra $(n-i-1)$ NS-branes, $(n-i-1)$ sets of D6-branes and $(n-i-1)$
sets of D4-branes  are present at the right hand side of the $NS5_R'$-brane
of Figure 17.
The magnetic superpotential can be written as
\bea
W_{dual} = \left(M_{i} q_i \widetilde{q}_i + f_{i-1} X_{i-1} q_i + 
\widetilde{f}_{i-1} \widetilde{q}_i
\widetilde{X}_{i-1} + \Phi_{i-1} f_{i-1} \widetilde{f}_{i-1} \right) + m_i M_{i}.
\nonu
\eea

$\bullet$ When the dual magnetic gauge group is $SU(\widetilde{N}_{c,n})$

When the Seiberg dual is taken for the last gauge group factor by
assuming that $\Lambda_n >> \Lambda_i$ where $i=1,2, \cdots, (n-1)$, 
one follows the procedure given in the subsection 5.5.
The gauge group is given by
\bea
Sp(N_{c,1}) \times \cdots \times 
SU(N_{c,n-1}) \times SU(\widetilde{N}_{c,n} \equiv N_{f,n} +N_{c,n-1}-N_{c,n}).
\nonu
\eea
The corresponding brane configuration can be 
obtained similarly and 
the extra $(n-3)$ NS-branes, $(n-3)$ sets of D6-branes and $(n-3)$
sets of D4-branes  
are present between the O6-plane and the $NS5_L'$-brane
of Figure 18.
The magnetic superpotential can be written as
\bea
W_{dual} = \left(M_n q_n \widetilde{q}_n + g_{n-1} X_{n-1} q_n + 
\widetilde{g}_{n-1} \widetilde{q}_n
\widetilde{X}_{n-1} + \Phi_{n-1} g_{n-1} \widetilde{g}_{n-1} \right) + m_n M_n.
\nonu
\eea

\section{Conclusions and outlook }

The meta-stable brane configurations we have found are summarized by
Figures 2B, 3B, 4B, and 5B for the first gauge theory described in section
2, by  Figures 
7B, 8B, 9B, and 10B for the second gauge theory given in section 3, by
Figures 12B, 13B, and 14B for the third gauge theory in section 4, and 
by Figures 16B, 17B, and 18B for the last gauge theory described in
section 5.

Some observations are found as follows:

$\bullet$ 
The nonsupersymmetric minimal energy brane configuration Figure 2B
with a replacement $N_f$ D6-branes by 
the NS5'-brane(neglecting  the
$NS5_R'$-brane, $N_f''$ D6-branes and $N_c''$ D4-branes 
and $N_f'$ D6-branes)
leads to 
the Figure 5B of \cite{Ahn07-7} with a rotation of NS5'-brane by 
$\frac{\pi}{2}$ angle.
The Figure 3B
with a replacement $N_f'$ D6-branes by 
the NS5'-brane(neglecting  the
$NS5_R'$-brane, $N_f''$ D6-branes and $N_c''$ D4-branes 
and $N_f$ D6-branes)
becomes  
the Figure 2B of \cite{Ahn07-7} with a rotation of NS5'-brane by 
$\frac{\pi}{2}$ angle.
Moreover, the Figure 5B
with a replacement $N_f''$ D6-branes by 
the NS5'-brane(neglecting  the
$NS5_L'$-brane, $N_f$ D6-branes and $N_c$ D4-branes 
and $N_f'$ D6-branes)
turns out to be 
the Figure 4B of \cite{Ahn07-7}.

$\bullet$ 
The nonsupersymmetric minimal energy brane configuration  Figure 7B
with a replacement $N_f''$ D6-branes by 
the NS5'-brane(neglecting  the
$NS5_R$-brane, $N_f''$ D6-branes and $N_c''$ D4-branes 
and $N_f'$ D6-branes)
leads to 
the Figure 7B of \cite{Ahn07-7} with a rotation of NS5-brane by
$\frac{\pi}{2}$
angle and 
the Figure 8B
with a replacement $N_f'$ D6-branes by 
the NS5'-brane(neglecting  the
$NS5_R$-brane, $N_f''$ D6-branes and $N_c''$ D4-branes 
and $N_f$ D6-branes)
reduces to 
the Figure 8B of \cite{Ahn07-7}.

$\bullet$ 
The nonsupersymmetric minimal energy brane configuration Figure 12B
with a replacement $N_f'$ D6-branes by 
the NS5'-brane(neglecting  the
$NS5_R$-brane, $N_f''$ D6-branes, $N_f$ D6-branes, and $N_c''$ D4-branes)
becomes 
the Figure 15B of \cite{Ahn07-6}.

$\bullet$ 
The nonsupersymmetric minimal energy brane configuration Figure 16B
with a replacement $N_f'$ D6-branes by 
the NS5'-brane(neglecting  the
$NS5_R'$-brane, $N_f''$ D6-branes, $N_c''$ D4-branes and $N_f$ D6-branes)
leads to 
the Figure 12B of \cite{Ahn07-6} while
the nonsupersymmetric minimal energy brane configuration Figure 18B
with a replacement of $N_f''$ D6-branes with NS5'-brane(neglecting  the
$NS5_L'$-brane, $N_f$ D6-branes, $N_f'$ D6-branes, 
and $N_c$ D4-branes)
gives rise to 
the Figure 14B of \cite{Ahn07-6}. 

It would be very interesting to find out
how the meta-stable brane configurations from 
type IIA string theory including the present work are related to
some  brane configurations 
found in recent works \cite{DHSV}-\cite{FU} where 
some of them are described in the type IIB string theory.

\vspace{.7cm}

\centerline{\bf Acknowledgments}

This work was supported by grant No.
R01-2006-000-10965-0 from the Basic Research Program of the Korea
Science \& Engineering Foundation.

\end{document}